\documentclass[preprint]{aastex}

\newcommand\be{\begin{equation}}
\newcommand\ee{\end{equation}}
\newcommand\noi{\noindent}
\newcommand\vb {{\bf v}} 
\newcommand\U{{\bf U}}
\newcommand\bb{{\bf b}}
\newcommand\ba{{\bf a}}
\newcommand\B{{\bf B}}
\newcommand\A{{\bf A}}
\newcommand\bS{{\bf S}}
\newcommand\x{{\bf x}}
\newcommand\y{{\bf y}}
\newcommand\w{{\bf w}}
\newcommand\z{{\bf z}}              
\newcommand\bk{{\bf k}}              
\newcommand\s{{\bf s}}              
              
\newcommand\f{{\bf f}}
\newcommand\bnabla{\mbox{\boldmath $\nabla$}}
\newcommand\bxi{\mbox{\boldmath $\xi$}}

\newcommand\smf{\scriptstyle\mathsf}

\def\refnew#1{(\ref{#1})}

\begin{document}

\title{\mbox{SIMULATIONS OF INCOMPRESSIBLE MHD TURBULENCE}}

\author{Jason Maron\altaffilmark{1} and Peter Goldreich\altaffilmark{2}}
\affil{California Institute of Technology, Pasadena, CA 91125, USA}

\altaffiltext{1}{Current Address, Department of Physics, UCLA, 
maron@tapir.caltech.edu}

\altaffiltext{2}{Caltech 150-21, pmg@gps.caltech.edu}

\begin{abstract}

We simulate incompressible, MHD turbulence using a 
pseudo-spectral code. Our major conclusions are as follows.

\noi 1) MHD turbulence is most conveniently described in terms of counter 
propagating shear Alfv\'en and slow waves. Shear Alfv\'en waves control the 
cascade dynamics. Slow waves play a passive role and adopt the spectrum set by 
the shear Alfv\'en waves. Cascades composed entirely of shear Alfv\'en 
waves do not generate a significant measure of slow waves.

\noi 2) MHD turbulence is anisotropic with energy cascading more rapidly along 
$k_\perp$ than along $k_\parallel$, where $k_\perp$ and $k_\parallel$ refer to
wavevector components perpendicular and parallel to the local magnetic field. 
Anisotropy increases with increasing $k_\perp$ such that excited modes are 
confined inside a cone bounded by $k_\parallel\propto k_\perp^\gamma$ where 
$\gamma<1$.  The opening angle of the cone, $\Theta(k_\perp)\propto 
k_\perp^{-(1-\gamma)}$, defines the scale dependent anisotropy. 

\noi 3) The 1D inertial range energy spectrum is well fit by a power law, 
$E(k_\perp)\propto k_\perp^{-\alpha}$, with $\alpha>1$. 

\noi 4) MHD turbulence is generically strong in the sense that the waves which 
comprise it suffer order unity distortions on timescales comparable to their 
periods. Nevertheless, turbulent fluctuations are 
small deep inside the inertial range. Their energy density is less than that of 
the background field by a factor $\Theta^{(\alpha-1)/(1-\gamma)}\ll 1$.  
 
\noi 5) MHD cascades are best understood geometrically. Wave packets suffer 
distortions as they move along magnetic field lines perturbed by counter 
propagating waves. Field lines perturbed by unidirectional waves
map planes perpendicular to the local field into each other. Shear Alfv\'en 
waves are responsible for the mapping's shear and slow waves for its 
dilatation. The amplitude of the former exceeds that of the latter by 
$1/\Theta(k_\perp)$ which accounts 
for dominance of the shear Alfv\'en waves in controlling the cascade dynamics.     

\noi 6) Passive scalars mixed by MHD turbulence adopt the same power spectrum as 
the velocity and magnetic field perturbations. 

\noi 7) Decaying MHD turbulence is unstable to an increase of the imbalance 
between the flux of waves propagating in opposite directions along the magnetic 
field. Forced MHD turbulence displays order unity fluctuations with respect to 
the balanced state if excited  at low $k_\perp$ by $\delta(t)$ correlated 
forcing. It appears to be statistically stable to the unlimited growth of 
imbalance. 

\noi 8) Gradients of the dynamic variables are focused into sheets aligned with 
the magnetic field whose thickness is comparable to the dissipation scale. 
Sheets formed by oppositely directed waves are uncorrelated. We suspect that
these are vortex sheets which the mean magnetic field prevents from rolling up. 

\noi 9) Items (1)-(6) lend support to the model of strong
MHD turbulence put forth by Goldreich \& Sridhar (GS). Results from our 
simulations are also consistent with the GS prediction $\gamma=2/3$, as are
those obtained previously by Cho \& Vishniac. 
The sole notable discrepancy is that 1D energy spectra determined from our 
simulations exhibit $\alpha\approx 3/2$, whereas the GS model predicts $\alpha= 
5/3$. Further investigation is needed to resolve this issue. 

\end{abstract}

%\nonumber
%\setcounter{equation}{0}

\section{INTRODUCTION}

\medskip

Most of the baryonic matter in the universe has such high electrical 
conductivity that magnetic fields diffuse very slowly through it. Thus fluid 
motions and motions of magnetic field lines are closely coupled. Large scale 
motions are generally turbulent, and incompressible MHD is the simplest 
approximation under which these complex coupled motions can be investigated. 

The inertial range of MHD turbulence is an essential ingredient in a variety of
astronomical phenomena. Cosmic rays are scattered by inertial range magnetic 
field fluctuations. This affects both their propagation and their acceleration 
in shock fronts \citep{bib:Blandford,bib:Berezinskii,bib:Chandran}. Reconnection 
of magnetic field lines is an important ingredient of flare activity and dynamo 
action. The rate at which it proceeds seems likely to depend upon the small 
scale structure of magnetic field lines \citep{bib:Parker,bib:Lazarian}. The 
scintillation of small angular diameter radio sources due to scattering by 
electron density fluctuations is almost certainly related to inertial range MHD 
turbulence \citep{bib:Higdon,bib:Rickett}.  

The organization of this paper is as follows. Relevant properties of MHD waves 
are described in \S \ref{sec:background}. In \S \ref{sec:MHDcascades} we 
introduce selected analytical models for the inertial range of MHD turbulence. 
The strategy we follow in designing our simulations is set forth in \S 
\ref{sec:strategy}. Results from these simulations are presented in \S 
\ref{sec:results}. In \S \ref{sec:discussion} we interpret our results and 
compare them with results from prior investigations. Technical details of our 
simulation method are relegated to the appendix. Two subsections merit special
mention.
 
A comparison between our simulations and those of earlier workers is given in \S 
\ref{subsec:comparison}. Here we merely point out that our simulations are most 
closely related to those presented by \citet{bib:Cho}. These authors demonstrate 
that under nearly isotropic forcing, MHD turbulence develops a scale dependent 
anisotropy which increases with increasing wave number in the manner suggested 
by \citet{bib:GSI} and \citet{bib:GSII}. Henceforth we refer to \citet{bib:GSI} 
and \citet{bib:GSII} separately as GSI and GSII, and together as GS. Our
simulations differ from those of \citet{bib:Cho} in that they are excited
anisotropically so that we can study the deep inertial range of MHD turbulence. 

Our most perplexing result, the shallow slope we find for the 1D energy 
spectrum, is discussed in \S \ref{subsec:Scalings}. Unfortunately we cannot
offer a definitive explanation. This will require further investigation. 

\section{BACKGROUND MATERIAL}
\label{sec:background}

\subsection{Basic Equations}

\medskip

\begin{deluxetable}{lllll}
\tabletypesize{}
\tablecaption{Notation \label{tab:notation}}
\tablewidth{0pt}
\tablehead{}
\startdata
$\vb   $   & fluid velocity & \hspace{0.1 cm} & $\B$    & magnetic field \\
$\rho$ & fluid density  & \hspace{0.1 cm} & $p$     & fluid pressure \\
$c$    & passive scalar concentration & \hspace{0.1 cm} & $\nu_v$ & momentum 
diffusivity \\
$\nu_B$ & magnetic diffusivity & \hspace{0.1 cm} & $\nu_c$ &
passive scalar diffusivity\\
\enddata
\end{deluxetable}

\medskip

The equations which govern magnetohydrodynamics (MHD) are written below using
notation defined in Table \ref{tab:notation}.
\begin{equation}
\rho\left(\partial_t \vb    + \vb    \cdot \bnabla \vb\right)
= - \bnabla\left(p + \frac{B^2}{8 \pi}\right) 
+ \frac{1}{4 \pi} \B \cdot \bnabla \B + \rho \nu_v \nabla^2 \vb,
\label{eq:momentum}
\end{equation}
\begin{equation}
\partial_t \B = \bnabla \times ( \vb    \times \B) + \nu_B \nabla^2 \B,
\label{eq:induction}
\end{equation}
\be
\partial_t \rho+\bnabla \cdot (\rho\vb) = 0, 
\label{eq:continuity}
\ee
\be
\bnabla \cdot \B = 0.
\label{eq:divB}
\ee
The concentration of a passive scalar advected by the fluid  evolves 
according to
\begin{equation}
\partial_t c + \bnabla \cdot (c \vb) = \nu_c \nabla^2 c.
\label{eq:passive}
\end{equation}

We simplify equations \refnew{eq:momentum}-\refnew{eq:passive} for applications 
in this paper.\footnote{Further steps are taken in \S 
\ref{subsec:spectralalgorithm} to cast these equations in a form suitable for 
computation.} Incompressibility is assumed throughout, so we set $\rho = 
1$ and define the total pressure $P=p+B^2/8\pi$. The magnetic field is measured 
in velocity units by $\bb \equiv \B/ \sqrt{4 \pi}$. 
Each diffusive term is replaced by a $n$'th order hyperdiffusivity with the
same coefficient $\nu_n$. With these modifications, equations 
\refnew{eq:momentum}-\refnew{eq:passive} transform to 
\begin{equation}
\partial_t \vb    = - \vb    \cdot \bnabla \vb- \bnabla P + \bb \cdot \bnabla 
\bb + \nu_n 
\nabla^{2n} \vb,
\label{eq:mhda}
\end{equation}
\begin{equation}
\partial_t\bb = - \vb    \cdot \bnabla \bb + \bb \cdot \bnabla \vb    + \nu_n 
\nabla^{2n}\bb,
\label{eq:mhdb}
\end{equation}
\begin{equation}
\bnabla \cdot \vb    = 0, 
\label{eq:mhdc}
\end{equation}
\be
\bnabla \cdot \bb = 0.
\label{eq:mhdd}
\ee
\be
\partial_t c+\vb\cdot\bnabla c=\nu_n \nabla^{2n} c. 
\ee

To relate $P$ to $\vb   $ and $\bb$, we take the divergence of equation 
\refnew{eq:mhda} which yields 
\be
\nabla^2 P=\bnabla\bb:\bnabla\bb-\bnabla\vb:\bnabla\vb.
\label{eq:PoissonP}
\ee
Thus
\be
P=\int \frac{d^3x^\prime}{4\pi}
\frac{(\bnabla\vb:\bnabla\vb-\bnabla\bb:\bnabla\bb)}{|\x^\prime-\x|}.
\label{eq:Pvb}
\ee

\subsection{Regimes}

We decompose the magnetic field into a uniform part plus 
fluctuations; 
\be
\bb = \langle\bb\rangle + \Delta\bb.
\ee
The Alfv\'en speed is defined by $v_{A}\hat{\z}= \langle\bb\rangle$; $v_A$ is 
taken to be constant in space and in time as is consistent with flux 
conservation. The energy densities of the mean magnetic field, the velocity 
field, and the magnetic fluctuations are denoted by $E_{\langle\bb\rangle}$, 
$E_{\vb}$, and $E_{\Delta\bb}$. The parameter
\begin{equation} 
\mu = \frac{E_{\vb} + E_{\Delta\bb}}{E_{\langle\bb\rangle}},
\end{equation}
which measures the relative importance of the fluctuations compared to the 
uniform field, determines the character of MHD turbulence. 

MHD turbulence with small $\mu$ can be described in terms of interacting 
waves. Kinetic and potential energy are freely interchanged so 
$E_{\vb}$ and $E_{\Delta\bb}$ have comparable magnitudes. Wavemode turbulence 
is the principal subject of this thesis. Analytic scalings are presented in \S 
\ref{sec:MHDcascades} to provide an intuitive feel for its dynamics. Results 
from our simulations are described in \S \ref{sec:results} and discussed in 
\S \ref{sec:discussion}.   

\subsection{Linear Waves In Incompressible MHD}
\label{subsec:eigenvectors}

Linear perturbations about a uniform background magnetic field 
can be decomposed into shear Alfv\'en and pseudo Alfv\'en 
waves. The pseudo Alfv\'en wave is the incompressible limit of the slow 
magnetosonic wave.\footnote{In the limit of incompressibility the fast 
magnetosonic wave has infinite phase velocity and cannot be excited.} 
As is well known, both waves conform to the dispersion relation
\begin{equation}
\omega^2=v_{A}^2 k_z^2.
\end{equation}

\noindent Eigenvectors for these modes take the form
\begin{equation}
{\hat\vb}_{A}(\bk,t) = \hat{\bf{a}}(\bk)
\exp{i\bk \cdot (\x \mp v_{A} t\hat{\z})}, \quad\quad
{\hat\bb}_{A}(\bk,t) = \mp\hat{\bf{a}}(\bk)
\exp{i\bk \cdot (\x \mp v_{A} t\hat{\z})},
\end{equation}
\begin{equation}
{\hat\vb}_{S}(\bk,t) = \hat{\bf{s}}(\bk)
\exp{i \bk \cdot (\x \mp v_{A} t\hat{\z})}, \quad\quad
{\hat\bb}_{S}(\bk,t) = \mp \hat{\bf{s}}(\bk)
\exp{i \bk \cdot (\x \mp v_{A} t\hat{\z})},
\end{equation}
where the unit polarization vectors are defined by
\begin{equation}
\hat{\bf a}\equiv {\hat{\bk}\times\hat{\z}\over 
[1-(\hat{\bk}\cdot\hat{\z})^2]^{1/2}}, \hspace{12mm} \hat{\bf s}\equiv 
{\hat{\z}-(\hat{\bk}\cdot\hat{\z})\hat{\bk}\over 
[1-(\hat{\bk}\cdot\hat{\z})^2]^{1/2}}.
\label{eq:polarization}
\end{equation}
We note that $\hat{\bk}$, $\hat{\bf s}$, and $\hat{\bf a}$ form a right-hand 
triad.

%hyphenated right-hand

MHD turbulence is anisotropic with power cascading more rapidly to high
$k_\perp$ than to high $k_z$. In the limit $k_\perp\gg k_z$, $\hat{\bf s}\to 
\hat{\z}$; displacements associated with slow modes align along the unperturbed 
magnetic field.

\subsection{Elsasser Variables} \label{subsec:Elsasser}

The Elsasser transformation
\begin{eqnarray}
\w_\uparrow = v_{A} \hat{\z} + \vb    - \bb
\hspace{1.5cm}
\w_\downarrow = - v_{A} \hat{\z} + \vb    + \bb
\label{eq:defw}
\end{eqnarray}
applied to equations (\ref{eq:mhda}) and (\ref{eq:mhdb}) with $\nu_n=0$ brings 
out the two wave characteristics
\begin{eqnarray} 
&
\partial_t \w_\uparrow + v_{A}\partial_z \w_\uparrow
= -\w_\downarrow \cdot \bnabla \w_\uparrow - \bnabla P,
& \label{eq:elsasserup} \\
&
\partial_t \w_\downarrow - v_A\partial_z \w_\downarrow
= -\w_\uparrow \cdot \bnabla \w_\downarrow - \bnabla P, & 
\label{eq:elsasserdown}
\end{eqnarray}
where from equation \refnew{eq:Pvb}, 
\be
P=\int \frac{d^3x^\prime}{4\pi}\frac{\bnabla\w_\uparrow:\bnabla \w_\downarrow}
{|\x^\prime - \x|}. 
\label{eq:Pww}
\ee
Linear waves propagate at the Alfv\'en speed $v_{A}$ either parallel 
$(\w_\uparrow)$ or anti-parallel $(\w_\downarrow)$ to the direction
of the background magnetic field. 

\subsection{Collisions Between Wave Packets}
\label{subsec:collisions}

A disturbance of the background field may be decomposed into
upward $(\w_\uparrow)$ and downward $(\w_\downarrow)$ propagating wave
packets.  In the special case of unidirectional propagation either 
$\w_\uparrow=0$ or $\w_\downarrow=0$, and an arbitrary nonlinear wave packet
is an exact solution of the equations of incompressible MHD \citep{bib:Parker}. 
To prove this, take the divergence of the equation for the nonzero $\w$. This 
yields $\nabla^2 P = 0$ which, since it applies globally, implies 
$\bnabla P = 0$, and hence that the wave packet propagates without 
distortion.\footnote{This conclusion remains valid for our simulations which are 
carried out in a computational box and employ periodic boundary conditions.} An 
important corollary is that nonlinear distortions occur only during collisions 
between oppositely directed wave packets.  

Collisions are constrained by the conservation laws of energy, 
\be
E={1\over 2}\int d^3x \left(|\vb|^2+|\bb|^2\right), 
\label{eq:E}
\ee
and cross helicity, 
\be
I={1\over 2}\int d^3x \, \vb    \cdot \bb.  
\label{eq:I}
\ee
These conservation laws follow directly from equations 
\refnew{eq:mhda}-\refnew{eq:mhdd} in the limit that $\nu_n=0$.
As a consequence, energy is not exchanged between colliding wave packets. A 
short proof follows. 

Take the dot product of equations \refnew{eq:elsasserup} and 
\refnew{eq:elsasserdown} with $\w_\uparrow$ and $\w_\downarrow$, respectively. 
The advective 
and pressure gradient terms reduce to total divergences. This establishes that 
\be
{d\over dt}\int d^3x\, |\w_\uparrow|^2={d\over dt}\int d^3x\, 
|\w_\downarrow|^2=0
\label{eq:conservew}
\ee
provided $\w_\uparrow$ and $\w_\downarrow$ either vanish at infinity or satisfy 
periodic 
boundary conditions. 
>From the defining equations for the Elsasser variables, we obtain
$|\w_\uparrow|^2/2=|\vb|^2+|\bb|^2$ for $\w_\downarrow=0$ and 
$|\w_\downarrow|^2/2=|\vb|^2+|\bb|^2$ for $\w_\uparrow=0$. Thus
\be 
E_\uparrow={1\over 4}\int d^3x \, |\w_\uparrow|^2 \quad {\rm and} \quad
E_\downarrow={1\over 4}\int d^3x \, |\w_\downarrow|^2 
\label{eq:dirE}
\ee
are the energies of isolated upward and downward propagating wave packets.
This completes the proof that wave packet collisions are elastic. 

\subsection{Wave Packets Move Along Field Lines}
\label{subsection:wavepackets}

To lowest nonlinear order in the wave amplitudes, distortions suffered in 
collisions between 
oppositely directed wave packets arise because each packet moves along
field lines perturbed by the other. The proof follows directly from equation 
\refnew{eq:elsasserup} written to second order in the amplitudes of the 
$\w_\uparrow$ and $\w_\downarrow$ fields. With the aid of equation 
\refnew{eq:Pww}, it can be shown that
\be
D_\uparrow\left(\w_\uparrow^{(2)} + 
\bxi^{(1)}\cdot\bnabla\w_\uparrow^{(1)} + \bnabla \int 
\frac{d^3x^\prime}{4\pi}\frac{\bnabla\bxi^{(1)}:\bnabla\w^{(1)}_\uparrow} 
{|\x^\prime - \x|}\right)=0.
\label{eq:wuptwo}
\ee
Here
\be
D_\uparrow\equiv \left(\frac{\partial}{\partial t} + 
v_A\frac{\partial}{\partial z}\right),  
\ee
and 
\be
\x(\x_0,t)\equiv\x_0+\bxi(\x_0,t). 
\label{eq:displacement}
\ee
The Lagrangian displacement, $\bxi$, connects the Lagrangian coordinate of a 
fluid particle, $\x_0$, to its Eulerian coordinate, $\x$.

Several steps are needed to establish equation \refnew{eq:wuptwo}. In terms of 
$\bxi$,
\be
\vb=\frac{\partial\bxi}{\partial t}\biggr\vert_{\x_0}, \quad\quad 
\bb=v_A{\hat\z}+v_A\frac{\partial\bxi}{\partial z_0}\biggr\vert_t.
\label{eq:bxivb}
\ee
To first order in the amplitudes of $\w_\uparrow$ and $\w_\downarrow$, we may 
replace $\x_0$ by $\x$ in the definition of $\bxi$ and write 
\be
\vb^{(1)}=\frac{\partial\bxi^{(1)}}{\partial t}\biggr\vert_\x, \quad\quad 
\bb^{(1)}= v_A\frac{\partial\bxi^{(1)}}{\partial z}\biggr\vert_t.
\label{eq:xivb}
\ee
It then follows from equations \refnew{eq:defw} and \refnew{eq:xivb} that 
\be
\w_\downarrow^{(1)}=D_\uparrow\bxi^{(1)}.
\label{eq:wdownbxi}
\ee
The final step is to verify that the linear operator $D_\uparrow$ passes 
through 
the integral sign and changes $\bxi^{(1)}$ to $\w_\downarrow^{(1)}$ while 
leaving the rest of the integrand unaltered.

Equation \refnew{eq:wuptwo} has a simple interpretation. Consider an upward 
moving wave packet for which $\w^{(2)}_\uparrow=0$ prior to its interaction 
with downward moving waves. Subsequent to this interaction suppose that 
$\bxi^{(1)}$ at fixed $z_\uparrow=z-v_At$ is changed by $\Delta\bxi^{(1)}$. 
Then as a function of $z_\uparrow$
\be
\Delta\w^{(2)}_\uparrow=-\Delta\bxi^{(1)}\cdot\bnabla\w^{(1)}_\uparrow - \bnabla 
\int 
\frac{d^3x^\prime}{4\pi}\frac{\bnabla\Delta\bxi^{(1)}:\bnabla\w^{(1)}_\uparrow}
{|\x^\prime - \x|}.
\label{eq:Delwtwo}
\ee
The first term on the right-hand side of this equation is the perturbation that 
would result from the unconstrained displacement of $\w^{(1)}_\uparrow$ by 
$\Delta\bxi^{(1)}$. The second term constrains the perturbation to preserve 
$\bnabla\cdot\w_\uparrow^{(2)}=0$.\footnote{This term arises from the gradient 
of the total pressure.} Since magnetic field lines are frozen
in the fluid, we conclude that, at least to second order, wave packets follow 
magnetic field lines. 

Two points are worth stressing in connection with equation \ref{eq:Delwtwo}.
\begin{itemize}
\item Downward propagating waves contribute the entire $\Delta\bxi^{(1)}$ since
it is measured at fixed $z_\uparrow=z-v_At$. This is consistent with the 
general rule that only oppositely directed wave packets interact. 
\item The turbulent energy cascade is associated with the shear 
of the $\Delta\bxi^{(1)}$ field. Uniform displacements, which arise as a 
consequence of the sweeping of small disturbances by larger ones, do not 
contribute to the transfer of energy across scales.   
\end{itemize}

The proof in this section has been couched in Eulerian coordinates. A 
technically simpler version in Lagrangian coordinates is given by 
\citet{bib:SG}. It consists of demonstrating that the third order 
Lagrangian 
density for incompressible MHD vanishes when written in terms of the transverse 
components of the displacement vector. Although simpler technically, the 
Lagrangian based result is more subtle conceptually. Its proper interpretation 
is provided in GSII. 

\section{MHD CASCADES}
\label{sec:MHDcascades}

A variety of models have been proposed for MHD turbulence. They share
the common feature that energy cascades from lower to higher wave number. 

\subsection{The Iroshnikov-Kraichnan Model}
\label{subsec:IK}

The standard model is that due to \citet{bib:Iroshnikov} and 
\citet{bib:Kraichnan}. Kraichnan's derivation of the IK spectrum
relies on the fact that only oppositely directed waves interact in
incompressible MHD. It assumes explicitly that the turbulence is isotropic and
implicitly that the dominant interactions are those which couple three waves. 

The above assumptions imply that the cascade time across
scale $\lambda$ is 
\be
t_c\sim \left({v_A\over v_\lambda}\right)^2 {\lambda\over v_A}.
\label{eq:tcIK}
\ee
Setting $v_\lambda^2/t_c$ equal to the dissipation rate per unit mass, 
$\epsilon$, then yields
\be
v_\lambda\sim \left(\epsilon v_A\lambda\right)^{1/4},
\label{eq:IKconfig}
\ee
which corresponds to the 1D power spectrum\footnote{Because the IK cascade is
isotropic, it is sufficient to specify its 1D power spectrum.} 
\be
E(k)\sim {\left(\epsilon v_A\right)^{1/2}\over k^{3/2}}.
\label{eq:IKE}
\ee
Nonlinearity is measured by $\chi\sim (v_\lambda/v_A)$, where $N\sim \chi^{-2}$ 
is the number of wave periods in $t_c$; 
\be
\chi\sim \left({\epsilon\lambda\over v_A^3}\right)^{1/4}.
\label{eq:NIK}
\ee
Since $\chi$ decreases with decreasing $\lambda$, only dissipation limits the 
length of the IK inertial range.

The IK model is flawed because the assumption of isotropy is inconsistent with 
the frequency and wavevector closure relations that resonant triads must 
satisfy \citep{bib:Shebalin}. These take the form
\begin{eqnarray}
\omega_1+\omega_2=\omega_3, 
\label{eq:freqclosure} \\
\bk_1+\bk_2=\bk_3.
\label{eq:kclosure}
\end{eqnarray}
But since $\omega=v_A|k_z|$, equation \refnew{eq:freqclosure} and the $z$ 
component of equation \refnew{eq:kclosure} yield the set
\begin{eqnarray}
|k_{1z}|+|k_{2z}|=|k_{3z}| 
\label{eq:abskz}\\
k_{1z}+k_{2z}= k_{3z}.
\label{eq:kz}
\end{eqnarray}
Because nonlinear interactions can only occur between oppositely directed
waves, the 3-mode coupling coefficient vanishes unless waves 1 and 2 propagate 
in opposite directions. In that case, equations \refnew{eq:abskz} and 
\refnew{eq:kz} imply that either $k_{1z}$ or $k_{2z}$ must vanish.
Since one of the incoming waves has zero frequency, 3-wave
interactions do not cascade energy 
along $k_z$.    

\subsection{Intermediate MHD Turbulence}

GSII propose an 
anisotropic MHD cascade 
based on scalings obtained from 3-wave interactions. It represents 
a new form of turbulence, which they term intermediate, because it 
shares some of the properties of both weak and strong turbulence. Although 
individual wave packets suffer small distortions in single 
collisions, interactions of all orders make comparable contributions to the 
perpendicular cascade.\footnote{This is a controversial claim.} 

To derive the scaling relations for the intermediate cascade, we repeat the
steps carried out in \S \ref{subsec:IK} for the IK model, but with 
$\lambda_\perp$ in place of $\lambda$ and $\lambda_\parallel$ held constant.
Here $\lambda_\perp$ and $\lambda_\parallel$ are correlation lengths in 
directions perpendicular and parallel to the local magnetic field.  
Thus 
\be
t_c\sim \left({v_A \lambda_\perp\over 
v_{\lambda_\perp}\lambda_\parallel}\right)^2 {\lambda_\parallel\over v_A}.
\label{eq:tcintermed}
\ee
Setting $\epsilon\sim v_{\lambda_\perp}^2/t_c$, we find
\be
v_{\lambda_\perp}\sim \left({\epsilon v_A \lambda_\perp^2\over 
\lambda_\parallel}\right)^{1/4},
\label{eq:intermedconfig}
\ee
and
\be
E(k_\perp)\sim {\left({\epsilon v_A k_\parallel}\right)^{1/2}\over k_\perp^2}.
\label{eq:intermedE}
\ee

Besides being anisotropic, the intermediate MHD cascade differs from the IK 
cascade in another important respect. The strength of nonlinear interactions, 
as 
measured by
\be
\chi\sim  \left({v_{\lambda_\perp}\lambda_\parallel\over 
v_A\lambda_\perp}\right)\sim \left({\epsilon\lambda_\parallel^3\over 
v_A^3\lambda_\perp^2}\right)^{1/4},
\label{eq:defchi}
\ee
increases along the cascade.
Thus, even in the absence of dissipation, the intermediate cascade has a finite 
inertial range. This suggests that a strong form of MHD turbulence must be the 
relevant one for most applications in nature. 

\subsection{Strong MHD Turbulence} 
\label{subsec:strongturb}

A cascade for strong MHD turbulence is described in GSI. Its defining property 
is that MHD 
waves suffer order unity distortions on time scales comparable to their 
periods. This state is referred to as one 
of critical balance. Motivation for the hypothesis of critical balance is given
in \citet{bib:GSI,bib:GSII} and summarized below. 
Our discussion of intermediate turbulence shows that $\chi$ increases if
it is less than unity. However, it cannot rise above unity since the frequency 
spread of the wave packets which emerge following a strong collision must 
satisfy the frequency-time uncertainty relationship. \citet{bib:Gruzinov} 
provides a  more physical explanation for the upper bound on $\chi$. He points 
out that for $\chi\gg 1$, 2D motions of scale $\lambda_\perp$ in planes 
perpendicular to the local magnetic field are uncoupled over separations greater 
than $\lambda_\parallel/\chi$ along the field direction. Thus, during a time 
interval of order $\lambda_\perp/v_{\lambda_\perp}\sim 
\lambda_\parallel/v_A\chi$ these motions reduce $\chi$ to order unity.

Crtical balance, and the assumption of a constant energy flux 
along the cascade as expressed by
\be
\epsilon\sim {v_{\lambda_\perp}^3\over \lambda_\perp},
\label{eq:contEflux}
\ee
imply
\be
\lambda_\parallel\propto \lambda_\perp^{2/3}.
\label{eq:anisotropy}
\ee 
Although there is a parallel cascade of energy in strong MHD turbulence, the 
degree of anisotropy increases along the cascade. 

Let us assume $v_L\sim v_A$ and isotropy on scale outer scale $L$. Then the 
3D energy spectrum of strong MHD turbulence takes the form
\be
E(k_\perp,\, k_\parallel)\sim {v_A^2\over 
L^{1/3} k_\perp^{10/3}}f\left({k_\parallel L^{1/3}\over k_\perp^{2/3}}\right),
\label{eq:3Dpower}
\ee
where $f(u)$ is a positive symmetric function of $u$ with the properties that 
$f(u)\approx 1$ for $|u|\lesssim 1$ and $f(u)$ negligibly small for $|u|\gg 1$. 
The power spectrum is flat as a function of $k_\parallel$ for 
$k_\parallel\lesssim k_\perp^{2/3}L^{-1/3}$ because the velocity and magnetic
perturbations on transverse scale $k_\perp^{-1}$ arise from independent wave 
packets whose lengths $\lambda_\parallel\sim \lambda^{2/3}_\perp L^{1/3}$.    
The 1D perpendicular power spectrum obtained from equation 
\refnew{eq:3Dpower} reads
\be
E(k_\perp)\sim {v_A^2\over  L^{2/3} k^{5/3}_\perp}.
\label{eq:1Dpower}
\ee
Thus the spectrum of strong MHD turbulence is an anisotropic version of the
\citet{bib:Kolmogorov} spectrum for hydrodynamic turbulence. 

Inertial range velocity differences and magnetic perturbations across 
perpendicular scale $\lambda_\perp$ satisfy 
\be
v_{\lambda_\perp}\sim b_{\lambda_\perp}\sim \left({\lambda_\perp\over 
L}\right)^{1/3}v_A.
\label{eq:scalings}
\ee
Thus even though the turbulence is properly classified as strong, deep in the 
inertial range magnetic field lines are nearly parallel across perpendicular 
separations $\lambda_\perp$ and nearly straight along parallel separations 
$\lambda_\parallel$; differential bending angles are of order 
$(\lambda_\perp/L)^{1/3}\sim (\lambda_\parallel/L)^{1/2}$. 

\subsubsection{Parallel Cascade}

It is interesting to examine the frequency changing interactions that
drive the parallel cascade. Referring back to the intermediate cascade,
we know that 3-wave interactions do not change frequencies. However,
interactions involving more than 3-waves can. For example, frequency changes 
arise in 4-wave interactions of the form
\begin{eqnarray}
\omega_1+\omega_2+\omega_3=\omega_4, 
\label{eq:freqclose} \\
\bk_1+\bk_2+\bk_3=\bk_4,
\label{eq:kclose}
\end{eqnarray}
where $k_{1z}$ and $k_{2z}$ have the same sign and $\omega_3=v_A|k_{3z}|=0$ 
\citep[GKII]{bib:NB96}.  
The parallel cascade they give rise to proceeds at a rate which is
smaller than that of the perpendicular cascade by a factor of order $\chi$. 
Because strong MHD turbulence is characterized by $\chi\sim 1$, it has a 
significant parallel cascade. 

\subsubsection{Field Line Geometry}
\label{subsubsec:geometry}

MHD turbulence is best understood geometrically. Field lines perturbed by waves 
propagating in one direction define two-dimensional mappings between $xy$ planes 
separated by distance $z$.  Shear Alfv\'en waves dominate the shear and slow 
waves the dilatation of these mappings. The magnitude 
of the shear exceeds that of the dilatation by a factor of order 
$\lambda_\parallel/\lambda_\perp\sim (L/\lambda_\perp)^{1/3}\gg 1$. These 
mappings describe the distortion that counter propagating waves would suffer
if they moved at uniform speed along the perturbed field lines. The dominance 
of 
the shear over the dilatation  explains why shear Alfv\'en waves control the 
perpendicular cascades of both types of wave. 

\begin{figure}

\plottwo{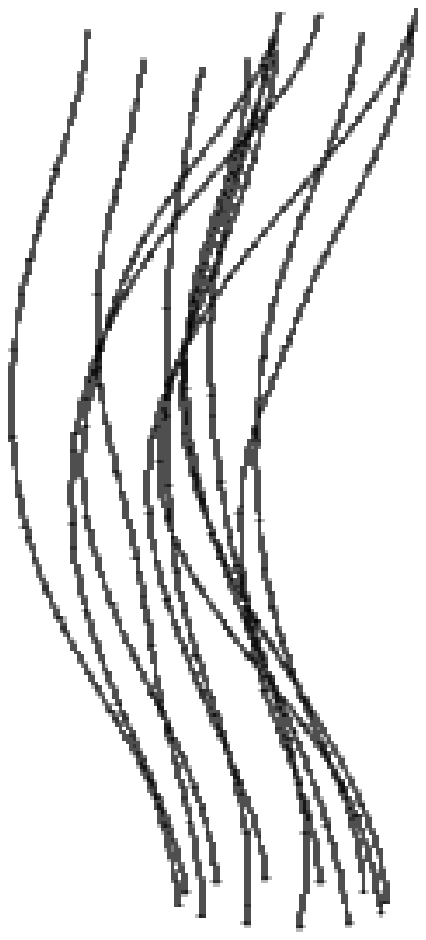}{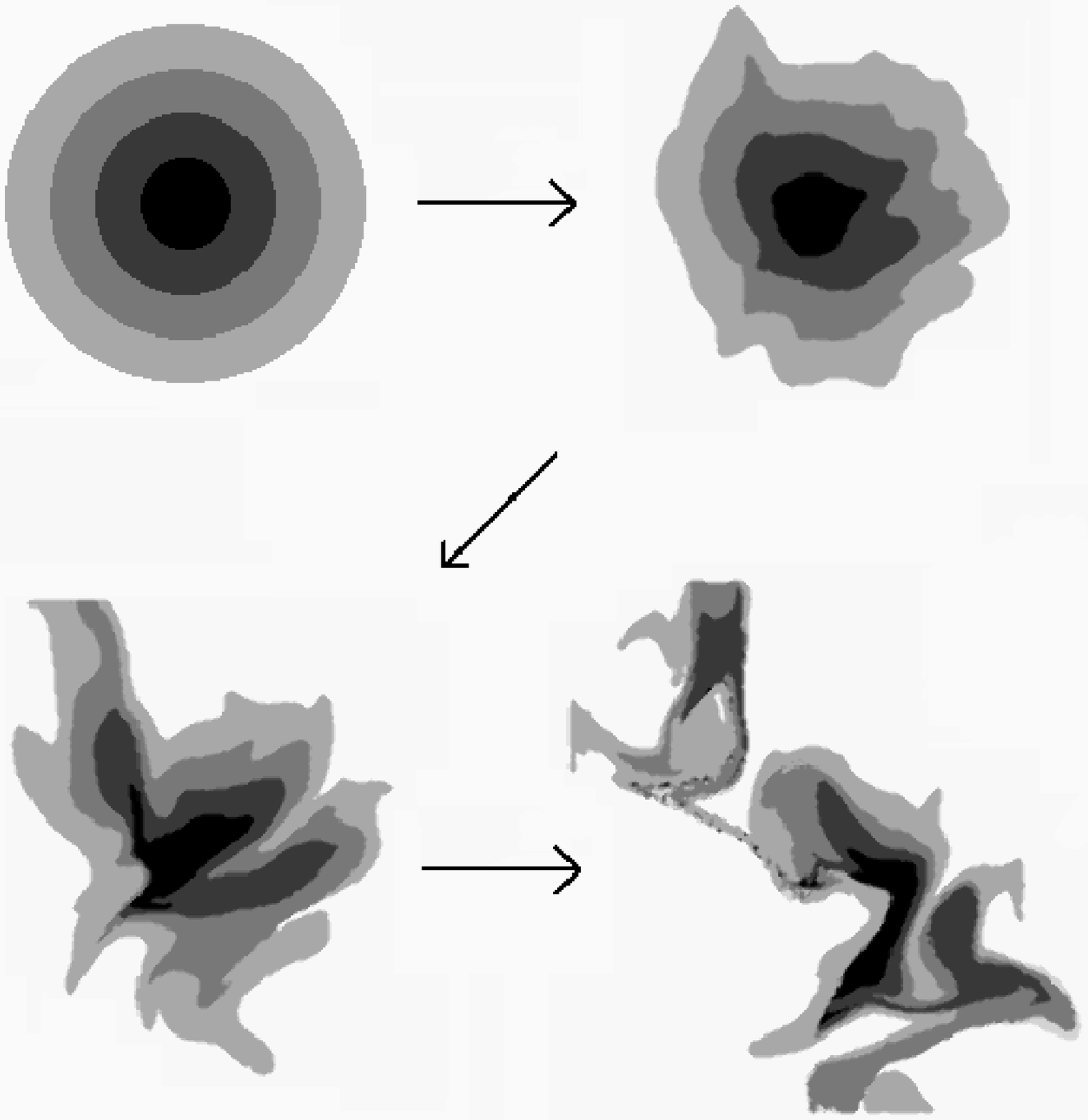}

\caption{\it Wavepacket Distortion Through Fieldline Wander. The left-hand panel 
displays a sample of field lines perturbed by 
downward propagating waves. The distortion of an originally circular bullseye 
pattern as it moves upward following these field lines is shown in the
right-hand panel. \label{figdistort}}
\end{figure}

The recognition that MHD waves tend to follow field lines is essential to
understanding their turbulent cascades. Figure \ref{figdistort} provides a
visual illustration of how this works. The left-hand panel displays a 
snapshot of field lines perturbed by downward propagating waves. In the
right-hand
panel we follow the evolution of a horizontal pattern as it propagates 
from the bottom to the top following these lines. The distortion of the 
initially circular bullseye is principally due to the shear in the
two-dimensional mapping defined by the perturbed field lines. The cascade time 
on 
the scale of the initial pattern is that over which the shear grows to order 
unity. 

%-----------------------------------------------------------------

%-----------------------------------------------------------------

This geometrical picture requires two qualifications. The first is that the 
propagation speed of MHD waves is not exactly constant but varies with 
the strength of the local magnetic field. Pressure perturbations associated 
with 
slow waves are balanced by perturbations of magnetic pressure. The resulting 
perturbations in propagation speed, of order $v_{\lambda_\perp}$, contribute to 
the nonlinear cascade. Over one wave period they lead to fractional distortions 
of order $v_{\lambda_\perp}/v_A\sim \lambda_\perp/\lambda_\parallel\ll 1$. Thus 
they are properly ignored. The second qualification is that MHD waves do not 
exactly follow field lines. The extent to which this effects their cascade 
remains to be quantified.  

The parallel cascade may also be viewed in geometrical terms. Consider an 
upward 
propagating wave packet of length $\lambda_\parallel$ and width $\lambda_\perp$ 
which is being distorted by downward moving wave packets of similar scale. 
Correlations along the parallel direction are shortened because the front and 
back of the wave packet undergo different 2-dimensional mappings. This happens 
because the upward propagating packet distorts each downward going packet as it 
passes through it. This distortion is of order $\chi$. For strong MHD 
turbulence 
$\chi\sim 1$ which accounts for its significant parallel cascade. 

Incidentally, the geometrical picture also aids the interpretation of results 
from
perturbation theory. For example, the 3-wave resonant interactions which 
dominate the perpendicular cascade and the 4-wave resonant interactions which 
cause the lowest order frequency changes each depend upon the amplitudes of 
modes with $k_z=0$. This is because the shear in the mapping between $xy$ 
planes 
separated by $\Delta z$ is proportional to the displacement amplitudes of modes 
with $k_z\lesssim 1/\Delta z$. Perturbation theory corresponds to the limit of 
vanishing cascade strength in which shears of order unity are achieved in the 
limit of infinite separation along the $z$-axis. 

\subsubsection{Relation to 2D Hydrodynamic Turbulence}

Magnetic field lines possess a tension which makes them ill-disposed
to bend, but they are easily shuffled. This accounts for the 2D
character of MHD turbulence. It also prompts an inquiry about the
relation of MHD turbulence to 2D hydrodynamic
turbulence. Each fluid elements conserves its vorticity in inviscid 2D
hydrodynamics. This results in a direct cascade of enstrophy (vorticity 
squared) toward high $k_\perp$ and an inverse cascade of energy
toward small $k_\perp$ \citep{bib:Lesieur}. As we now
demonstrate, MHD turbulence does not share these characteristics.

The vorticity equation in MHD, obtained by taking the curl of equation 
\refnew{eq:mhda} with $\nu_n=0$, reads
\be
{\partial (\bnabla\times \vb)\over \partial t} = \bnabla\times\left[ 
\vb\times(\bnabla\times \vb) -  \bb\times(\bnabla\times \bb)\right].
\label{eq:vorticity}
\ee
We concentrate on shear Alfv\'en waves since they dominate the field aligned 
vorticity for nearly perpendicular cascades. Scaling the terms in equation
\refnew{eq:vorticity} shows that 
\be
{\partial\omega_\parallel\over \partial t}\sim {v_{\lambda_\perp}\over 
\lambda_\perp}\omega_\parallel.
\label{eq:MHDvorticity}
\ee
Thus $\omega_\parallel$ changes on the cascade time scale; it is not even 
approximately conserved. Consequently, there is no enstrophy constraint to 
prevent energy from cascading toward larger $k_\perp$.

\section{SIMULATION STRATEGY} 
\label{sec:strategy}

What follows is a comprehensive discussion of the techniques used in
our simulations.Technical aspects of the spectral method are presented in
the Appendix.  

\subsection{Spectral Wave Mode Decomposition}
\label{subsec:spectraldecomp}

Separation of $\tilde{\vb}(\bk)$ and $\tilde{\bb}(\bk)$ into upward and downward 
propagating components is accomplished by forming Fourier coefficients of the 
Elsasser variables $\tilde{\w}_\uparrow(\bk)$ and $\tilde{\w}_\downarrow(\bk)$ 
according to
\be
\tilde{\w}_\uparrow(\bk)=\tilde{\vb}(\bk)-\tilde{\bb}(\bk) \hspace{12mm} 
\tilde{\w}_\downarrow(\bk)=\tilde{\vb}(\bk)+\tilde{\bb}(\bk).
\label{eq:tildeElsasser}
\ee
Projections of $\tilde{\w}_\uparrow(\bk)$ and $\tilde{\w}_\downarrow(\bk)$ along 
the polarization directions of the linear incompressible MHD eigenmodes given 
by equation \refnew{eq:polarization} yield amplitudes of upward and downward 
propagating Alfv\'en and slow waves. In obvious notation

\be
A_{\uparrow}(\bk) \equiv\hat{\ba}\cdot\tilde{\w}_\uparrow(\bk)= {\tilde 
v}_A(\bk) - 
{\tilde b}_A(\bk) 
\quad\quad
S_{\uparrow}(\bk) \equiv\hat{\s}\cdot\tilde{\w}_\uparrow(\bk)= {\tilde v}_S(\bk) 
- 
{\tilde b}_S(\bk)
\label{eq:tildeASup}
\ee
\be
A_{\downarrow}(\bk) \equiv\hat{\ba}\cdot\tilde{\w}_\downarrow(\bk)= 
{\tilde v}_A(\bk)+{\tilde b}_A(\bk) 
\quad\quad
S_{\downarrow}(\bk) \equiv\hat{\s}\cdot\tilde{\w}_\downarrow(\bk)={\tilde 
v}_S(\bk)+{\tilde b}_S(\bk)
\label{eq:tildeASdown}
\ee
where
\be
\tilde{v}_A(\bk)=\hat{\ba}\cdot{\tilde \vb}(\bk) \quad\quad 
\tilde{b}_A(\bk)=\hat{\ba}\cdot{\tilde \bb}(\bk) \quad\quad
\tilde{v}_S(\bk)=\hat{\s}\cdot{\tilde \vb}(\bk) \quad\quad
\tilde{b}_S(\bk)=\hat{\s}\cdot{\tilde \bb}(\bk).
\ee

The eigenmode frame, $(\hat{\bk}, \hat{\s}, \hat{\ba})$, is tied to the 
direction 
of the mean field, $\hat{\bb}_0=\hat{\z}$, to which the local field direction, 
$\hat{\bb}$, is inclined by an angle $\theta\sim v_{L_\perp}/v_A$. Consequently, 
our method for spectral decomposition erroneously mixes Alfv\'en and slow 
modes. However, for nearly transverse cascades the mixing is only of order 
$\theta^2\ll 1$.

Field line tilt also causes $k_r$ and $k_z$ to differ from $k_\perp$ and 
$k_\parallel$;
\be
k_r=\cos\theta k_\perp + \sin\theta k_\parallel \hspace{15mm}
k_z=-\sin\theta k_\perp +\cos\theta k_\parallel.
\ee
Thus 
$k_r\approx k_\perp[1+O(\theta^2)]$. However, $k_z \approx k_\parallel + \theta 
k_\perp\approx \theta k_\perp$, where the final relation applies because the 
degree of anisotropy increases with increasing $k_\perp$ along MHD cascades.
Thus, $k_\perp$ can be represented to acceptable accuracy by $k_r$. However, 
$k_\parallel$ cannot be obtained from $k_z$. Henceforth, we treat as equivalent
$k_r$ and $k_\perp$ and $L_x=L_y$ and $L_\perp$. However, we are always
careful to distinguish $k_z$ from $k_\parallel$ and to note that
\be
k_z\approx \frac{v_{L_\perp}}{v_A}k_\perp.  
\label{eq:kzkperp}
\ee

\subsection{Power Spectra}
\label{subsec:powerspectra}

Three-dimensional power spectra of field quantities, $E_{\rm 3D}(\bk)$, are 
azimuthally symmetric functions of $\bk_r=k_x{\hat \x}+k_y{\hat \y}$ at fixed 
$k_z$.\footnote{In any specific realization this is true only in a statistical 
sense.} Accordingly, we define the 2D integrated power spectrum by
\be
E_{\rm 2D}(k_r,\, k_z) = k_r \int_0^{2\pi}d\phi E_{\rm 3D}(\bk) . 
\label{eq:E2D}
\ee

It is important to note that $E_{\rm 2D}(k_r,\, k_z)$ is not equivalent to
$E_{\rm 2D}(k_\perp,\, k_\parallel)$. Moreover, the latter cannot be derived 
from the former. This shortcoming is due to the failure of the spectral 
decomposition procedure described in \S \ref{subsec:spectraldecomp} to 
determine 
$k_\parallel$. It means that the 2D power spectrum is not a useful 
quantity.\footnote{We obtain 2D information from structure functions.}
However, we make so much use of the 1D integrated power spectrum defined by
\be
E_{\rm 1D}(k_r)=\int_{-\infty}^\infty dk_z E_{\rm 2D}(k_r,\, k_z),  
\label{eq:E1D}
\ee
that henceforth we drop the subscript 1D.  

\subsection{Structure Functions}
\label{subsec:structurefunctions}

The 3D behavior of MHD turbulence is best captured in real space using 
second order structure functions tied to the local magnetic field. 
We define transverse and longitudinal structure functions for the vector field 
$\U$ by
\be
SFT_U(x_\perp)\equiv <[\U(\x^\prime + \x_\perp) - \U(\x^\prime)]\cdot
[\U(\x^\prime + \x_\perp) - \U(\x^\prime)]>,
\label{eq:SFT}
\ee
where $\x_\perp\cdot \bb=0$, and 
\be
SFL_U(x_\parallel)\equiv <[\U(\x^\prime + \f(x_\parallel)) - \U(\x^\prime)]\cdot
[\U(\x^\prime + \f(x_\parallel)) - \U(\x^\prime)]>,
\label{eq:SFL}
\ee
with $\f(x_\parallel)=\int_0^{x_\parallel}\, ds\, \hat{\bb}(s)$. The
integral is taken along the field direction with $s$ measuring arc length from 
$\x^\prime$. Averaging over $\x^\prime$ is done with random volume sampling.
Since the vector fields of interest possess statistical axial symmetry about 
${\hat\bb}$, we include an axial averaging of the direction of $\x_\perp$ at 
fixed $x_\perp\equiv |\x_\perp|$ in the computation of $SFT_U(x_\perp)$. 

\subsection{Timestep And Hyperviscosity}
\label{subsec:timestep}

\begin{figure}

\plotone{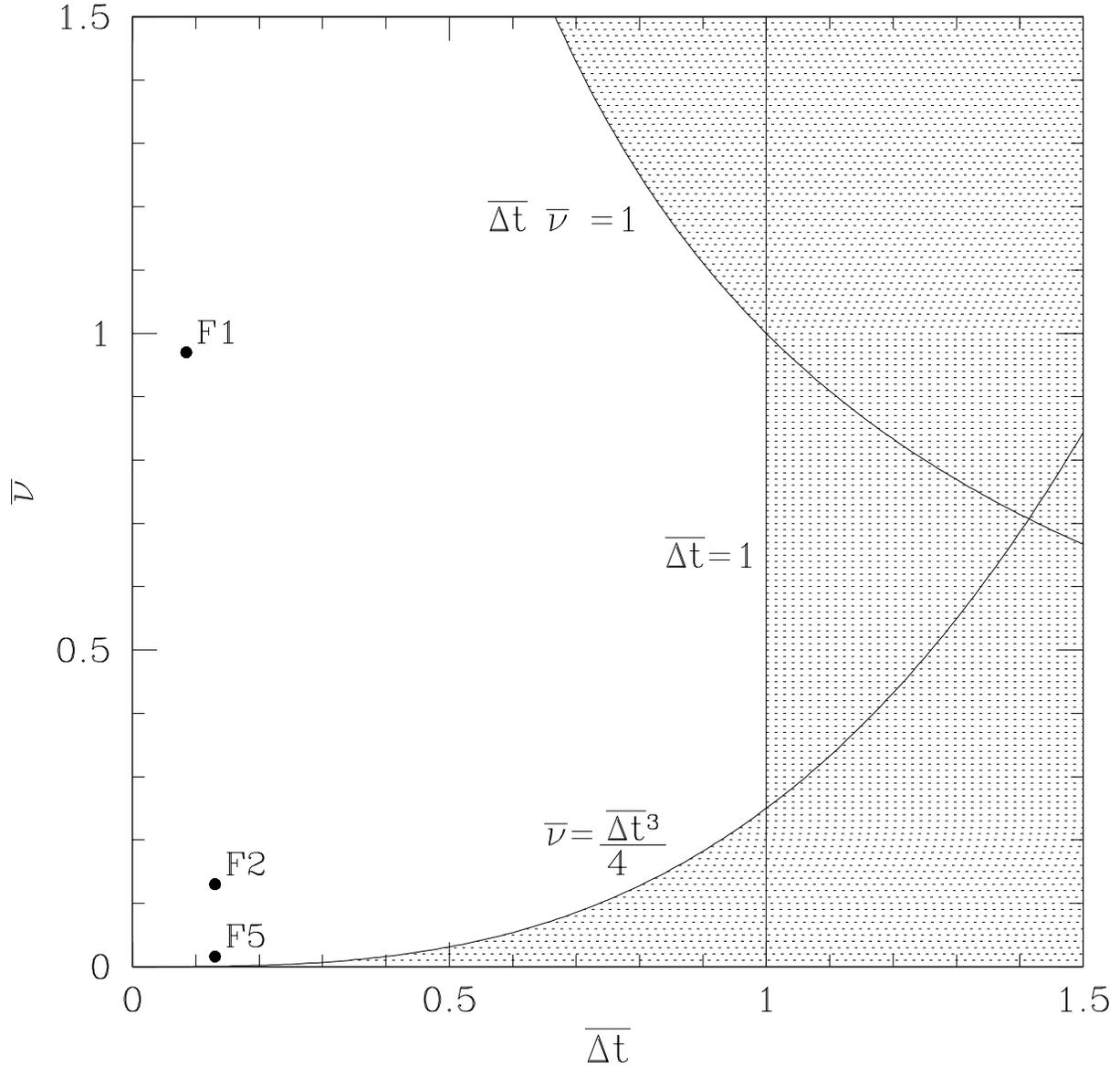}

\caption{\it Timestep and Hyperviscosity. We illustrate constraints on the 
dimensionless timestep and hyperviscosity 
as described in \S \ref{subsec:timestep}. The constraint given by
equation \refnew{eq:numin} is not shown because it depends upon an additional 
parameter. 
Allowed choices lie in the unshaded part of the figure. 
Each plotted point represents values used in an individual simulation. 
\label{fig:viscosity}} 
\end{figure}

The anisotropy of MHD turbulence complicates the discussion of constraints on 
the timestep and hyperviscosity. Accordingly, we begin by discussing
the simpler case of spectral simulation of isotropic hydrodynamic turbulence.  
These constraints are summarized in Figure \ref{fig:viscosity}.

\subsubsection{Isotropic Hydrodynamic Turbulence}
\label{subsubsec:HDturb}

We assume the Kolmogorov scaling. Given velocity $v_L$ on outer scale $L$, 
inertial range velocity differences across $\lambda\lesssim L$ scale as 
$v_\lambda\sim (\lambda/L)^{1/3}v_L$ down to inner scale $\ell\sim 
(\nu_n/v_LL^{2n-1})^{3/(6n-2)}L$.  

Four conditions constrain the values of the timestep, $\Delta t$, and 
hyperviscosity, $\nu_n$, suitable for a spectral simulation of isotropic 
hydrodynamic turbulence. Each refers to the behavior of modes with the largest 
wavevectors, $k_M$. We express these constraints in terms of the dimensionless 
variables ${\overline{\Delta t}}\equiv v_Lk_M\Delta t$ and 
${\overline{\nu}}\equiv (\nu_nk_M^{2n})/(v_Lk_M)$. 

\begin{itemize}
\item Conditions 1) and 2) are concerned with computational accuracy. 
\begin{description}
\item [1.] Advection by outer scale eddies gives rise to fractional changes of 
order $v_Lk_M \Delta t$ in the Fourier components of the smallest scale 
modes during one timestep.\footnote{Changes caused by interactions which are 
local in Fourier space 
are smaller by a factor $(k_ML)^{-1/3}$.}
Accurate computation requires
\be
\overline{\Delta t}\lesssim 1,
\label{eq:Deltadvec}
\ee
which is the spectral equivalent of the Courant condition in real space. 
\item [2.] Hyperviscosity causes a fractional decay of order $\nu_n 
k_M^{2n}\Delta t$ in the amplitudes of the smallest scale modes during a single 
timestep. Thus
\be
\overline{\Delta t}\lesssim \frac{1}{\overline{\nu}_n}.
\label{eq:Deltvis}
\ee
\end{description}
\item Conditions 3) and 4) are required to maintain stability. 
\begin{description}
\item [3.] This constraint depends upon the algorithm used to advance the 
variables in time. Integration with RK2 results in an unphysical transfer of 
energy from large to small scale modes. Consider 1D uniform advection at speed 
$v_L$ of the single Fourier mode $v(k_M, t)$. RK2 yields $v(k_M,\Delta t)=
[1-iv_Lk_M\Delta t-(v_Lk_M\Delta t)^2/2]v(k_M,0)$. Thus $|v(k_M,\Delta t)|^2=
[1+(v_Lk_M\Delta t)^4/4]|v(k_M,0)|^2$. In order that hyperviscosity maintain 
stability,
\be
\overline{\Delta t}\lesssim \left(4\overline{\nu}_n\right)^{1/3}.
\label{eq:DeltRK2}
\ee
\item [4.] A turbulent cascade transfers energy from large to small scales 
where it is dissipated by viscosity.  Spectral simulations of turbulence must 
include a mechanism which is able to dispose of the energy carried by the 
cascade before it reaches $k_M$.  Otherwise it would reflect back to smaller 
$k$ and the high $k$ Fourier modes would approach energy equipartition with 
those of lower $k$.\footnote{The energy per Fourier mode scales as $k^{-11/3}$ 
in Kolmogorov turbulence.} Hyperviscosity suffices provided the inner scale it 
sets is larger than the grid resolution. This requires
\be
\overline{\nu}\gtrsim \left(k_ML\right)^{-1/3}=(\pi N/2)^{-1/3}.
\label{eq:numin}
\ee
Dealiasing also involves a loss of energy and can stabilize simulations run 
with a sufficiently small timestep even in the absence of hyperviscosity. 
Further investigation is needed to clarify the manner in which energy is lost 
due to dealiasing.
\end{description}
\end{itemize}

%-----------------------------------------------------------------

%-----------------------------------------------------------------------

\subsubsection{Anisotropic Magnetohydrodynamic Turbulence}
\label{subsubsec:MHDturb}

We restrict our discussion of MHD turbulence to cases in which the
energy in the mean magnetic field greatly exceeds that in the kinetic and 
magnetic fluctuations. As discussed in Chapter \ref{sec:MHDcascades}, analytic 
arguments indicate that the Kolmogorov scaling is obeyed in planes 
perpendicular to the mean magnetic field. Thus constraints on the timestep and 
hyperviscosity deduced in \S \ref{subsubsec:HDturb} for hydrodynamic turbulence 
pertain to MHD turbulence in the $xy$ plane provided we take ${\overline{\Delta 
t}}\equiv v_{L_\perp}k_{M_\perp}\Delta t$ and ${\overline{\nu}}\equiv 
(\nu_nk_{M_\perp}^{2n})/(v_{L_\perp}k_{M_\perp})$. 

Different constraints arise from motion along the direction of the mean magnetic 
field. Strong MHD turbulence is anisotropic with energy cascading more rapidly 
along $k_\perp$ than along $k_\parallel$. Analytic arguments imply that 
the anisotropy at perpendicular scale $k_\perp^{-1}$ is determined by the 
condition that the nonlinearity parameter $\chi = (v_\lambda k_\perp)/(v_A 
k_\parallel)\sim 1$, where $k_\parallel$ is the wavevector component in the 
direction of the local magnetic field. It is important to maintain the 
distinction between $k_\parallel$ and $k_z$. As discussed in \S 
\ref{subsec:spectraldecomp}, $k_z\approx (v_{L_\perp}/v_A)k_\perp\sim (k_\perp 
L_\perp)^{1/3}k_\parallel$. 

We control the value of $(v_{L_\perp}L_z)/(v_AL_\perp)$ in our simulations. 
Typically this quantity is set somewhat larger than unity 
in order to ensure that the largest scale structures cascade on a time scale 
shorter than the Alfv\'en crossing time $L_z/v_A$. As a consequence, 
\be
v_Ak_{M_z}\approx 
\frac{v_AL_\perp}{v_{L_\perp}L_z}v_{L_\perp}k_{M_\perp}\lesssim 
v_{L_\perp}k_{M_\perp}.
\label{eq:vkcomp}
\ee
After this preparation, we are ready to examine the constraints placed on
$\Delta t$ and $\nu_n$ by evolution in the $z$ direction.
\begin{description}
\item [1.] Advection in the $z$ direction is dominated by propagation at the 
Alfv\'en speed since $v_{L_z}\ll v_A$ in our simulations. Thus computational 
accuracy demands $v_Ak_{M_z}\Delta t\lesssim 1$. Since $v_Ak_{M_z}\lesssim 
v_{L_\perp}k_{M_\perp}$, this constraint is
less severe than that imposed by equation \refnew{eq:Deltadvec}.
\item [2.] Hyperdiffusivity is not important in the $z$ direction because 
$k_{M_z}\ll k_{M_\perp}$ and we use a scalar hyperdiffusivity.
\item [3.] Integration with RK2 leads to an unphysical transfer of energy from 
large to small scales due to advection at the Alfv\'en speed. Provided equation 
\refnew{eq:DeltRK2} is satisfied, this does not cause any difficulty 
because $v_Ak_{M_z}\lesssim v_{L_\perp}k_{M_\perp}$.  
\item [4.] The maximum wave number in the $z$ direction, $k_{M_z}$, must be 
larger than that at the inner scale of the cascade. From equation 
\refnew{eq:kzkperp}
\be
k_{M_z}L_z\sim 
\left(\frac{v_{L_\perp}L_z}{v_AL_\perp}\right)k_{M_\perp}L_\perp.
\label{eq:kmaxz}
\ee
As mentioned above, the factor preceding $k_{M_\perp}L_\perp$ is typically
larger than unity. Thus adequate resolution along $z$ generally requires 
$N_z>N_\perp$. 
\end{description}

\subsection{Simulation Design}
\label{subsec:design}

We carry out simulations of both forced and decaying MHD
turbulence. Amplitudes of Fourier modes within 3 lattice units of the
origin, normalized wave vector $|\s| \leq 3$, are incremented at each timestep 
in simulations of forced turbulence and assigned initial values in
simulations of decaying turbulence. Each component of these amplitudes
receives an addition of a complex number with random phase and
absolute value drawn from a Boltzmann distribution with specified mean
subject to the constraint that $\bk\cdot\vb(\bk)=0$ and $\bk\cdot\bb(\bk)=0$.
Thus we are forcing both velocity and magnetic fluctuations.
Forcing $\vb$ alone would artificially correlate the power received by
$\w_\uparrow$ and $\w_\downarrow$. With our technique, fluctuations in
the energy input to waves moving in opposite directions are independent.   

The aspect ratio of our simulation box is chosen to
match the anisotropy of the turbulence.
In most of our simulations, $L_z\gg L_x=L_y$. 
We scale lengths to $L_z=1$ and velocities to $v_A=1$. Thus waves take $\Delta 
t=1$ to propagate the length of the box. The excitation level, set by the 
parameter $(v_{L_\perp}L_z)/(v_AL_\perp)$, is chosen so that the longest waves 
cascade in less than $\Delta t=1$. An equivalent statement is that a 
typical fieldline wanders by more than $L_\perp$ in the transverse direction 
during its passage across the length $L_z$ of the box. This requires the 
excitation parameter to be somewhat larger than unity. Typical values in our 
simulations are of order $5$. We generally run our simulations for a few 
crossing times. Thus nonlinear interactions are fully expressed on all scales. 

Our basic procedure comes with a variety of refinements. The fields can 
be decomposed into their $\w_\uparrow$ and $\w_\downarrow$ components as given 
by equations \refnew{eq:tildeElsasser}. Each of these may be further separated 
into shear Alfv\'en and slow modes according to equations \refnew{eq:tildeASup} 
and \refnew{eq:tildeASdown}. In this manner we can selectively input and remove 
waves of any type and with any direction of propagation. 

\section{SIMULATION RESULTS}
\label{sec:results}

Parameters of the simulations referred to in this section are
listed in Table \ref{tab:simulationparameters}. 
Each simulation is carried out in a box of dimensions $L_x=L_y=2\times 10^{-3}$
and $L_z=1$, includes an external magnetic field of unit strength aligned with 
the $z$-axis, and uses a fourth order hyperviscosity. The
dimensionless forcing power is denoted by ${\cal P}$. It is chosen so
that the rms dimensionless fluctuations of $v$ and $b$ have magnitude
$3\times 10^{-3}$. These values also characterize the initial states
of simulations of decaying turbulence. 

\subsection{Power Spectra}
\label{subsec:powerspec}

We obtain power spectra from our simulations as described in \S 
\ref{subsec:powerspectra}.  

\subsubsection{1D Power Spectra}
\label{subsubsec:1Dpowerspectra}

\begin{figure}

\plotone{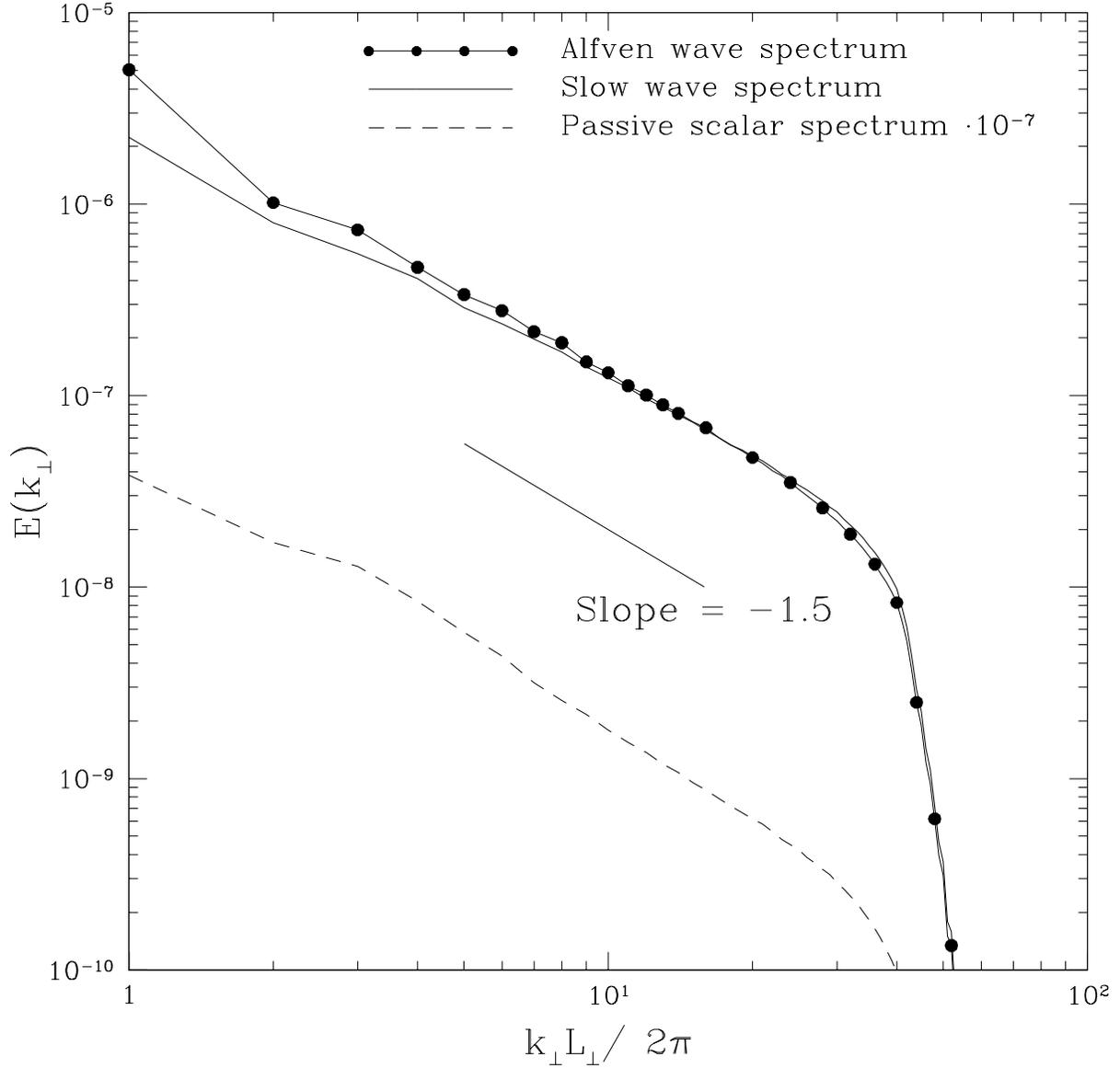} 
\caption{\it 1D Averaged Power Spectra. Energy spectra obtained by averaging 
results from simulations F2, F3, 
and F4 with resolution $128\times 128\times 512$. \label{fig:1Dpsave}}
\end{figure}

\begin{figure}
\plotone{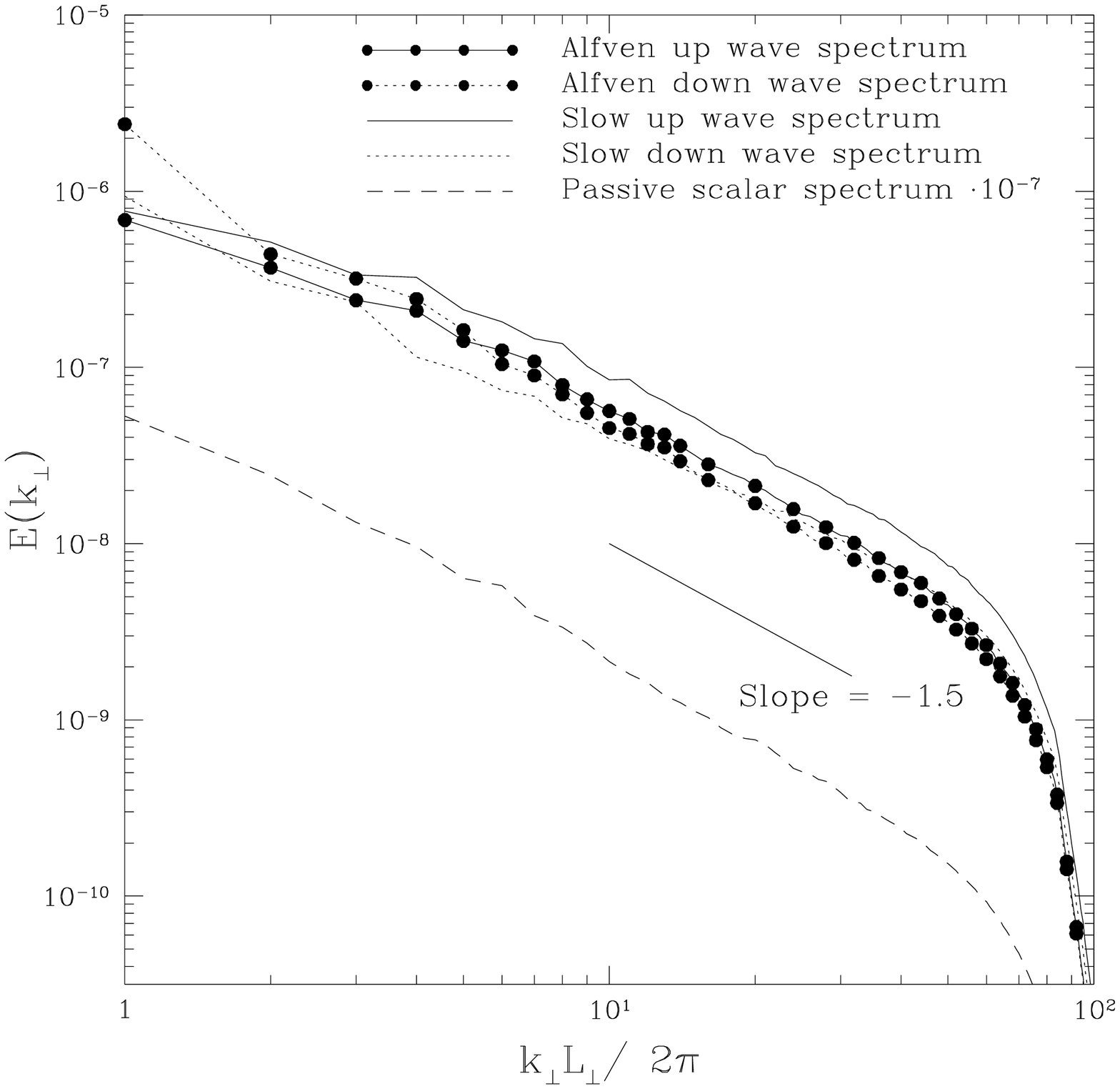}

\caption{\it Highest Resolution 1D Power Spectra. Energy spectra obtained from 
simulation F5 with resolution $256\times 
256\times 512$. \label{fig:1Dps117}} 
\end{figure}

Examples of 1D power spectra obtained by averaging results from three
simulations (F2, F3, and F4) of resolution $128\times 128\times
512$ are presented in Figure \ref{fig:1Dpsave}. Each spectrum has an
inertial range slope of approximately 1.5. The power spectra displayed
in Figure \ref{fig:1Dps117} come from a single simulation (F5) with
resolution $256\times 256\times 512$. Aside from their extended
inertial ranges, they look similar to those plotted in Figure
\ref{fig:1Dpsave}.

%-----------------------------------------------------------------

%-----------------------------------------------------------------

%-----------------------------------------------------------------
\subsubsection{2D Power Spectra}

\begin{figure}
\plotone{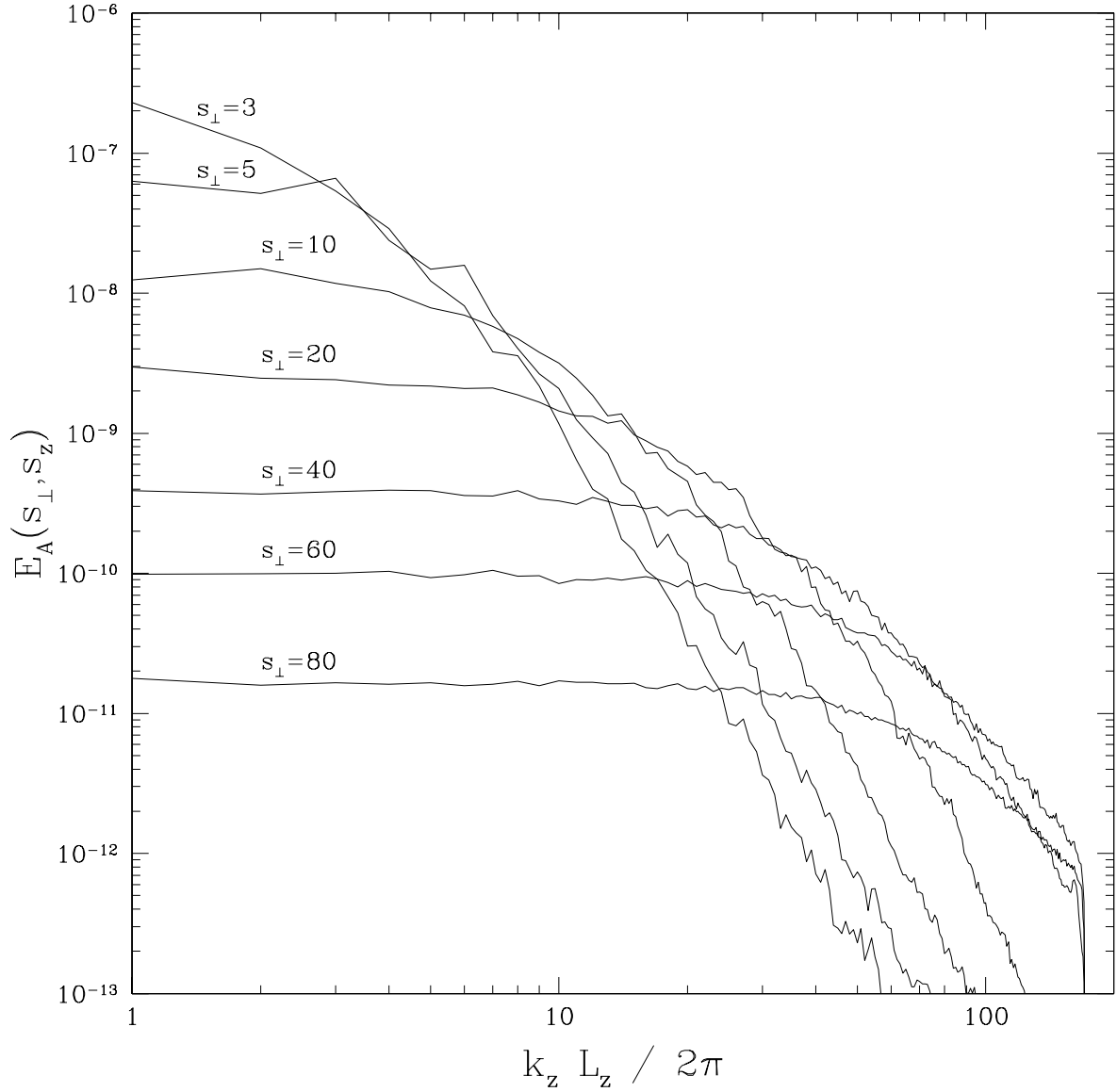}

\caption{\it Cuts Across 2D Power Spectrum. The Alfv\'en energy spectrum as a 
function of $s_z$ at fixed $s_\perp$ 
from simulation F5. 
\label{fig:dzaag}} 
\end{figure}

Figure \ref{fig:dzaag} displays a sequence of 1D power spectra made by taking
cuts parallel to the $s_z$ axis across the 2D power spectrum of shear 
Alfv\'en waves from simulation F5. Note that there is negligible power at the 
highest $s_z$ even for the highest $s_\perp$. Thus this simulation has adequate 
resolution along the $z$ direction, something we verify for each of our 
simulations. As we emphasize in \S \ref{subsec:spectraldecomp} and 
\S \ref{subsec:powerspectra}, these cuts do {\it not} suffice to determine the 
structure of turbulence parallel to the local magnetic field. For that we
need to use structure functions (see \S \ref{subsec:structure}). 

%------------------------------------------------------------

%----------------------------------------------------------------

\subsection{Structure Functions}
\label{subsec:structure}

\begin{figure}
\plotone{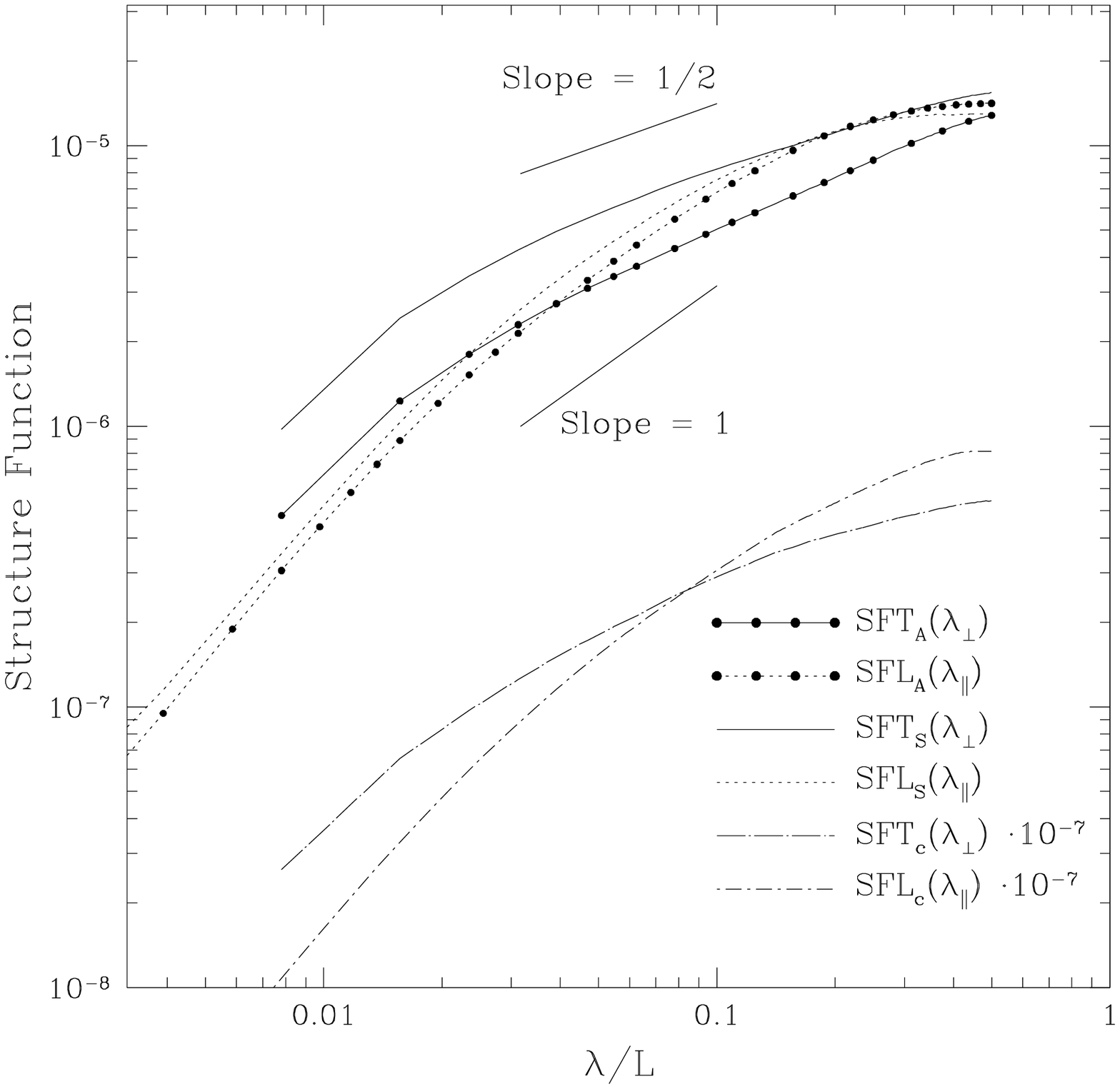}

\caption{\it Transverse and Longitudinal Structure Functions. Structure 
functions transverse and longitudinal to the local magnetic 
field direction are obtained by averaging results from simulations F2, F3, 
and F4 with resolution $128\times 128\times 512$.
\label{fig:structure}} 
\end{figure}

Figure \ref{fig:structure} displays transverse and longitudinal 
structure functions for both shear Alfv\'en and slow waves calculated as 
described in \S \ref{subsec:structurefunctions} from data obtained by
averaging results from simulations 
F2, F3, and F4. The plots are truncated at
$\lambda/L=0.5$ because at greater separations the structure functions are 
affected by the application of periodic boundary conditions. 

%-----------------------------------------------------------------

%-----------------------------------------------------------------
\subsubsection{Anisotropy}
\label{subsubsec:anisotropy}

\begin{figure}
\plotone{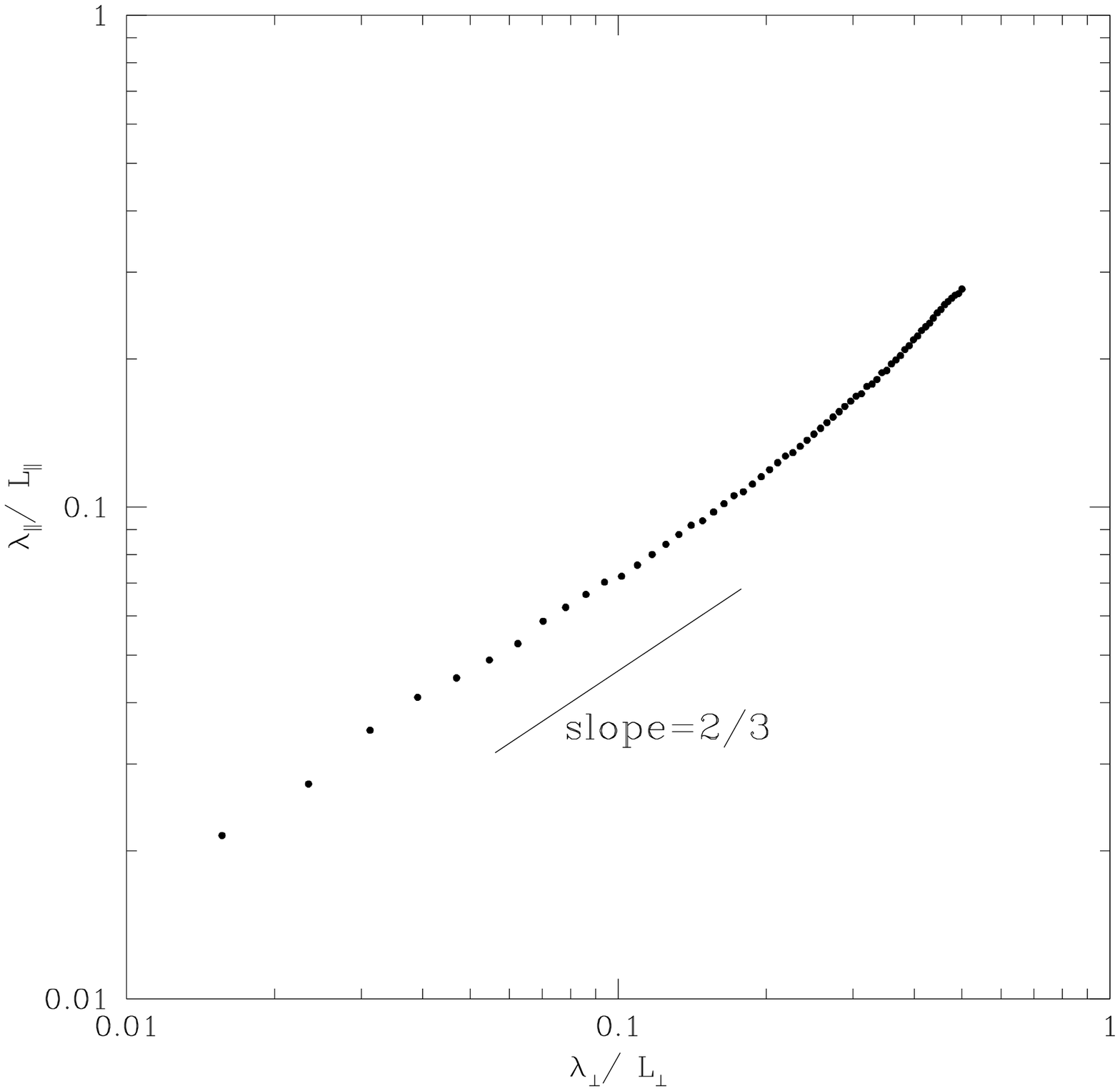}
\caption{\it Ordered Pairs of $\lambda_\perp$ and $\lambda_\parallel$. 
Anisotropy of MHD turbulence is quantified by plotting values of
$\lambda_\perp$ and $\lambda_\parallel$ obtained by setting 
$SFT_A(\lambda_\perp) = SFL_A(\lambda_\parallel)$ using the data displayed in 
Figure \ref{fig:structure}. \label{fig:anisotropy}} 
\end{figure}

Structure functions are the best measure of the scale dependent
anisotropy of an MHD cascade. Ordered pairs of $\lambda_\parallel$ and
$\lambda_\perp$ obtained by equating the longitudinal and transverse
structure functions for shear Alfv\'en waves shown in Figure \ref{fig:structure} 
are plotted in Figure \ref{fig:anisotropy}. We leave it to the reader to judge 
the degree to which this supports the prediction by GS that 
$\lambda_\parallel\propto \lambda_\perp^{2/3}$ in the inertial range.  

%---------------------------------------------------------------------

%-----------------------------------------------------------------

\subsubsection{Ratio of Nonlinear to Linear Time Scales}
\label{subsubsec:chi}

\begin{figure}
\plotone{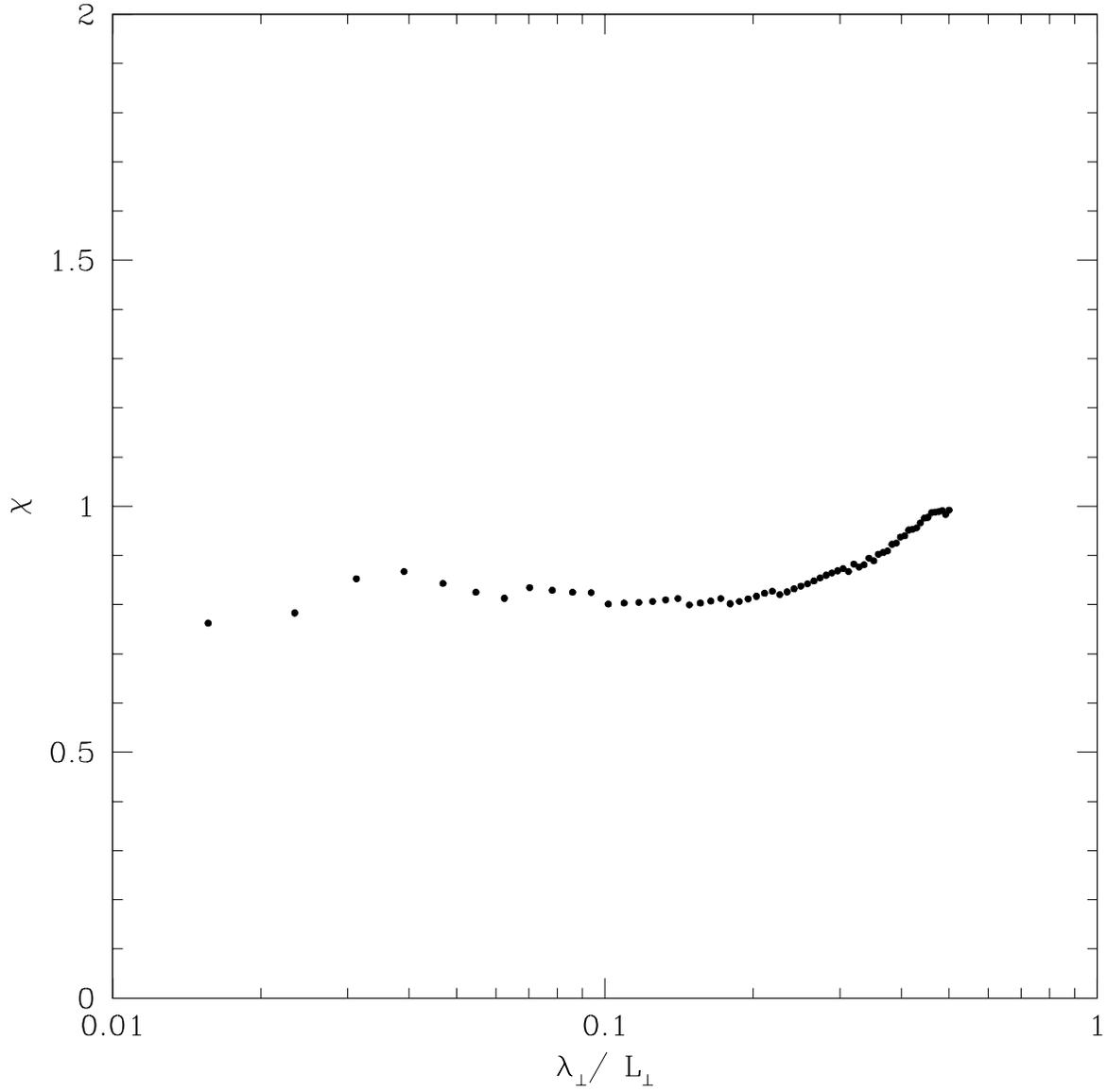}
\caption{\it Critical Balance. Data from Figures \ref{fig:structure} and 
\ref{fig:anisotropy} are
combined to form $\chi$, the ratio of linear to nonlinear time 
scales. Note that $\chi$ has a nearly constant value close to unity
throughout the inertial range. This confirms that MHD turbulence maintains
a state of critical balance.
\label{fig:chi}} 
\end{figure}

The quantity $\chi=(\lambda_\perp v_A)/(\lambda_\parallel
v_{\lambda_\perp})$ is the ratio of the nonlinear to linear timescale 
associated with wave packets of dimensions $(\lambda_\perp,
\lambda_\parallel)$. To evaluate $\chi$ we take 
$v_{\lambda_\perp}=SFT_A^{1/2}(\lambda_\perp)$
from Figure \ref{fig:structure} and $\lambda_\perp/\lambda_\parallel$ from
Figure \ref{fig:anisotropy}. The plot in Figure \ref{fig:chi} establishes 
that $\chi$ maintains a value near unity throughout the inertial range.

%--------------------------------------------------------------------

%-----------------------------------------------------------------

\subsection{Energy Loss}
\label{subsec:energyloss}

\begin{figure}
\plotone{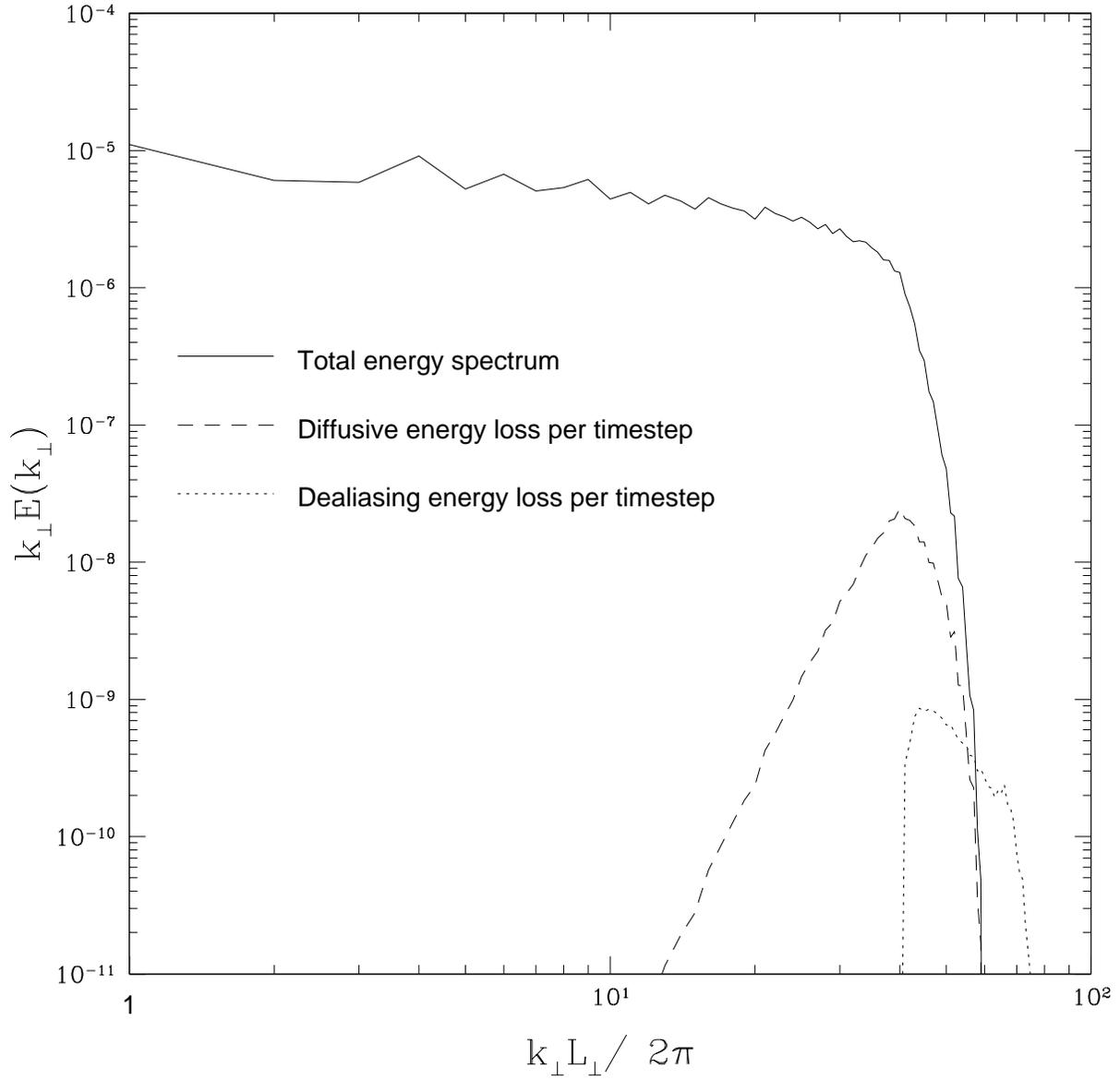}
\caption{\it Energy Loss Per Timestep by Hyperviscous Dissipation and 
Dealiasing. Energy is lost from high $k_\perp$ modes by hyperviscous 
dissipation and dealiasing. The former dominates the latter when integrated over 
the spectrum. Neither is significant in the inertial range. The results shown 
here are from simulation F2. 
\label{fig:energyloss}} 
\end{figure}

Total, mechanical plus magnetic, energy is conserved in the inertial range of 
MHD turbulence. It is lost from high $k_\perp$ modes by a combination of 
hyperviscous dissipation and dealiasing.\footnote{Energy is lost during
dealiasing when we set the amplitudes of modes with $|s_\alpha|>N_\alpha/3$ to 
zero.}
Neither represents reality, but we hope that their effects do not compromise 
inertial range dynamics. Figure 
\ref{fig:energyloss} includes plots from simulation F2 of the hyperviscous 
and dealiasing energy losses per computational timestep. For reference, the 
total power spectrum is also displayed. Hyperviscous dissipation dominates 
dealiasing except at the highest $k_\perp$ where the residual power is 
negligible. This situation is typical of all our simulations. 

%-----------------------------------------------------------------

%-----------------------------------------------------------------

\subsection{Imbalance}
\label{subsec:imbalance}

Because only oppositely directed waves interact, turbulent cascades
tend to become unbalanced. By unbalanced, we mean that unequal
fluxes of energy propagate in opposite directions along the magnetic
field. 

\subsubsection{Forced Turbulence}
\label{subsubsec:forced}

\begin{figure}
\plotone{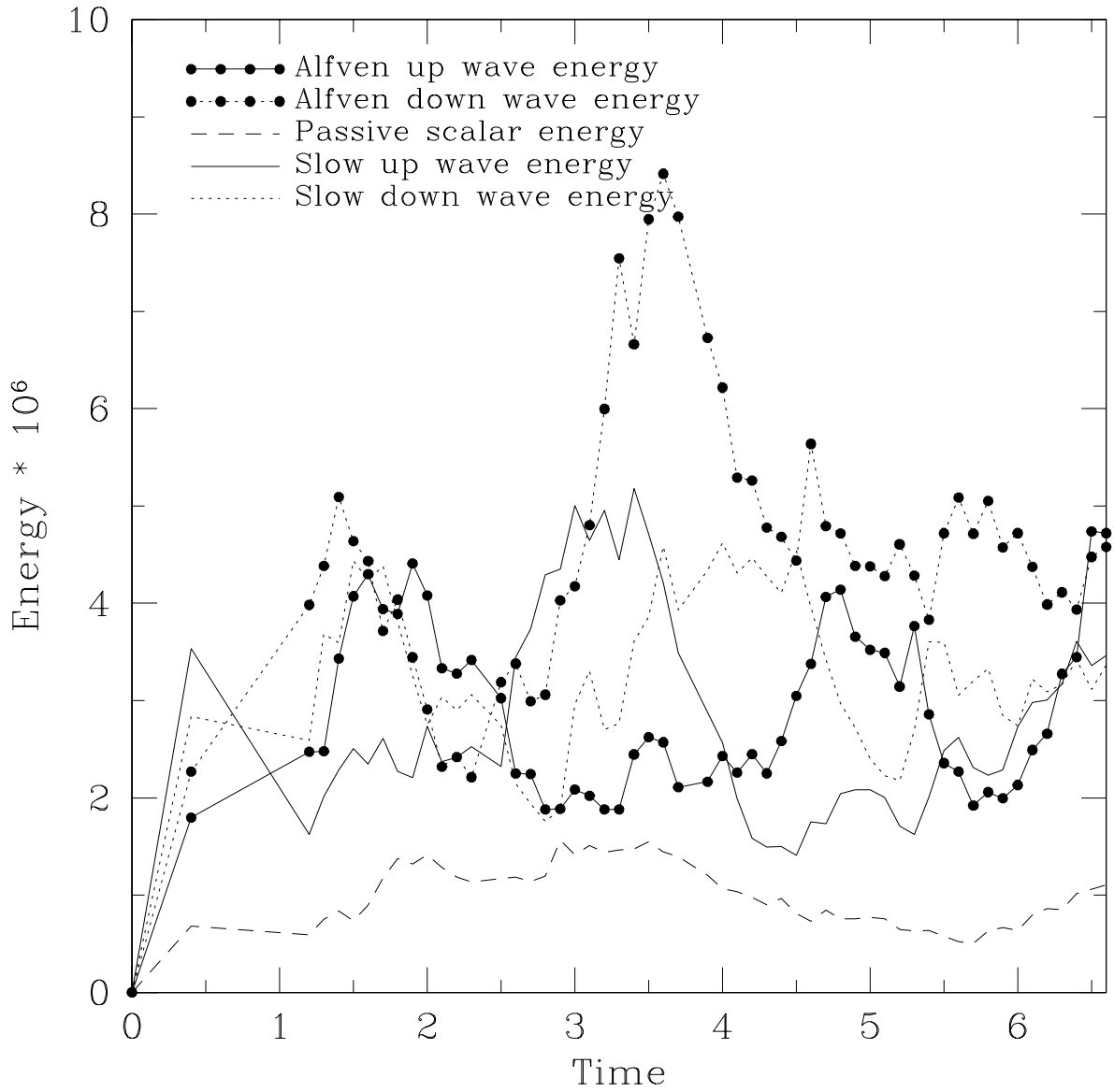}
\caption{\it Forced Turbulence. Mode energies as a function of time for forced 
turbulence from simulation 
F1 of resolution $64\times 64\times 256$.
\label{fig:Evstforced}}
\end{figure}

Mode energies from simulation F1 of forced turbulence with resolution 
$64\times 64\times 256$ are plotted as a function of time in 
Figure \ref{fig:Evstforced}.\footnote{This simulation is the source of initial 
conditions for many higher resolution simulations.} Characteristic fluctuations 
of order unity occur on a time scale $\Delta t=1$. Imbalance appears to 
saturate on longer time scales.

%-----------------------------------------------------------------

%-----------------------------------------------------------------

\subsubsection{Decaying Turbulence}
\label{subsubsec:decay}

\begin{figure}
\plotone{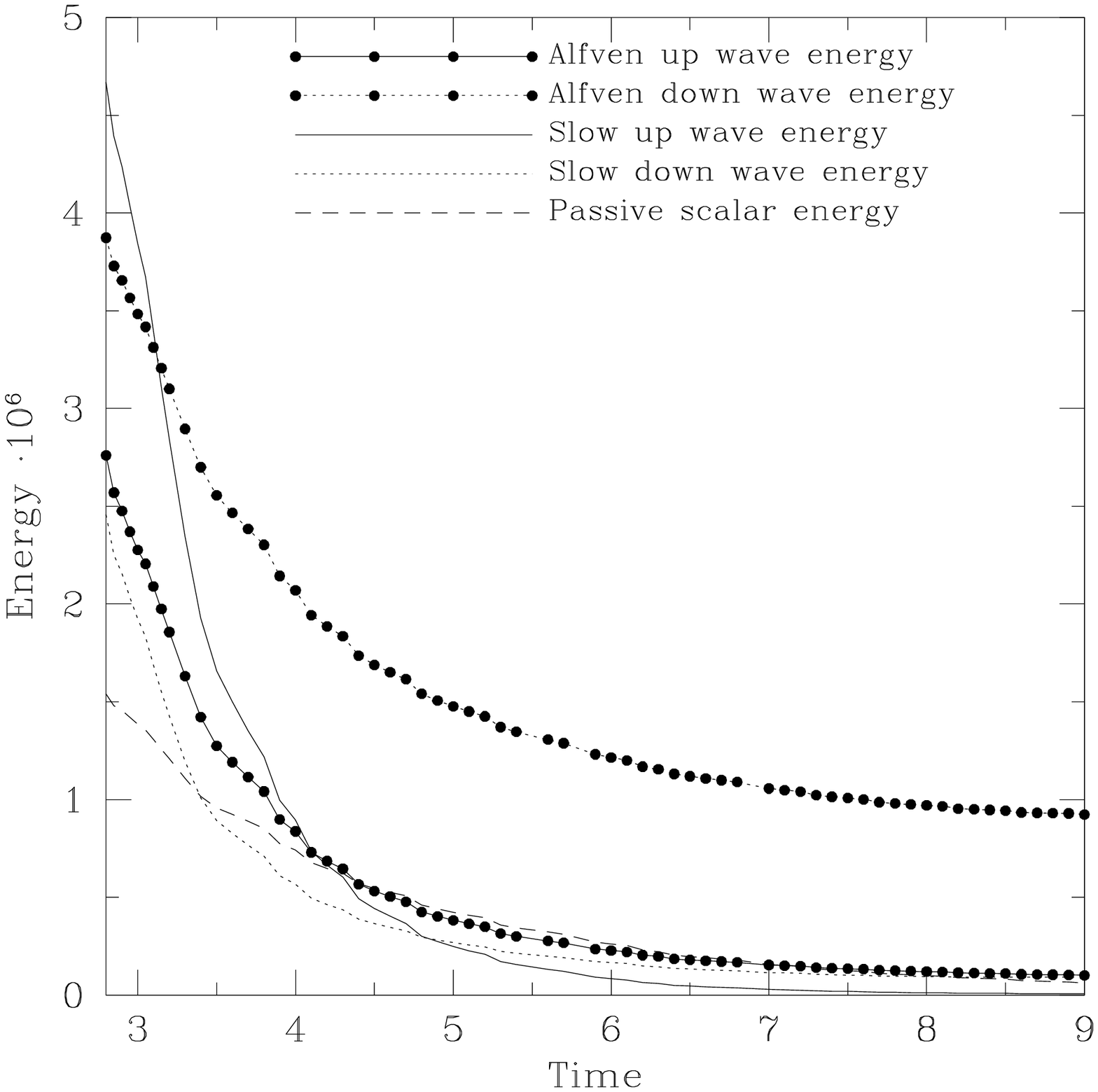}
\caption{\it Decaying Turbulence. Energy as a function of time for shear 
Alfv\'en and slow modes in 
decaying turbulence. The simulation is D1 with resolution $64\times 64\times 
256$.
\label{fig:Evstdecay}} 
\end{figure}

Imbalance is more severe in decaying turbulence. Figure \ref{fig:Evstdecay}
displays energies of individual modes as a function of time obtained from
simulation D1 of resolution $64\times 64\times 256$. The initial imbalance 
increases without limit.  

%-----------------------------------------------------------------

%-----------------------------------------------------------------
\subsection{Passive Role Of Slow Waves}
\label{subsec:slow}

\subsubsection{Cascading Of Slow Waves By Shear Alfv\'en Waves}
\label{subsubsec:cascade}

\begin{figure}
\plotone{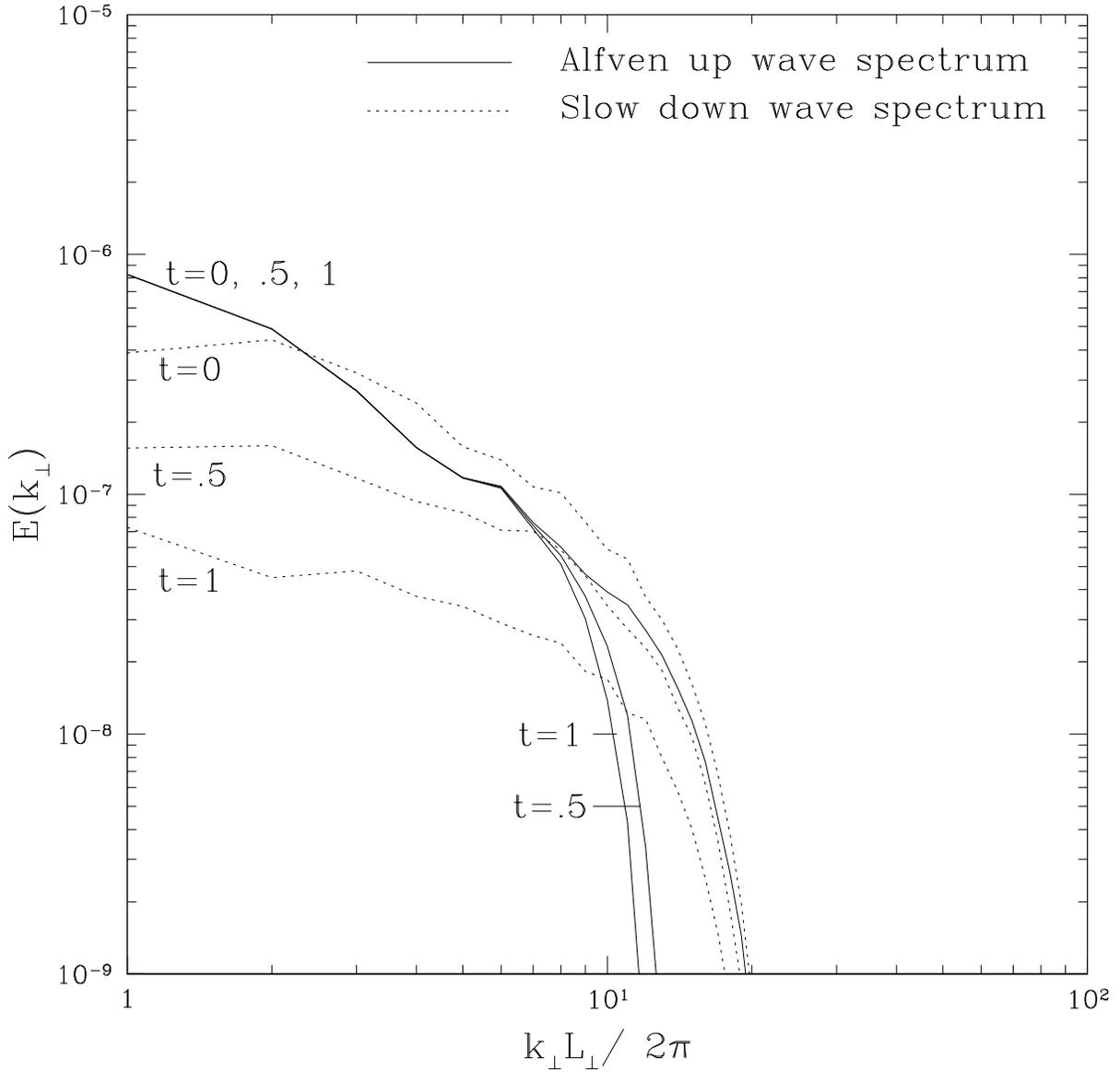}
\caption{\it Passive Role of Slow Waves. Upward moving shear Alfv\'en waves 
interact with downward moving slow 
waves. The power spectrum of the former decays only at large $k_\perp$ whereas
that of the latter decays at all $k_\perp$. Results are taken from simulation
D2 of decaying turbulence with resolution $64\times 64\times 256$.
\label{fig:slowpassive} }
\end{figure}

Simulation D2 of decaying turbulence with resolution $64\times 64\times 256$ 
is designed to assess the mutual effects of shear Alfv\'en waves on slow waves 
and vice versa. We initialize it by removing the upward propagating slow waves 
and the downward propagating shear Alfv\'en waves from simulation F1 at 
$t=6.6$. It is then run for $\Delta t=1$. Figure \ref{fig:slowpassive} 
illustrates the evolution of the energies in upward propagating shear Alfv\'en 
waves and downward propagating slow waves. The only change in the spectrum of 
shear Alfv\'en waves is a decay at large $k_\perp$ which is entirely 
attributable to energy loss by hyperviscosity and dealiasing. By contrast, the 
spectrum of slow waves decays at all $k_\perp$ at a rate consistent with that
shown in Figure \ref{fig:Evstdecay}. These findings demonstrate that shear 
Alfv\'en waves control the MHD cascade and that the slow waves play a passive
role.

%-----------------------------------------------------------------

%-----------------------------------------------------------------
\subsubsection{Conversion Of Shear Alfv\'en Waves To Slow Waves}
\label{subsubsec:convert}

\begin{figure}
\plotone{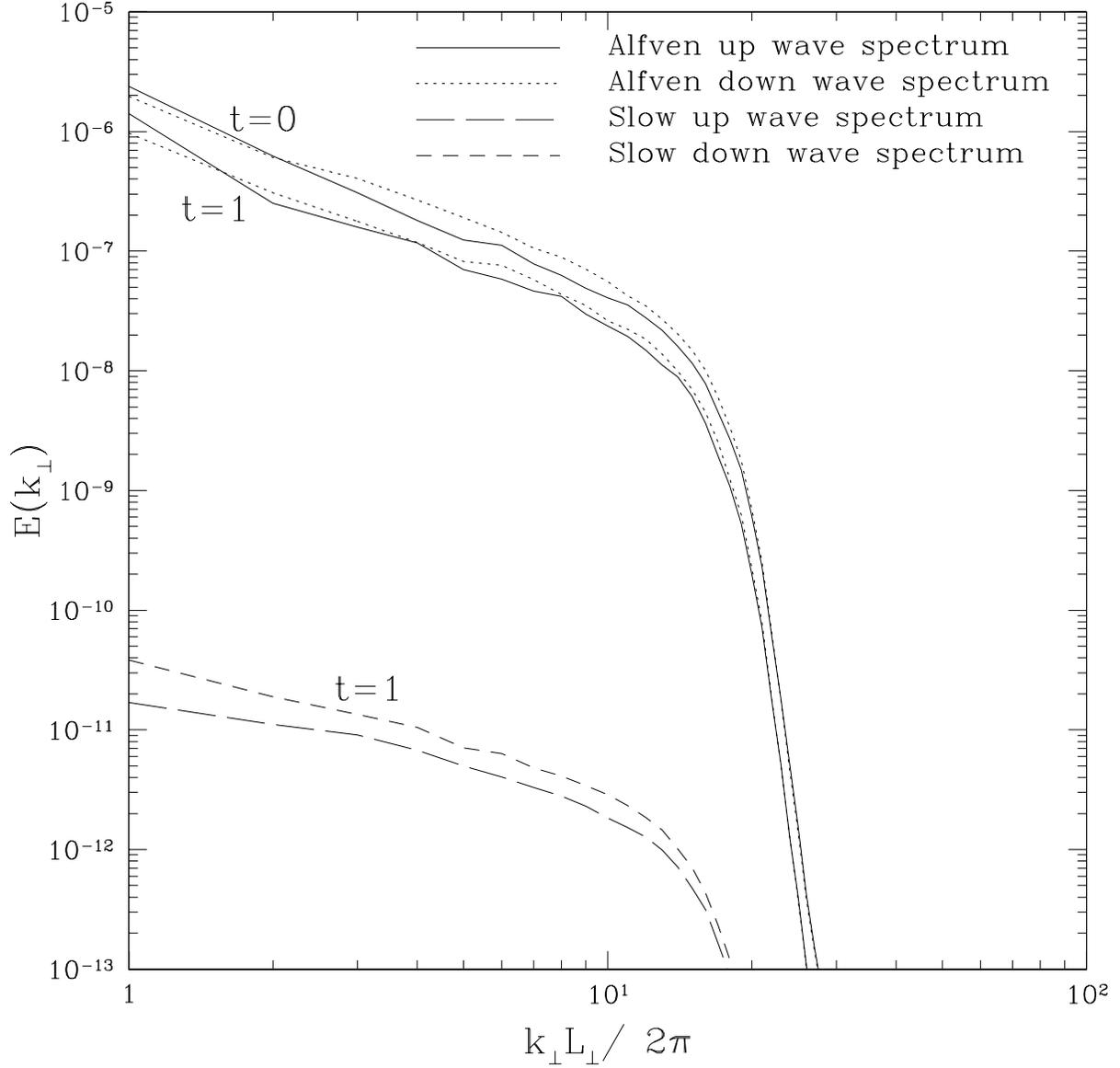}
\caption{\it Negligible Conversion of Shear Alfv\'en to Slow Waves. Slow wave 
production in a simulation of decaying turbulence initialized 
with pure shear Alfv\'en waves. After one decay time the slow waves contain 
$\lesssim 10^{-4}$ of the
total energy. This amount is indistinguishable from the false slow
waves that our spectral decomposition procedure would report due to
the tilt of the local magnetic field relative to the global $z$ axis.
Data plotted comes from simulation D3 with resolution $64\times 64\times 
256$. Simulation F1 provides the initial conditions for simulation D3.
\label{fig:D3} }
\end{figure}

Simulation D3 of decaying turbulence with resolution $64\times 64\times 256$
is tailored to measure the rate at which shear Alfv\'en waves 
are converted into slow waves. It is initialized from F1 at $t=6.6$ by 
removing all slow waves and then run for $\Delta t=1$. As demonstrated by 
Figure \ref{fig:D3}, at the end of this interval, which corresponds to 
about a decay time at the outer scale (see Fig. \ref{fig:Evstdecay}), the slow 
waves carry negligible energy. The small admixture shown may result from
the limited ability of our scheme of spectral decomposition to distinguish 
slow waves from shear Alfv\'en waves as discussed in \S 
\ref{subsec:spectraldecomp}.  

%-----------------------------------------------------------------

%-----------------------------------------------------------------

\subsection{Cascade Diagnostics}
\label{subsec:diagnostics}

We design special simulations to exploit the fact that only oppositely directed
waves interact. Each of these is initialized by removing all but a narrow band 
in $k_\perp$ of up waves from a fully developed forced simulation. These are 
then run without forcing so that we can observe the evolution of the energy in 
the up band as it spreads into adjacent bands. Since the down waves evolve 
weakly, we restrict the lengths of these runs to $\Delta t=1/2$ so that 
interactions do not repeat. 

Initial conditions for simulations D4, D5, D6, and D7 are
provided by band-filtering simulation F2 at $t=2.8$ with up modes
retained from $2 \leq s_\perp \leq 4$, $4 \leq s_\perp \leq 8$, $8
\leq s_\perp \leq 16$, and $16 \leq s_\perp \leq 32$,
respectively. Each of these simulations has resolution $128\times
128\times 512$. Resolution $256\times 256\times 512$ simulations D8
and D9 are initialized from simulation F5 by band-filtering at
t=2.95 with up modes retained from $16 \leq s_\perp \leq 32$ and $32
\leq s_\perp \leq 64$, respectively.

\subsubsection{Absence Of An Inverse Cascade}
\label{subsubsec:inverse}

\begin{figure}
\plotone{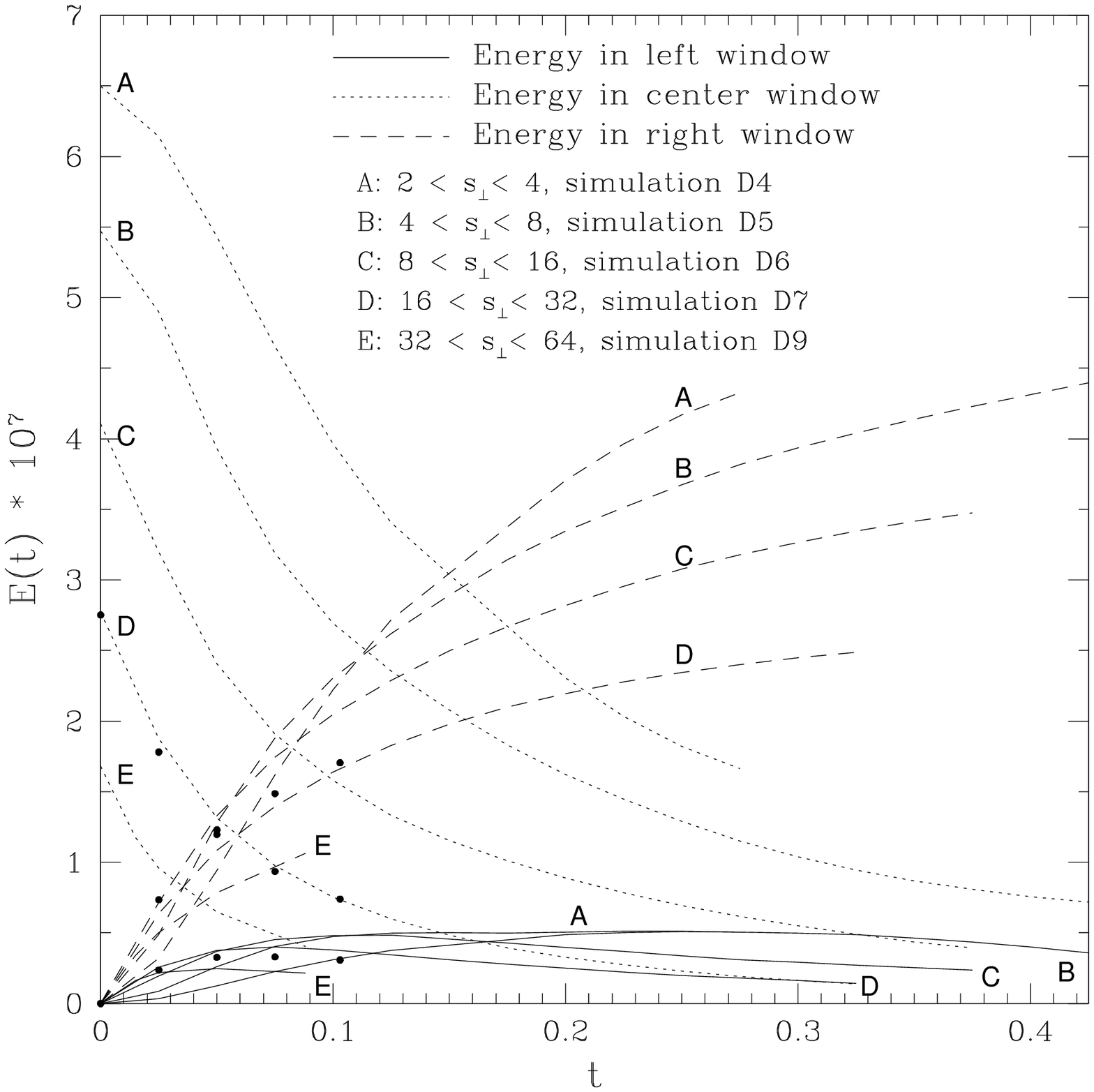}

\caption{\it Summary of Bandpass Filtered Simulations. We plot the energy as a 
function of time in the center band and in the 
bands immediately to its left and right for each bandpass filtered simulation.  
Initially all of the energy is in the central band. As time passes it spreads
into adjacent bands.
Points correspond to simulation D8, which is initialized with the same up 
mode band as simulation D7 but with twice the transverse resolution in the 
down modes.  There is good agreement between simulations D7 and D8 as
shown 
in more detail in Figure \ref{fig:reflect}.
\label{fig:band}}
\end{figure}

\begin{figure}

\plotone{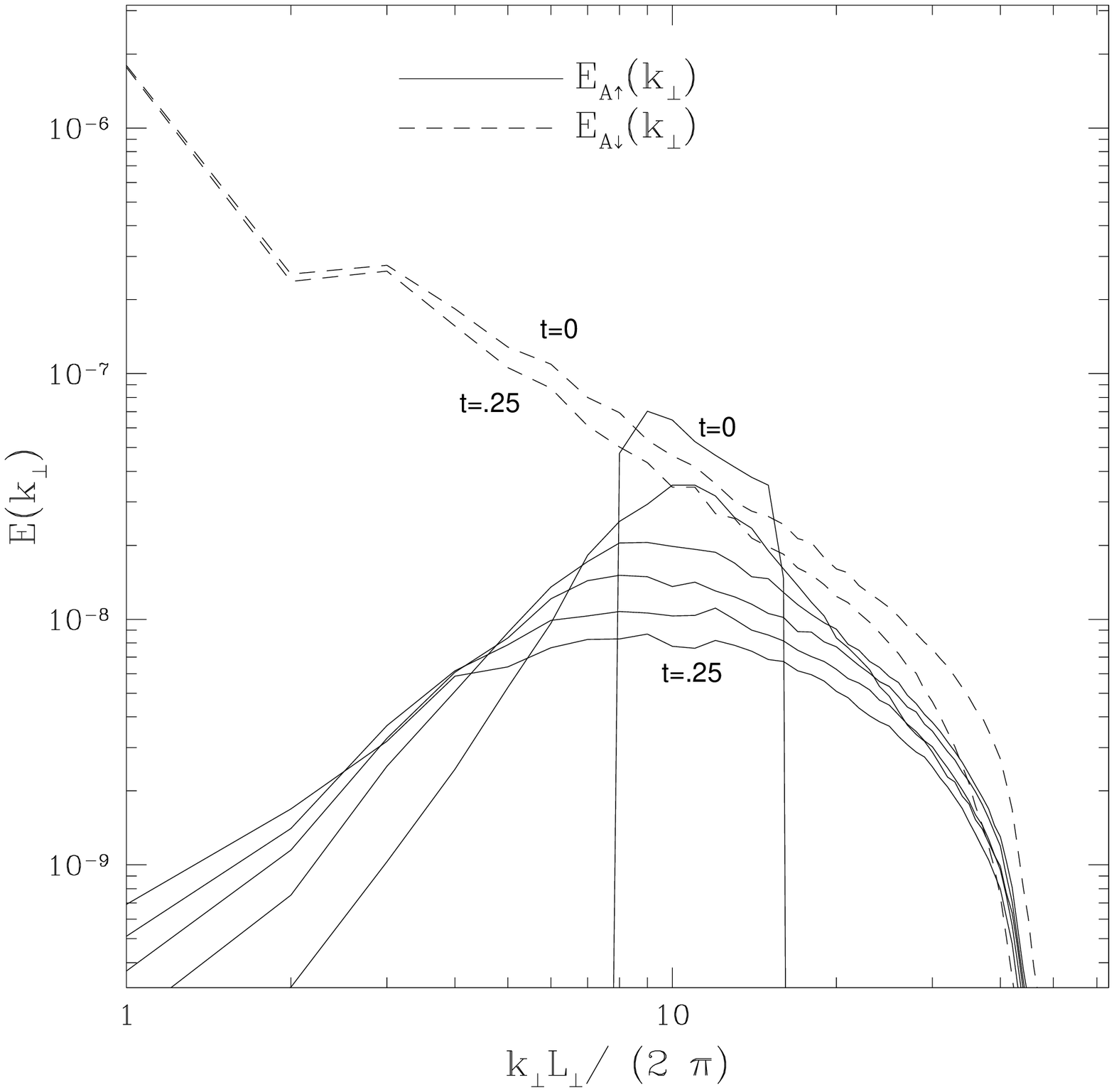}

\caption{\it Absence of an Inverse Cascade. Dashed lines depict the initial and 
final down wave spectra. The up wave 
spectra are plotted at a succession of times differing by $\Delta t=.05$. 
Almost all of the energy that leaves the band $8 \leq s_\perp \leq 16$ moves to 
higher $s_\perp$. These data are taken from simulation D6.
\label{fig:inverse}}
\end{figure}

Figure \ref{fig:band} summarizes how energy spreads from each of selected 
band into adjacent bands. It establishes that the predominant movement is 
toward higher $k_\perp$. There is no evidence for an inverse cascade. A more
detailed demonstration for the selected band $8 \leq s_\perp \leq 16$ is 
provided in Figure \ref{fig:inverse} which is based on simulation D6. 

%-----------------------------------------------------------------

%-----------------------------------------------------------------
%----------------------------------------------------------------

\subsubsection{Resolution Dependence}
\label{subsubsec:resolution}

\begin{figure}
\plotone{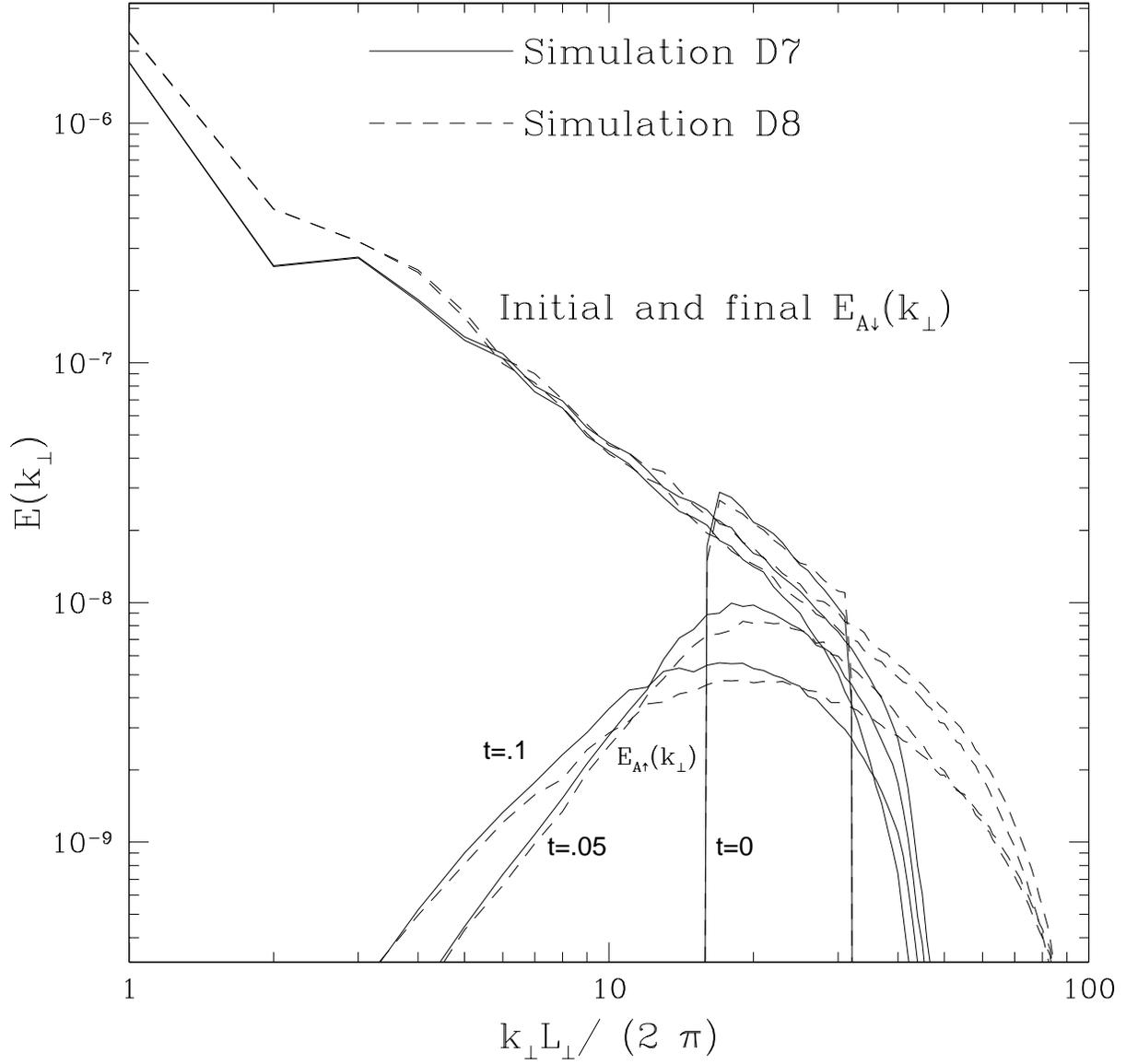}

\caption{\it Comparison of Simulations at Different Resolution. Increased 
resolution has little effect on the evolution of energy
in the left-hand and central bands. Thus the latter resides in the
inertial range in even the lower resolution simulation. Energy which
moves into the right-hand band is more rapidly dissipated in the lower
resolution simulation and more effectively stored in the higher
resolution one. 
\label{fig:reflect}} 

\end{figure}

Figure \ref{fig:reflect} compares results from simulation D7 of resolution
$128\times 128\times 512$ with those from simulation D8 of resolution 
$256\times 256\times 512$. Each simulation is initialized with energy in
up waves confined to the band $16 \leq s_\perp \leq 32$. Note how well the
energies in the central and left-hand bands from the two simulations match as 
they evolve. This establishes that the largest $k_\perp$ band in simulation 
D7 is a valid part of the inertial range.

%-----------------------------------------------------------------

%-----------------------------------------------------------------
\subsubsection{Cascade Time}
\label{subsubsec:cascadetime}

As a standard measure of the time scale for energy transfer across 
$\lambda_\perp$, we take $t_c \sim \lambda_\perp / v_{\lambda_\perp}$, where 
$v_{\lambda_\perp}$ is obtained from the transverse structure functions of 
downward propagating Alfv\'en waves according to $2v^2_{\lambda_\perp} = 
SFT_{A\downarrow}(\lambda_\perp)$. 
Banded simulations also permit a more 
direct measure of the cascade time as that at which the energy in the
right-hand band matches that in the central band. We identify this version by 
the symbol $t_h$. 

Values for the different types of cascade time are given in Table 
\ref{tab:cascadetimes}. Even for 
the lowest $k_\perp$ band, each is substantially smaller than the time, $\Delta 
t=1$, that waves take to cross the computational box.

\medskip

\begin{deluxetable}{llllll}
\tabletypesize{}
\tablecaption{Cascade Times \label{tab:cascadetimes}}
\tablewidth{0pt}
\tablehead{
\colhead{$s_\perp$} &
\colhead{$\lambda_\perp/L_\perp$} &
\colhead{$v_\perp$} &
\colhead{$t_c = \lambda_\perp / v_\perp$} &
\colhead{$t_h$} &
\colhead{$t_h / t_c$}
}
\startdata
2-4 &.188  &$3.28\cdot10^{-3}$&\quad .115 &.152&1.32\\
4-8 &.094  &$2.67\cdot10^{-3}$&\quad .070 &.115&1.64\\
8-16&.047  &$2.10\cdot10^{-3}$&\quad .045 &.080&1.78\\
16-32&.023 &$1.58\cdot10^{-3}$&\quad .029 &.058&2.00\\
32-64&.0117&$1.29\cdot10^{-3}$&\quad .0181&.044&2.43\\

\enddata

\end{deluxetable}

\begin{figure}
\plotone{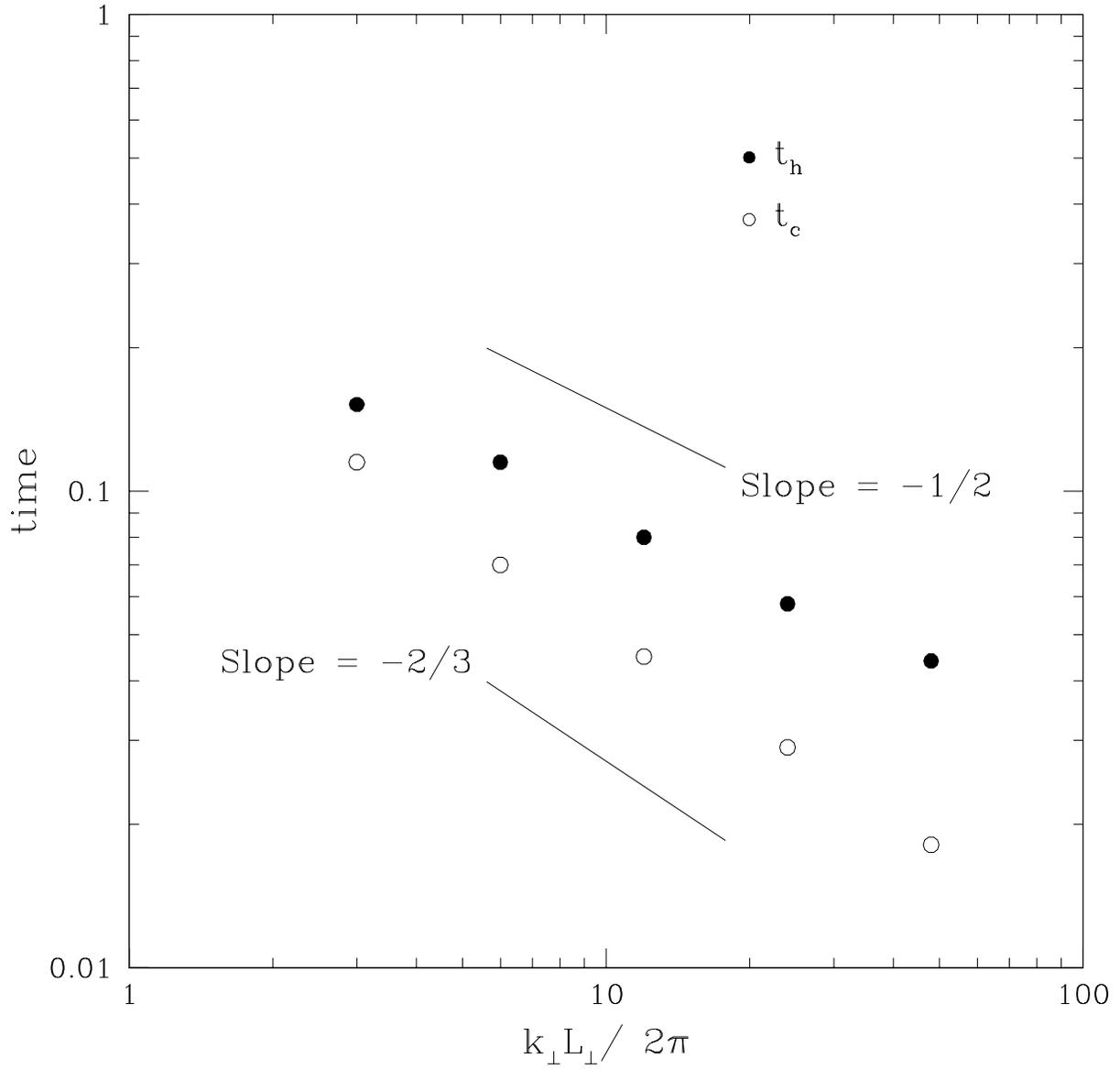}

\caption{\it Cascade Times. A comparison of cascade times based on different 
definitions. See 
text for details. 
\label{fig:cascadetime}}

\end{figure}

For ease of comparison, we plot the tabulated values against $k_\perp 
L_\perp/(2\pi)$ in Figure 
\ref{fig:cascadetime}. 
Note that $t_h$ declines much more slowly with increasing $k_\perp$ than
$t_c$ does. 

%-------------------------------------------------------------------

%-----------------------------------------------------------------

\subsection{Intermittency}
\label{subsec:intermittency}

\begin{figure}

\plottwo{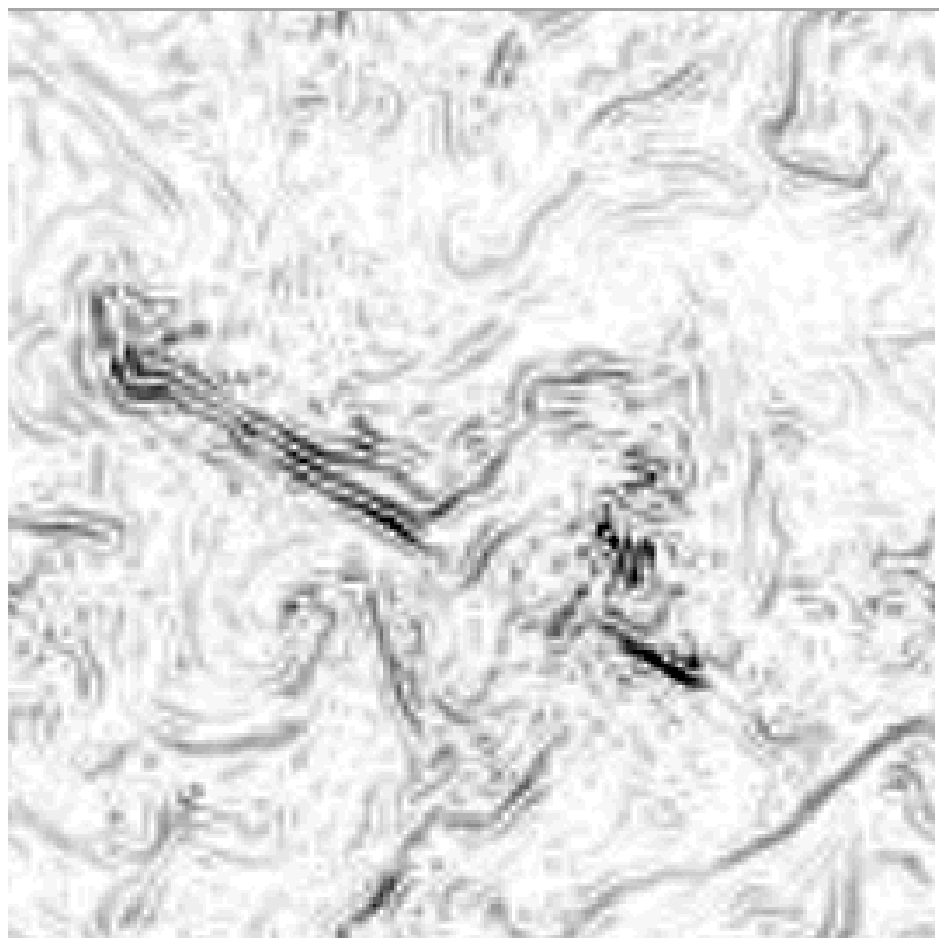}{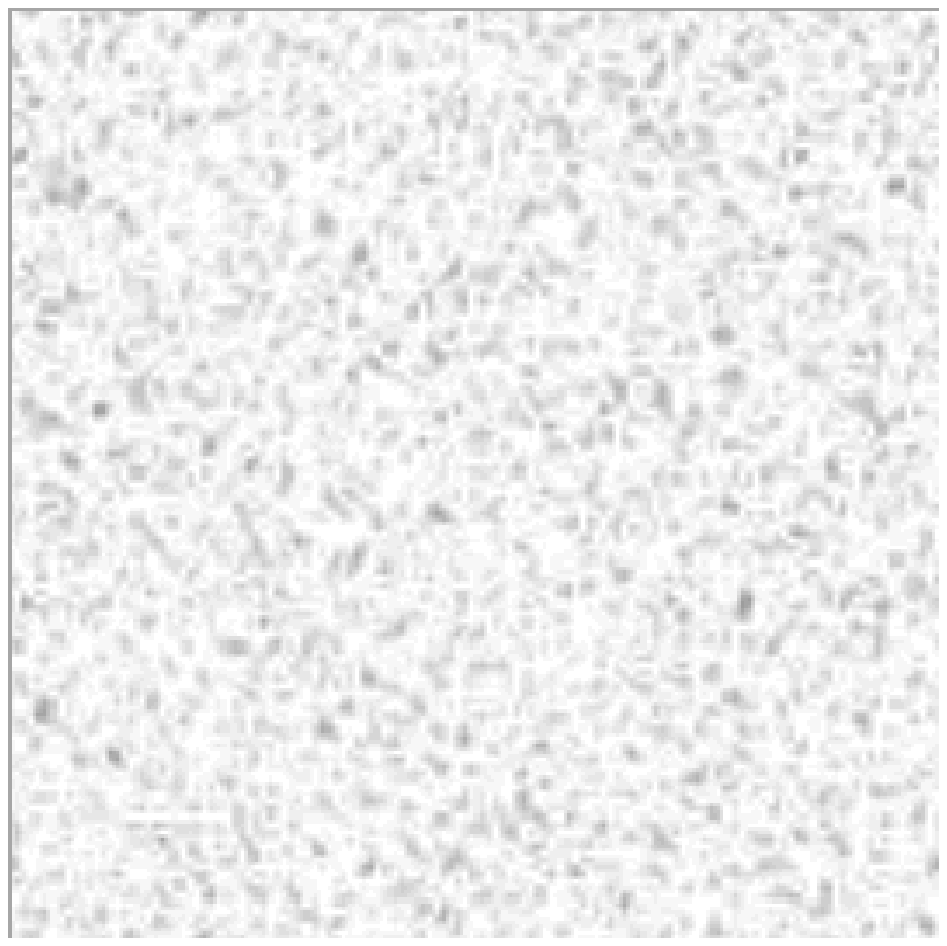} 

\caption{\it Alfv\'en Up Modes. The left-hand panel
shows a grey scale image of $|\bnabla\times \A_\uparrow|$ in 
a $(x,y)$ slice at $z=0$. For comparison, an image based on the same Fourier 
coefficients with random phases is shown in the right-hand panel.
\label{fig:Aup}}
\end{figure}

\begin{figure}

\plottwo{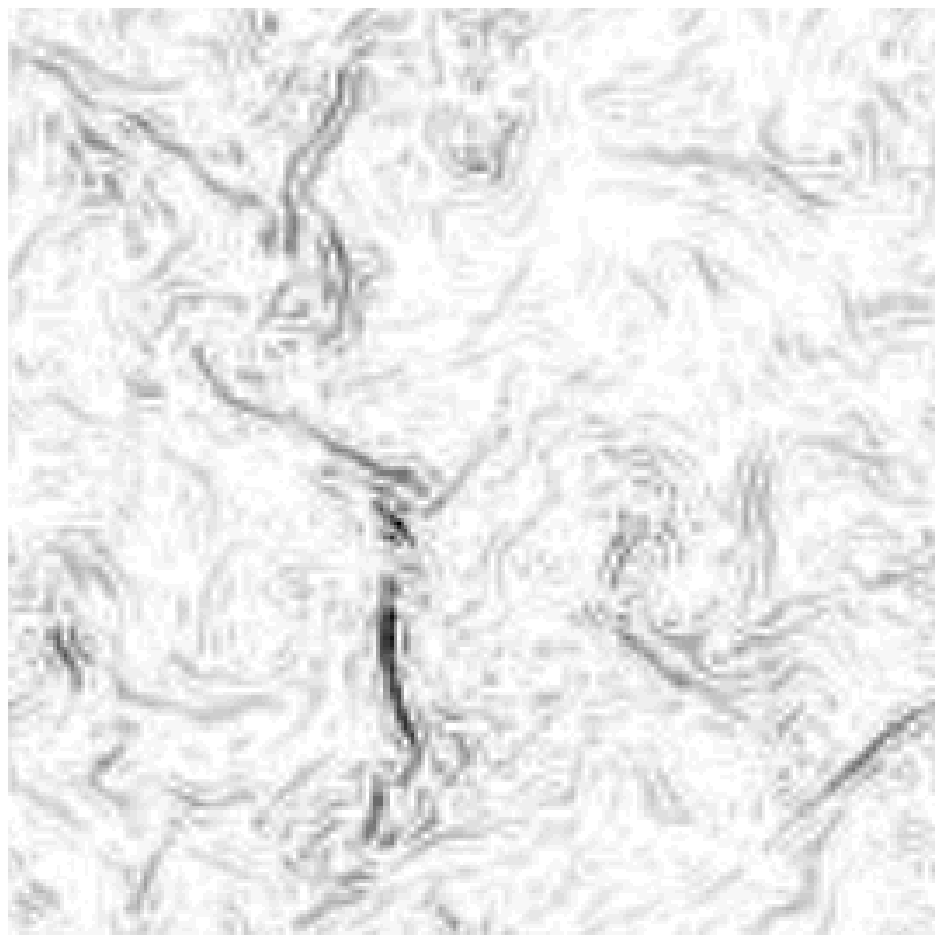}{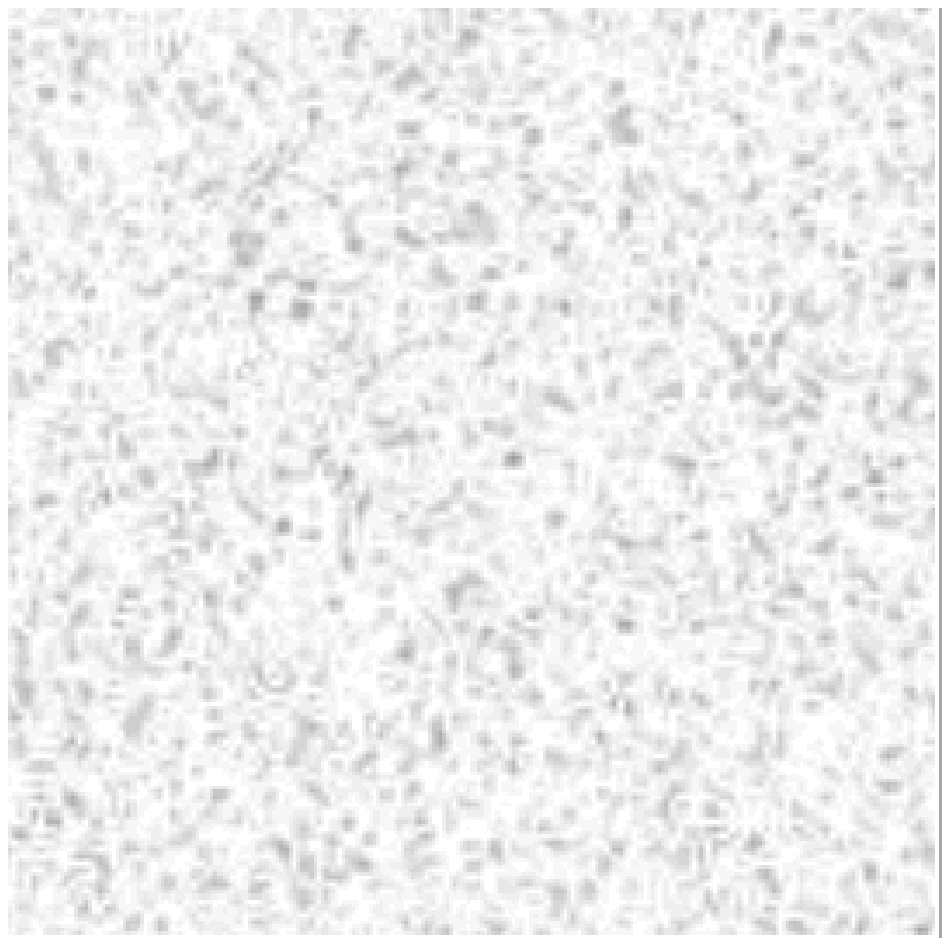}

\caption{\it Alfv\'en Down Modes. The left-hand panel shows a grey scale image 
of $|\bnabla\times
\A_\downarrow|$ in a $(x,y)$ slice at $z=0$. For comparison, an image based on
the same Fourier coefficients with random phases is shown in the
right-hand panel.
\label{fig:Adw}}
\end{figure}

\begin{figure}

\plottwo{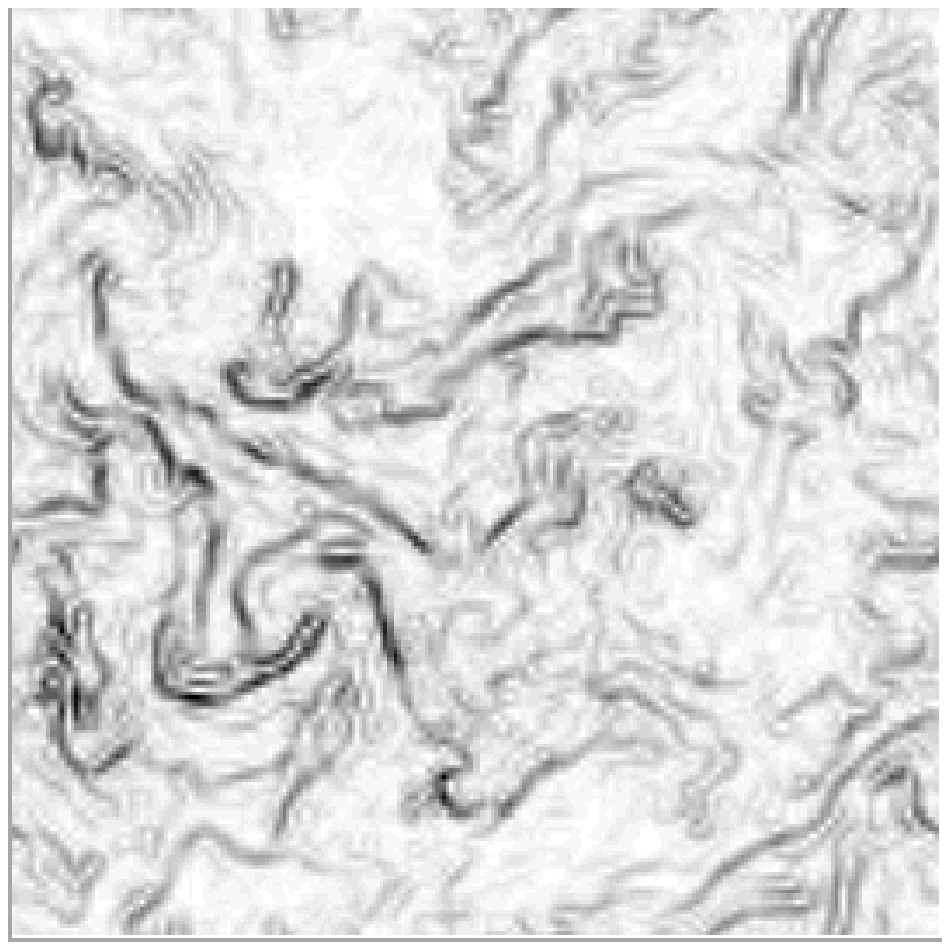}{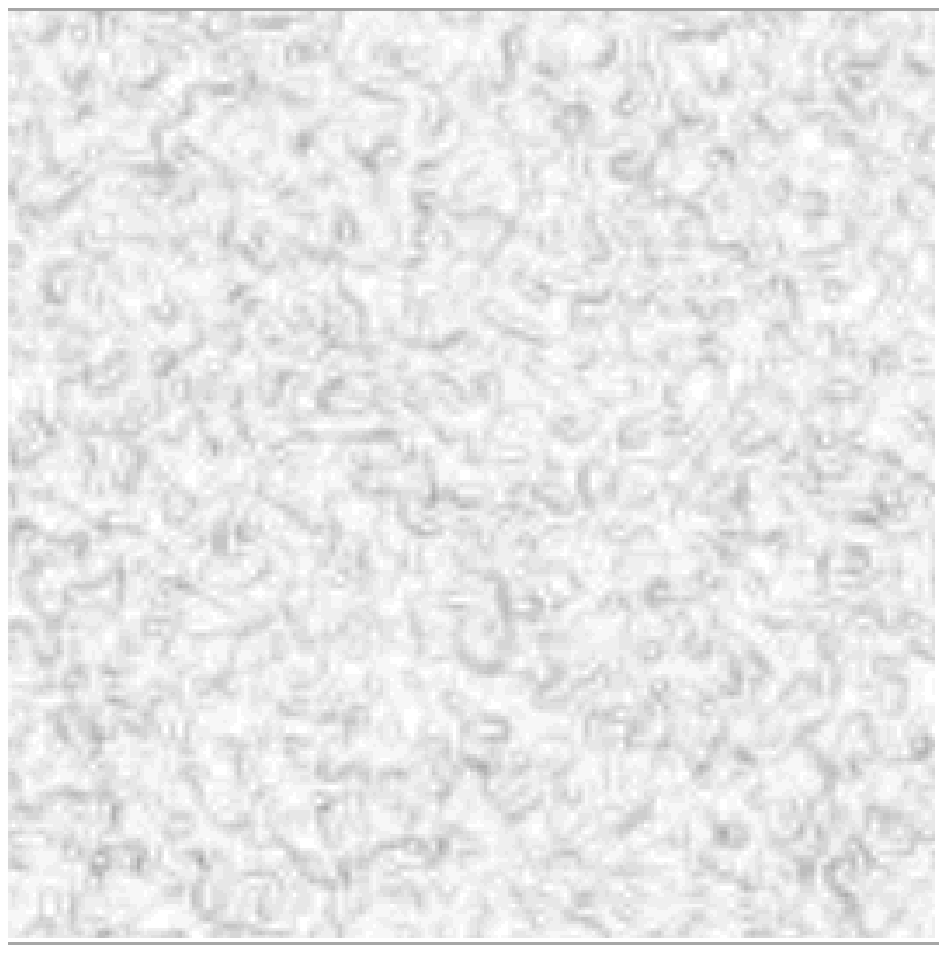}

\caption{\it Slow Up Modes. The left-hand panel shows a grey scale image of 
$|\bnabla\times
\bS_\uparrow|$ in a $(x,y)$ slice at $z=0$. For comparison, an image based
on the same Fourier coefficients with random phases is shown in the
right-hand panel.
\label{fig:Sup}}
\end{figure}

\begin{figure}

\plottwo{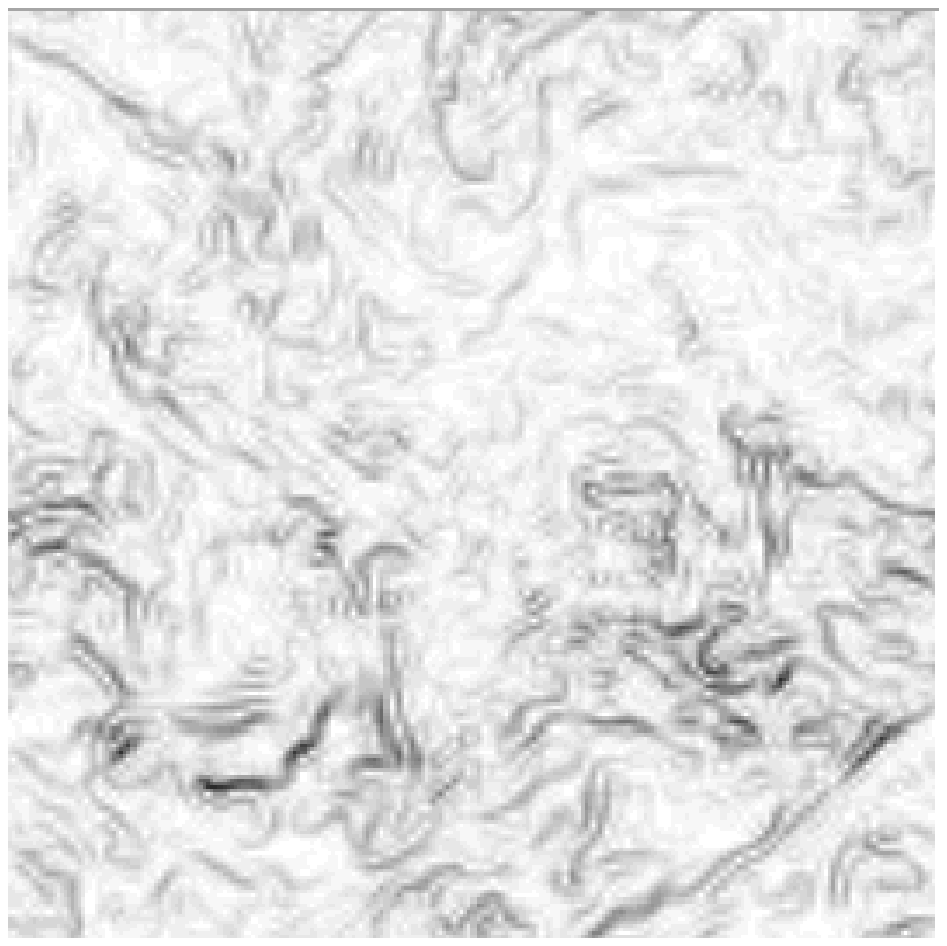}{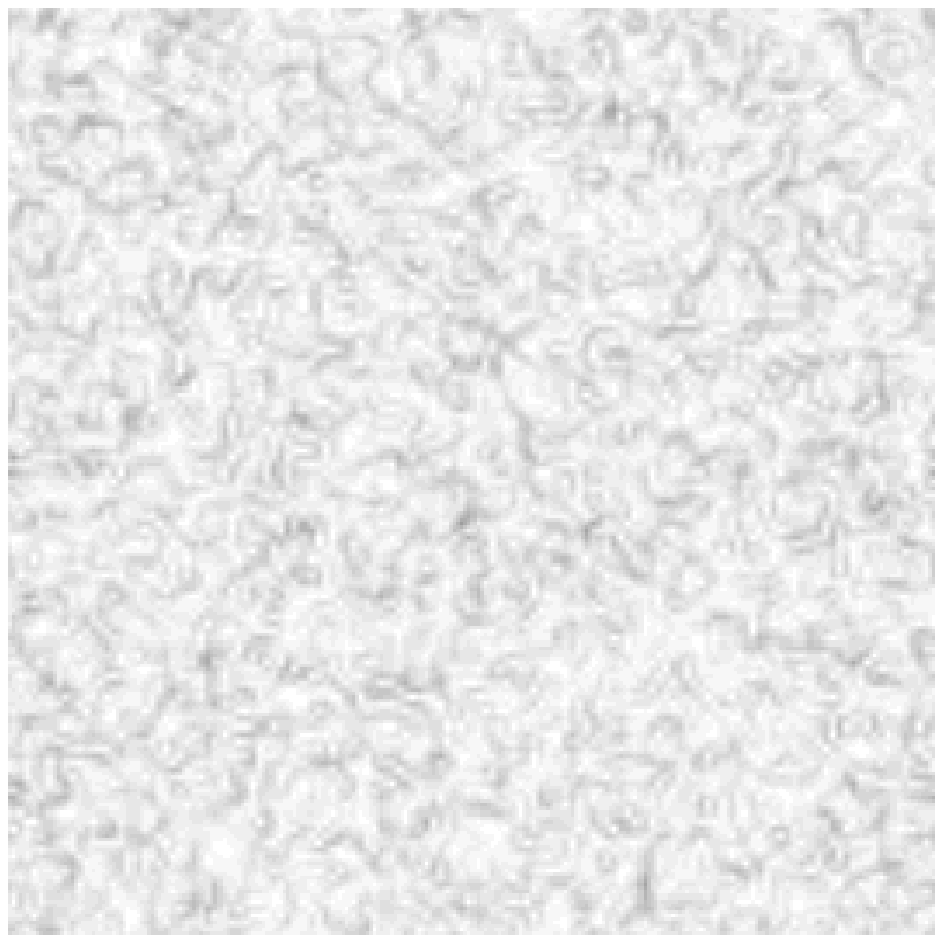}

\caption{\it Slow Down Modes. The left-hand panel shows a grey scale image of 
$|\bnabla\times
\bS_\downarrow|$ in a $(x,y)$ slice at $z=0$. For comparison, an image based
on the same Fourier coefficients with random phases is shown in the
right-hand panel.
\label{fig:Sdw}}
\end{figure}

Simulations of hydrodynamic turbulence exhibit structure that is not
seen in random phase realizations of velocity fields with identical
power spectra \citep{bib:Jimenez}. We find the same to
be true for MHD turbulence. This is illustrated in Figures
\ref{fig:Aup}-\ref{fig:Sdw}. The left-hand panels display magnitudes
of the curls of upward and downward propagating shear Alfv\'en and
slow waves in a $(x,y)$ slice at $z=0$ taken from simulation
F5. Randomizing the phases of the Fourier coefficients used to
generate the left-hand panels yields the images shown in the
right-hand panels. Coherent structures, which are conspicuous in the
former, are absent in the latter.

While the eye does an excellent job of recognizing intermittency, it is helpful
to have a quantitative measure. To accomplish this, we apply a sequence of
high-pass filters to the Fourier coefficients of the Elsasser fields and a
sequence of low-pass filters to their gradients.  A filter is identified by a
value of $s_\perp$.  High-pass filters remove modes with smaller $s_\perp$ and
low-pass filters remove those with larger $s_\perp$.  Transverse structure in
the Elsasser fields is dominated by low $s_\perp$ modes and that in their
gradients by high $s_\perp$ modes.  Applying a sequence of high pass filters
with increasing $s_\perp$ to the Fourier coefficients of the Elsasser fields
emphasizes structure of decreasing scale. Similarly, applying a sequence of low
pass filters with increasing $s_\perp$ to the Fourier coefficients of the
gradients of the Elsasser fields targets structure of increasing scale.

Filtered data is obtained from the simulations used to produce Figures
\ref{fig:Aup}-\ref{fig:Sdw}. Normalized fourth order moments of relevant 
quantities $q$ are computed according to 
\be
M_4(q) = \frac{\langle q^4\rangle}{\langle q^2\rangle^2},
\label{eq:moment4}
\ee
where angular brackets denote volume average.

\begin{figure}
\plotone{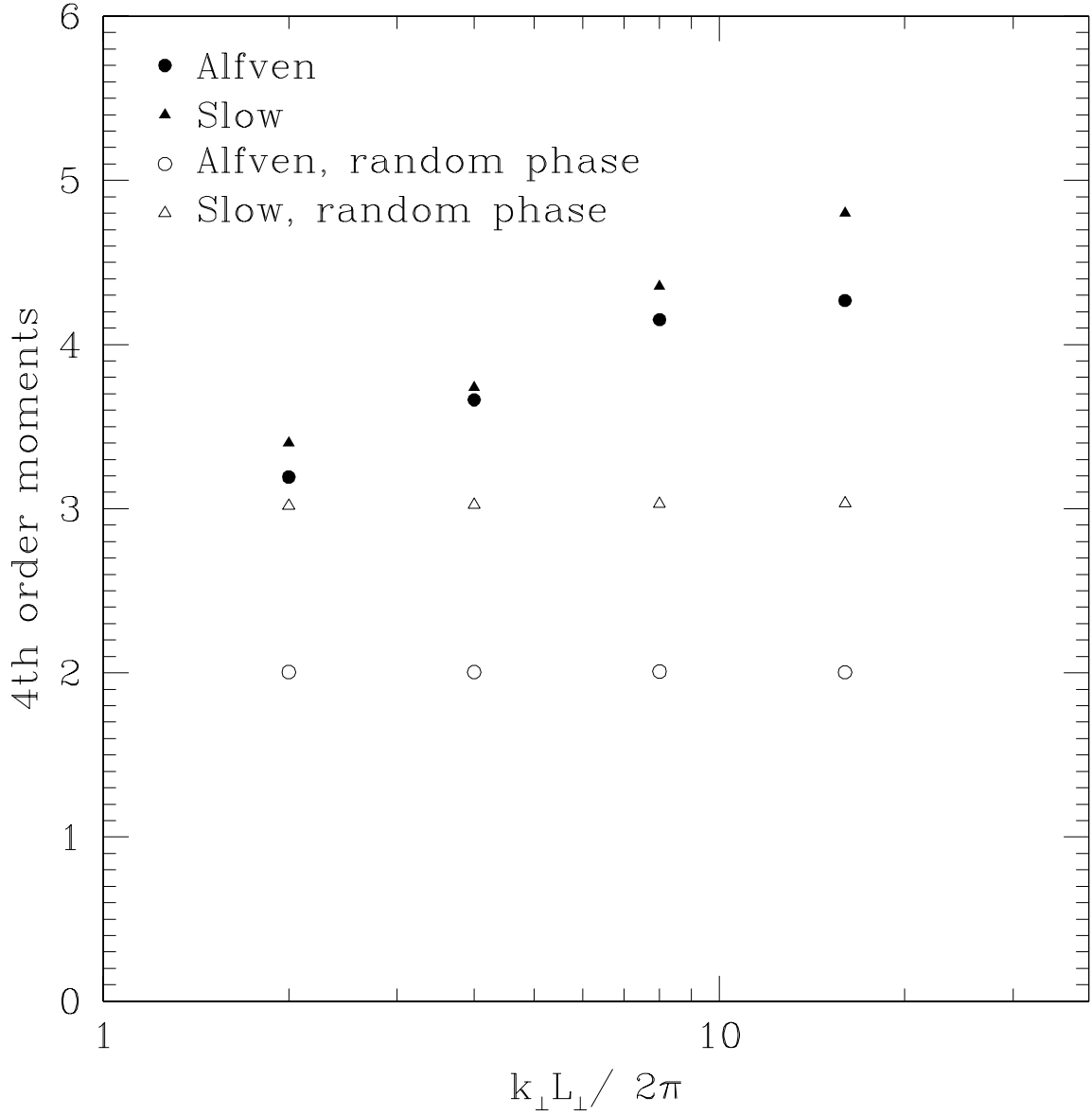}
\caption{\it Normalized 4th Order Moments for Alfv\'en and Slow Modes. Plotted 
points are average values of moments obtained from upward and 
downward propagating waves. The location of the high-pass filter is denoted by 
$k_\perp$.
\label{fig:highpass}}
\end{figure}

\begin{figure}
\plotone{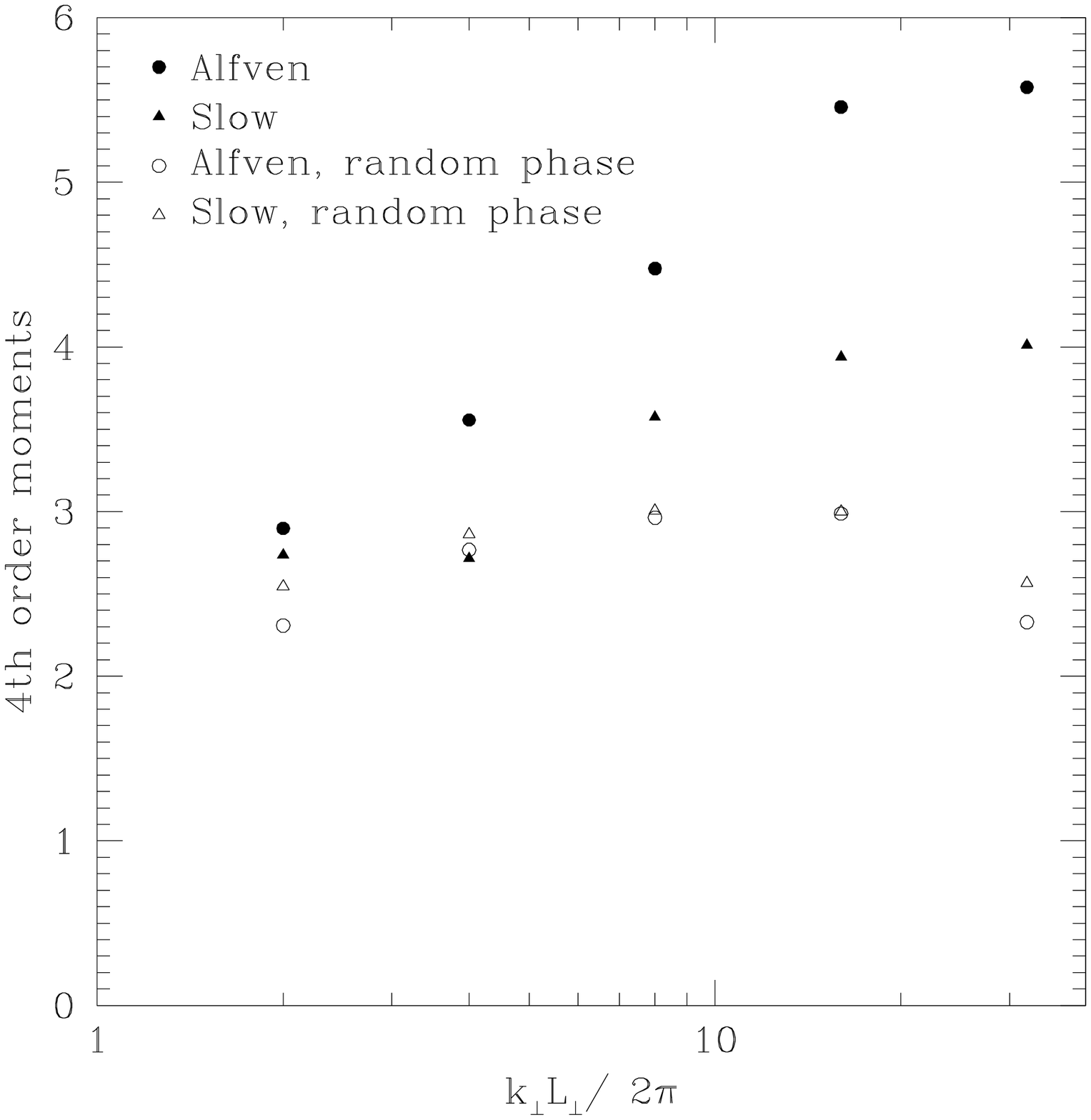}
\caption{\it Normalized 4th Order Moments from Gradients of Alfv\'en and Slow 
Modes. Gradients are defined as $\partial_x q$, where $q$ is one of the Elsasser 
fields.  Moments obtained from upward and downward propagating waves are 
averaged. The location of the low-pass filter is denoted by $k_\perp$.
\label{fig:lowpass}}
\end{figure}

Figure \ref{fig:highpass} displays moments of the Elsasser fields as a function 
$k_\perp$ for high-pass filtered data. Moments of gradients of the
Elsasser fields as a function of $k_\perp$ for low-pass filtered data are
plotted in Figure \ref{fig:lowpass}. For comparison, each figure includes 
moments obtained from the random phase versions of the corresponding 
simulations. It is worth noting that $M_4(q)=1+2/n$ for data obeying 
n-dimensional Gaussian statistics, and that slow waves correspond to $n=1$ and 
shear Alfv\'en waves to $n=2$ in the limit $k_\perp\gg k_z$. 

%-----------------------------------------------------------------

%-----------------------------------------------------------------

%-----------------------------------------------------------------

%-----------------------------------------------------------------

%-----------------------------------------------------------------

%-----------------------------------------------------------------

%-----------------------------------------------------------------

\subsubsection{Passive Scalar}
\label{subsubsec:passivescalar}

\begin{figure}
\plottwo{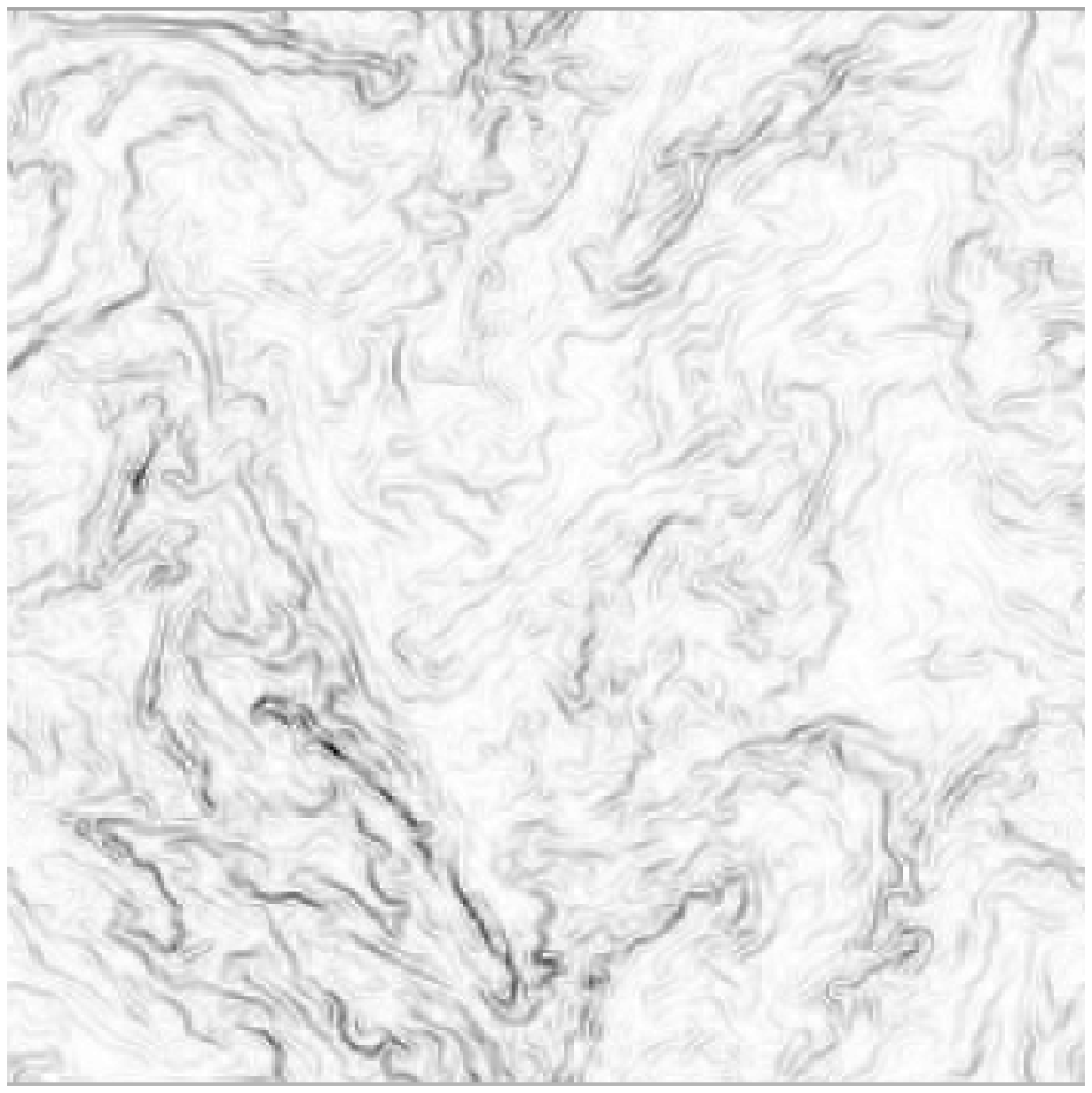}{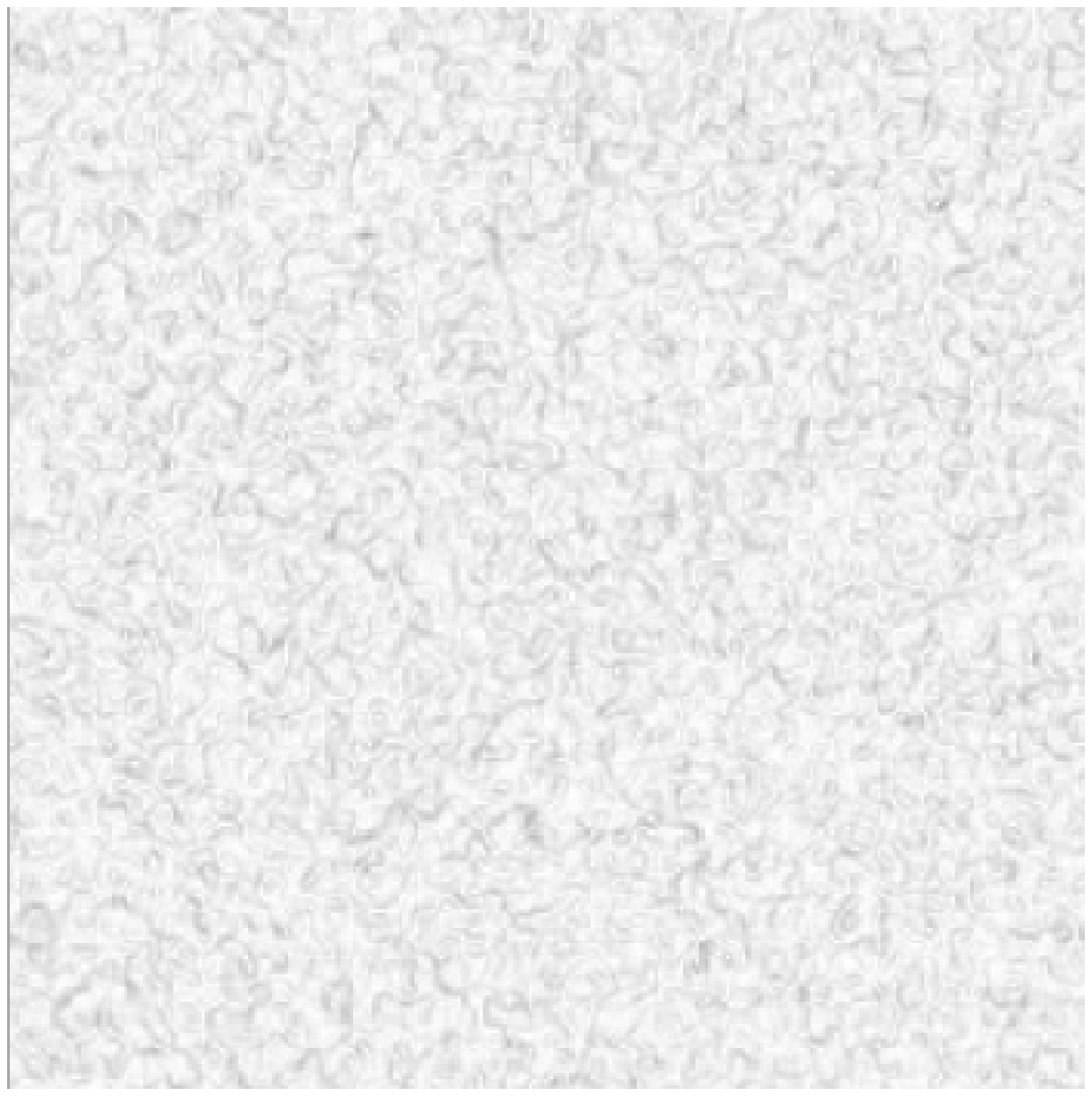}
\caption{\it Passive Scalar Gradient Magnitude. Passive scalar gradient 
magnitude, $|\nabla c|$, from our highest 
resolution, $256\times 256\times 512$, simulation F5. The image plane is 
$z=0$. Simulation data and its random phase transform are plotted in the left 
and right hand panels, respectively.
\label{fig:scalar.grad}}
\end{figure}

Intermittency also characterizes the concentration of the passive scalar.
In the left and right hand panels of Figure \ref{fig:scalar.grad}, we plot the 
magnitude of the gradient of the passive scalar computed in our highest 
resolution simulation F5. The contrast between the simulation and random
phase data is striking. Coherent structures which are prominent in the former 
are absent from the latter. 

%-----------------------------------------------------------------

\clearpage

%-----------------------------------------------------------------------

\section{DISCUSSION OF RESULTS} 
\label{sec:discussion}

\subsection{Comparison With GS Model}

The GSI model for the inertial range cascade of strong MHD turbulence is based 
on two assumptions.
\begin{enumerate} 
\item Energy transfer is local in wavenumber space.
\item Linear and nonlinear time scales maintain near equality. 
\end{enumerate}
These assumptions lead to two predictions.
\begin{enumerate}
\item The 1D energy spectrum $E(\bk_\perp)\propto k_\perp^{-5/3}$. 
\item The cascade is anisotropic with energy confined within a cone 
$k_\parallel\propto k_\perp^{2/3}$.
\end{enumerate}
Results from simulations presented in \S \ref{subsec:powerspec} and \S 
\ref{subsec:structure} agree with some aspects of the GS model and differ with 
others. An analysis of the reality and meaning of the departures from the GS 
scalings is presented in \S \ref{subsec:comparison}.

\subsubsection{Power Spectra and Structure Functions}

The 1D power spectra displayed in Figures \ref{fig:1Dpsave} and 
\ref{fig:1Dps117} exhibit inertial range slopes, $m_{\rm ps}$, closer to $-3/2$ 
than to the $-5/3$ predicted in GSI. This is consistent with the slopes of
the transverse structure functions $m_{\rm sf}$ shown in Figure 
\ref{fig:structure} being close to $1/2$. Since power spectra and structure 
functions are related by Fourier transforms, these slopes should satisfy $m_{\rm 
ps}+m_{\rm sf}=-1$.

A clear increase of anisotropy with decreasing scale is demonstrated in Figure 
\ref{fig:anisotropy}. It is consistent with the prediction by GSI that 
$\lambda_\parallel\propto \lambda_\perp^{2/3}$. \citet{bib:Cho} contains the 
initial confirmation of this relation.
 
\subsubsection{Critical Balance}

Equality of linear and nonlinear time scales, also known as critical balance,
predicts that $\lambda_\parallel/v_A\approx \lambda_\perp/v_{\lambda_\perp}$.
Figure \ref{fig:chi} shows that the ratio
$\chi=(\lambda_\parallel v_{\lambda_\perp})/(\lambda_\perp v_A)$ maintains a
value near unity throughout the inertial range as predicted in GSI.
However, there is a marginal problem 
of consistency. Together, $\lambda_\parallel\propto \lambda_\perp^{2/3}$ and 
$\chi=\,$constant imply $v_{\lambda_\perp}\propto \lambda_\perp^{1/3}$. But the
transverse structure function from which we obtain $v_{\lambda_\perp}$ to use
in forming $\chi$ yields $v_{\lambda_\perp}\propto \lambda_\perp^{1/4}$. 

\subsubsection{Cascade Times}
\label{subsubsec:CascadeTime}

Two measures of the cascade time are plotted against $k_\perp$ in Figure 
\ref{fig:cascadetime}; $t_c\sim \lambda_\perp/v_\perp$ and $t_h\sim 
v_\perp^2/\epsilon$. Each exhibits a power law dependence upon $k_\perp$ with
the former's slope being steeper than the latter's. For $v_\perp\propto 
\lambda_\perp^{1/3}$, both $t_c$ and $t_h$ would be proportional to 
$\lambda_\perp^{-2/3}$. However, they are not. Clearly, $v_\perp\propto 
\lambda_\perp^{1/4}$, which yields $t_c\propto \lambda_\perp^{3/4}$ and 
$t_h\propto\lambda_\perp^{1/2}$, provides a better, although still imperfect
fit.  A speculative explanation for the difference between $t_c$ and $t_h$
is offered in \S \ref{subsubsec:speculations}.
 
\subsection{Slow Modes}

\subsubsection{Their Passive Role In Cascade}

The passive role played by slow waves in nearly transverse MHD cascades is 
neatly illustrated by Figure \ref{fig:slowpassive}. GSII anticipates this 
behavior and offers a brief motivation. We provide an 
intuitive explanation in terms of field line geometry in \S 
\ref{subsubsec:geometry}. A mathematical derivation based on the equations of 
motion written in terms of Elsasser variables (eqn. [\ref{eq:defw}]) is 
outlined below. 
Consider the evolution of upward directed waves in a cascade whose anisotropy 
is measured by the scale dependent angle $\Theta\approx k_\parallel/k_\perp\ll 
1$. The nonlinear terms $\w_\downarrow\cdot\bnabla\w_\uparrow$ and $\bnabla P$ 
in equation 
\refnew{eq:elsasserup} are responsible for their cascade. For comparable 
magnitudes of slow and shear Alfv\'en waves, $\w_d\cdot\bnabla\w_u$ is smaller 
by a factor $\Theta$ if $\w_d\propto \s$ than if $\w_d\propto \ba$. Here $\s$ 
and $\ba$ are the unit polarization vectors of slow and shear Alfv\'en waves as
defined in equation \refnew{eq:polarization}. Since the $\w_d\cdot\bnabla\w_u$ 
term is the sole source of $P$, the same comparison applies to the $\bnabla P$
term. Note that these comparisons hold for both shear Alfv\'en and slow $\w_u$ 
waves. As they are independent of the degree of nonlinearity, they apply to
the intermediate MHD cascade as well as to the strong one. 

\subsubsection{Lack Of Conversion Of Shear Alfv\'en Waves To Slow Waves}

Figure \ref{fig:D3} demonstrates that the conversion of shear Alfv\'en waves 
to slow waves is of negligible significance in MHD cascades. GSI contains the 
original prediction. A modified version of the argument given there
is described below. It compares the rate at which slow waves are created in a 
balanced cascade composed entirely of shear Alfv\'en waves to the rate at which 
the shear Alfv\'en cascade. 

Our starting point is the Fourier transformed equation of motion for upward 
propagating waves written in terms of Elsasser variables
\begin{eqnarray}
&{}&\left({\partial\over \partial t} - i\omega(\bk)\right){\tilde 
\w}_\uparrow(\bk) =  
\nonumber \\ &-&{i\over 8\pi^3}\int
d^3k_1\,d^3k_2\left\{{\tilde \w}_\uparrow(\bk_1)-{\hat \bk}\left[{\hat 
\bk}\cdot{\tilde 
\w}_\uparrow(\bk_1)\right]\right\}\left[{\bf k}\cdot{\tilde 
\w}_\downarrow(\bk_2)\right]\delta(\bk_1 +\bk_2 - \bk) 
\label{eq:dtildewdt}
\end{eqnarray}
\noi where $\omega(\bk)=k_zv_A $ is the linear frequency of the shear Alfv\'en 
and slow waves. The rates of change of the amplitudes of slow and shear Alfv\'en 
waves with wavevector $\bk$ in a cascade of pure shear Alfv\'en waves are given 
by
\begin{eqnarray}
&{}&\left({\partial\over \partial t} - i\omega(\bk)\right){\tilde 
\bS}_\uparrow(\bk) =  
\nonumber \\ &-&{i\over 8\pi^3}\int
d^3k_1\,d^3k_2\left[{\hat \s}(\bk)\cdot{\tilde \A}_\uparrow(\bk_1)\right] 
\left[\bk_1\cdot{\tilde \A}_\downarrow(\bk_2)\right]\delta(\bk_1 +\bk_2 - \bk), 
\label{eq:dtildeSdt}
\end{eqnarray}
and
\begin{eqnarray}
&{}&\left({\partial\over \partial t} - i\omega(\bk)\right){\tilde 
\A}_\uparrow(\bk) =  
\nonumber \\ &-&{i\over 8\pi^3}\int
d^3k_1\,d^3k_2\left[{\hat \ba}(\bk)\cdot{\tilde \A}_\uparrow(\bk_1)\right] 
\left[\bk_1\cdot{\tilde \A}_\downarrow(\bk_2)\right]\delta(\bk_1 +\bk_2 - \bk). 
\label{eq:dtildeAdt}
\end{eqnarray}
Let us compare these two rates.\footnote{The net growth rate of shear Alfv\'en
waves vanishes in a steady state cascade. Restricting the integral to 
$k_1\leq k$ yields the rate at which the amplitude of $\A_\uparrow$ grows due to 
the 
cascading of longer ($k_1<k$) upward propagating shear Alfv\'en waves.} The 
magnitude of ${\hat \s}(\bk)\cdot{\tilde \w}_\uparrow(\bk_1)$ is smaller than 
that of 
${\hat \ba}(\bk)\cdot{\tilde \w}_\uparrow(\bk_1)$ by the scale dependent 
anisotropy 
factor $\Theta \sim k_z/k_\perp\ll 1$. Thus only a fraction 
$\Theta^2(k_\perp)\ll 1$ of the energy in shear Alfv\'en waves is converted into 
slow waves as the shear Alfv\'en waves cascade across $k_\perp$. This accounts 
for the negligible production of slow waves as shown in Figure \ref{fig:D3}.

\subsection{Dynamics Of Imbalance}

The proclivity of MHD cascades for imbalance is a consequence of nonlinear
interactions being restricted to collisions between oppositely directed waves.  

\subsubsection{Forced Turbulence}  

Large fluctuations are observed in the energies of different wave types in 
simulations of forced turbulence. Nevertheless, the imbalance appears to be
bounded. Figure \ref{fig:Evstforced} provides an excellent example of this
behavior.

A simple dynamical model suffices to capture the essence of imbalance in forced 
MHD turbulence. It consists of the coupled equations
\be
{dE_\uparrow\over dt}=-{E_\uparrow  E_\downarrow^{1/2}\over L}+\Lambda, \quad 
{\rm and} \quad 
{dE_\downarrow\over dt}=-{E_\downarrow  E_\uparrow^{1/2}\over L}+\Lambda.
\label{eq:Edot}
\ee
Here $E$ denotes the energy density of the shear Alfv\'en waves, $L$ the 
transverse outer scale, and $\Lambda$ the excitation rate. The equilibrium 
energy density and the nonlinear cascade time scale are defined by $E_{\rm 
eq}=(\Lambda L)^{2/3}$ and $t_c=L^{1/3}/\Lambda^{2/3}$. 

To investigate the stability of forced balanced cascade, we set
\be
E_\uparrow=E_{\rm eq}+\Delta E_\uparrow  \quad {\rm and} \quad 
E_\downarrow=E_{\rm eq}+\Delta E_\downarrow,
\ee
and then substitute these expressions into equations \refnew{eq:Edot} to 
obtain
\be
{d\Delta E_\uparrow\over dt}
=-{1\over 2t_c}\left(2\Delta E_\uparrow+\Delta E_\downarrow\right) \quad 
{\rm and} \quad {d\Delta E_\downarrow\over dt}=-{1\over 2t_c}\left(2\Delta 
E_\downarrow+\Delta 
E_\uparrow\right).
\label{eq:lineardEdt}
\ee
Assuming a time dependence proportional to $e^{st}$, we find eigenvalues
\be
s_1=-{1\over 2t_c} \quad {\rm and} \quad s_2=-{3\over 2t_c}.
\ee
This establishes the stability of the forced balanced cascade. It also shows
that fluctuations associated with the $s_1$ eigenmode decay rather slowly. 
These
characteristics accord well with the runs of Alfv\'en wave energy densities 
displayed in Figure \ref{fig:Evstforced}.

A more sophisticated analysis would include a proper statistical treatment of 
forcing and an investigation of the spectrum of fluctuations.  

\subsubsection{Decaying Turbulence}

Simulations of decaying MHD turbulence exhibit large imbalances. Thus a 
perturbation analysis is inappropriate. Fortunately, for $\Lambda=0$ equations 
\refnew{eq:Edot} admit an analytic solution. As is easy to verify by direct 
substitution, $\Delta E^{1/2}\equiv E_\uparrow^{1/2}-E_\downarrow^{1/2}=\Delta 
E_0^{1/2}$ is 
a constant, and 
that
\be
{d\over dt}\ln\left({E_\uparrow\over E_\downarrow}\right)={\Delta E^{1/2}_0\over 
L}.
\label{eq:imbalance}
\ee
Thus imbalance grows exponentially in decaying turbulence. This accounts 
qualitatively for the behavior seen in Figure \ref{fig:Evstdecay}. 

\citet{bib:Dobro} propose the growth of imbalance in decaying 
MHD turbulence as an explanation for the fact that the preponderance of shear 
Alfv\'en waves in the solar wind propagate outward along the interplanetary 
magnetic field. Support for this proposal is provided by simulations described 
in \citep{bib:Pouquet}. 

\subsubsection{Axial Asymmetry}

Axial asymmetry refers to a net polarization of Shear Alfv\'en waves. This can 
occur even in a balanced cascade. MHD cascades have a tendency to develop axial 
asymmetry because the strength of nonlinear interaction between oppositely 
directed shear Alfv\'en waves 1 \& 2 is proportional to ${\hat 
\ba(\bk_1)}\cdot{\hat \ba(\bk_2)}$ (see eqn. [\ref{eq:dtildeAdt}]). Thus the 
interaction vanishes for parallel polarizations and is strongest for orthogonal 
polarizations. 

Decaying turbulence is unstable to the growth of axial asymmetry. Waves with
the subdominant polarization cascade more rapidly that those with the dominate 
polarization. Axial asymmetry is bounded in forced turbulence. However, 
correlated fluctuations occur across the inertial range within regions of 
spatial scale comparable to the outer scale since the cascade tends to preserve 
polarization alignment.     

\subsection{Intermittency}

Tubes of high vorticity, often referred to as worms, are prominent features in
simulations of hydrodynamic turbulence. Worms have diameters of order the
dissipation scale and lengths approaching the outer scale. They are thought to
form from the rolling up of vortex sheets. In spite of their prominence, worms 
do not affect the inertial range dynamics \citep{bib:Jimenez}. 

Coherent structures are also evident in MHD simulations. Examples are shown in
\S \ref{subsec:intermittency} where the magnitudes of the curls of the dynamical 
fields and of the gradient of the passive scalar are plotted in $(x,y)$ slices.      
These regions have narrow dimensions comparable to the dissipation scale and 
lengths approaching the outer scale $L_\perp$. In these respects they resemble 
worms. 
%pmg: I've modified the remainder of this paragraph.
We suspect that these structures are vortex sheets which extend along the $z$ 
axis and that the magnetic field prevents them from rolling up. Their 
correlation lengths along $z$ are unresolved. However, this is not a strong 
constraint since $k_{M_z}L_\perp\approx 0.3$. 

\subsection{Comparison With Previous Simulations}
\label{subsec:comparison}

\citet{bib:Shebalin} report the development of anisotropy in 
isotropically excited MHD. Their simulations are two-dimensional, with one axis 
parallel to the direction of the mean magnetic field. Thus they are composed 
entirely of slow waves. An isotropic distribution of slow waves will initiate an 
anisotropic cascade. However, nonlinear  interactions among slow waves weaken as 
the cascade becomes more transverse because their strength is proportional to 
coefficients such as $\bk_2\cdot \bS_\downarrow(\bk_1)$, and for $\bk\to 
\bk_\perp$, ${\hat 
\s}\to {\hat\z}$.\footnote{This remains true in 3D} Convincing demonstrations of 
the development of anisotropy in fully three-dimensional MHD simulations are 
presented in \citet{bib:Oughton} and in \citet{bib:Matthaeus}. Each of these 
papers provides evidence that anisotropy 
increases at smaller scales. Each also claims that up to a saturation limit, 
anisotropy is more pronounced the larger the ratio of mean to fluctuating 
magnetic field strength. Matthaeus et al. note that this latter trend is 
inconsistent with the scaling for anisotropy proposed by GSI. 

A analysis by \citet{bib:Cho} clears up the confusion 
regarding the scale dependence of anisotropy in MHD turbulence. Unlike previous 
workers, who measure anisotropy in coordinate systems fixed to the sides of 
their computational boxes, Cho and Vishniac compute anisotropy in local 
coordinate frames tied to the direction of the total magnetic field. We follow 
their technique of using structure functions computed in directions parallel and 
perpendicular to that of the local magnetic field (see \S 
\ref{subsubsec:anisotropy}). Both they and we find results that are 
consistent with the relation $\lambda_\parallel\propto \lambda_\perp^{2/3}$ 
proposed by GSI. However, the ratio of the fluctuating to mean field is
$\sim 0.5$ in their simulations and $\sim 0.01$ in ours.

\subsection{Departure From GS Scalings}
\label{subsec:Scalings}

The GS scalings lead to the unambiguous prediction of a scale dependent 
anisotropy, $k_\parallel\propto k_\perp^{2/3}$, and a 1D Kolmogorov spectrum, 
$E(k_\perp)\propto k_\perp^{-5/3}$. While our simulations are in accord with the 
former, they consistently indicate that the 1D power spectrum has an index 
closer to 3/2 than 5/3. 

We do not know whether our simulation method is producing an anomalous power law 
or whether we have discovered a feature of MHD turbulence that is not
incorporated in the GS scalings. We first examine several effects which might 
produce anomalous power laws in numerical simulations. Then we offer some 
speculations about the role of intermittency.  

\subsubsection{Forcing}

\citet{bib:Borue} attribute the $k^{-1.85}$ inertial range they find for a
simulation of hydrodynamic turbulence to temporal intermittency associated with 
forcing. We are unable to discern any difference in the inertial range slopes 
from our simulations of forced and decaying MHD turbulence. As an example, 
compare the shear Alfv\'en wave spectra from both forced simulation F1 and 
decaying simulation D3 which are plotted in Figure \ref{fig:D3}. 

\subsubsection{Dealiasing}

Energy loss in our simulations is due to a combination of dealiasing and
hyperviscosity as illustrated in Figure \ref{fig:energyloss} for F2, one of our 
intermediate resolution simulations. Their relative importance depends upon the 
parameter ${\overline\nu}(\pi N/2)^{1/3}$ defined in equation \refnew{eq:numin}.
With our choice of viscosity parameters, dealiasing contributes relatively less 
of the energy loss in our lowest resolution simulation F1 and relatively more in 
our highest resolution simulation F5. However, all of our simulations show 
similar inertial range slopes as a comparision of Figures \ref{fig:1Dpsave}, 
\ref{fig:1Dps117}, and \ref{fig:D3} verifies. If there is any significant 
difference, it is that F5, the highest resolution simulation, for which 
dealiasing is least important, has the steepest inertial range slope. 

\subsubsection{Hyperviscosity}

Hyperviscosity applied in hydrodynamic simulations is known to produce a 
spurious flattening of the 1D inertial range slope over a range of $k$ 
approaching the viscous cutoff \citep{bib:Borue}. A weaker form of this 
bottleneck effect is apparent in the simulations of MHD turbulence described by 
\citet{bib:Cho}. 
Both \citet{bib:Borue} and \citet{bib:Cho} employ an 8'th order hyperviscosity. 
Our simulations, which use a 4'th order hyperviscosity, show no indication of a
bottleneck effect. This can be seen from the absence of flattening of the
inertial range slope at high $k_\perp$ in Figures \ref{fig:1Dpsave} and 
\ref{fig:1Dps117}.

\begin{figure}

\plotone{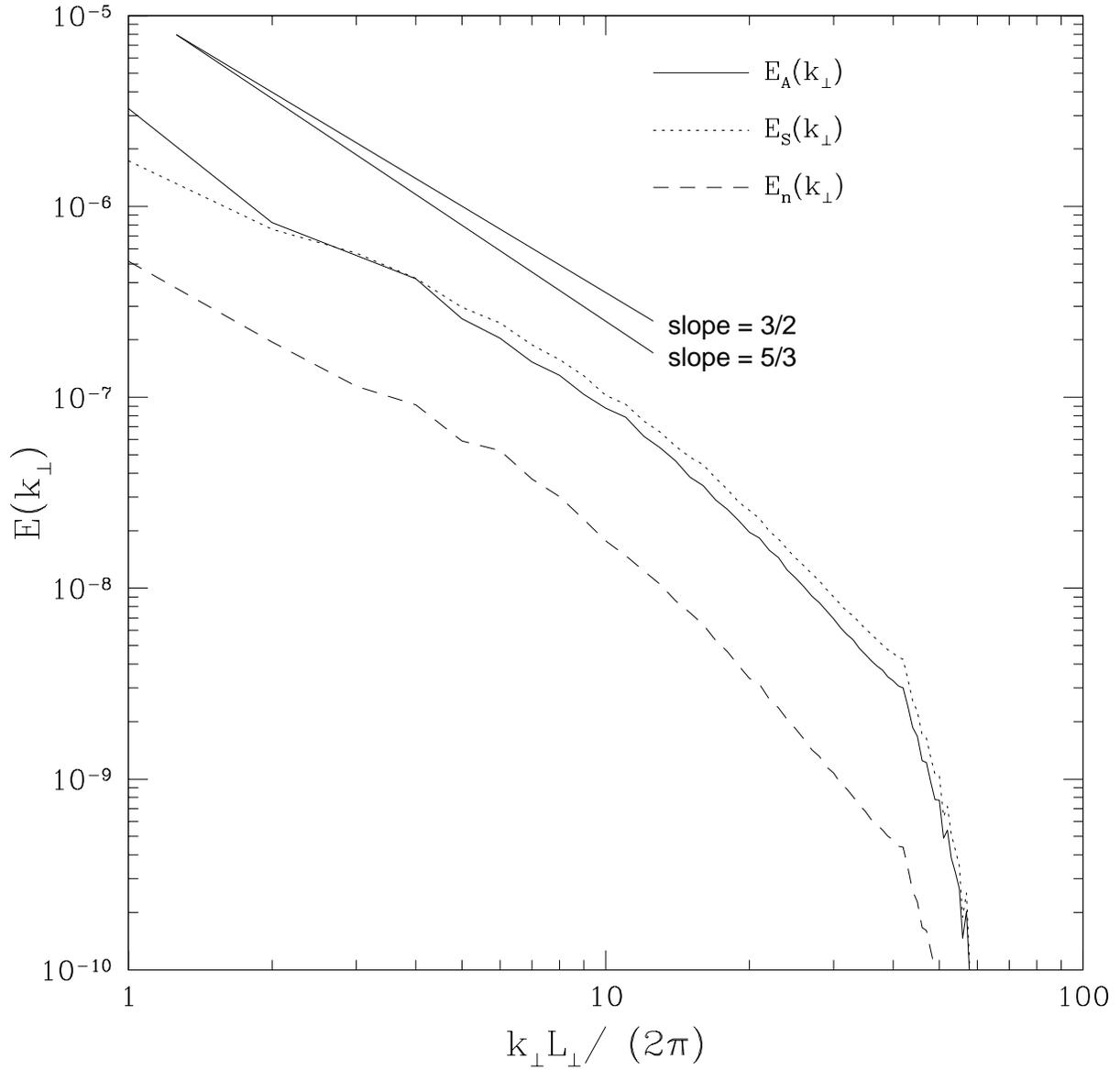} 
\caption{\it Inertial Range Computed With Ordinary Viscosity. Simulation F6 with
$\nu_1=2\times 10^{-4}$ and resolution $128\times 128\times 512$ is
similar to simulation F2 but uses ordinary viscosity instead of hyperviscosity.
\label{fig:F6}}
\end{figure}

Could hyperviscosity flatten the slope across the entire inertial range? 
\citet{bib:Muller} suggest that it does. They present the results of a $512^3$ 
simulation with nearly isotropic forcing which uses 1'st order (physical) 
viscosity. It exhibits an inertial range slope slightly steeper than 5/3. Then
they mention that a similar calculation done with 2'nd order hyperviscosity 
results in a flatter inertial range. To test the effect of hyperviscosity on the 
inertial range slope, we carry out
simulation F6 which uses physical viscosity but is otherwise similar to 
simulation F2. The 1D spectrum from this simulation is plotted in Figure 
\ref{fig:F6}. Its inertial range appears to have a slope closer to 3/2 than 5/3. 
This is not entirely conclusive because the inertial range is truncated at the
low $k_\perp$ end by forcing and at the high $k_\perp$ end by viscosity. Higher 
resolution simulations which are beyond our current computational resources 
are needed to settle this issue.

%-----------------------------------------------------------------

%-----------------------------------------------------------------

\subsubsection{Speculations}
\label{subsubsec:speculations}

For the moment, let us accept the shallow inertial range slope as a real
feature of MHD turbulence. How might it be accounted for? An intriguing 
possibility is that the nonlinear interactions responsible for the cascade 
become increasingly intermittent with decreasing scale. 

A given degree of spatial intermittency in the energy density is likely to have 
more serious consequences for the turbulent cascade in MHD than in HD. 
In HD energy cascades according to the local value of $\lambda/v$.
But in MHD nonlinear interactions are restricted to collisions between 
oppositely directed wave packets. Thus if the spatial filling factor of
the energy density, $f$, is small, that of the interactions, $f^2$, is smaller
still. 
This may account for the shallower slope of $t_h\sim v_\perp^2/\epsilon$ as 
compared to $t_c\sim \lambda_\perp/v_\perp$ seen
in Figure \ref{fig:cascadetime} and discussed in \S \ref{subsubsec:CascadeTime}.

\acknowledgments

Research reported in this paper was supported by NSF grant 94-14232 and a
NSF Graduate Research Fellowship held by JM. We thank Ben Chandran, Steve 
Cowley, Andrei Gruzinov, Reuben Krasnopolsky, Russell Kulsrud, and Yoram 
Lithwick for informative discussions. Our simulations were carried out on 
supercomputers operated by the helpful staff at the Caltech Center for Advanced 
Computing Resources. Edith Huang provided especially valuable advice on their 
use. 

\appendix

\section{APPENDIX}

\subsection{Spectral Notation}
\label{subsec:spectralnotation}

Cartesian coordinates are distinguished by Greek indices which run from $1-3$. 
Simulations are carried out in boxes whose sides have lengths $L_\alpha$ 
which are partitioned by $N_\alpha$ grid points. Integer coordinate components, 
$l_\alpha$, and integer wavevector components, $s_\alpha$, are defined
through the relations
\be
x_\alpha=\frac{L_\alpha}{N_\alpha}l_\alpha, \,\, {\rm where} \,\, 0\leq 
l_\alpha 
< N_\alpha,
\label{eq:defl}
\ee
and
\be 
k_\alpha=\frac{2\pi}{L_\alpha}s_\alpha, \,\, {\rm where} \,\, -{N_\alpha\over 
2}\leq s_\alpha \leq {N_\alpha\over 2}. 
\label{eq:defs}
\ee

The discrete Fourier transform in 1D is given by
\be
\tilde{q}(s) = \frac{1}{N} \sum_{l} q(l)e^{2\pi i sl/ N },
\label{eq:discreteFT}
\ee
where the tilde, $\tilde{}$, denotes Fourier transform. Generalization to 3D is 
trivial.

\subsection{Spectral Algorithm} 
\label{subsec:spectralalgorithm}

\subsubsection{Fourier Space Equations}
\label{subsubsection:Fourier}

We evolve the incompressible MHD equations in Fourier space where they
take the form \citep{bib:Lesieur} 
\begin{eqnarray} 
&
\partial_t \tilde{v}_\alpha
= -i k_\gamma \left( \delta_{\alpha\beta} - \frac{k_\alpha k_\beta}{k^2} 
\right)
  \left( \widetilde{v_\beta v}_\gamma - \widetilde{b_\beta b}_\gamma \right) - 
\nu_n k^{2n} \tilde{v}_\alpha& \label{eq:Fmomentum}, \\
&\partial_t \tilde{b}_\alpha = -i k_\beta
(\widetilde{v_\beta b}_\alpha - \widetilde{b_\beta v}_\alpha) - \nu_n k^{2n} 
\tilde{b}_\alpha, & 
\label{eq:Finduction}\\
& k_\alpha\tilde{v}_\alpha=0, &\label{eq:Fdivv}\\
& k_\alpha\tilde{b}_\alpha=0, &\label{eq:Fdivb}\\
&\partial_t \tilde{c} = -i k_\beta \widetilde{v_\beta c} - \nu_n
k^{2n} \tilde{c}. \label{eq:Fpassive}&
\end{eqnarray}

\subsubsection{Integration Method}
\label{subsubsection:integ}

Equations \refnew{eq:Fmomentum}, \refnew{eq:Finduction}, and
\refnew{eq:Fpassive} constitute a system of ordinary differential
equations with time as the dependent variable and the Fourier
coefficients $\{\tilde{v}_\alpha,\, \tilde{b}_\alpha,\, \tilde{c}\}$
as the independent variables. We employ a modified version of the
second order Runge-Kutta algorithm (RK2) to advance the variables in
time. First order algorithms are substantially less stable than RK2
at the same timestep. 

RK2 advances the variables across an interval $\Delta t$ in two stages. 
Derivatives evaluated at the initial time are used to compute trial values
of the variables at the midpoint $\Delta t/2$. Then derivatives computed at 
$\Delta t/2$ with these trial values are used to advance the variables from 
$t = 0$ to $\Delta t$.
In symbolic form
\be 
\tilde{q}_{\smf{trial}}(\Delta t / 2) = \tilde{q}(0) + 
\partial_t \tilde{q}(0)\Delta t/2 \label{eq:RK2trial}     
\ee
is followed by 
\be
\tilde{q}(\Delta t)=\tilde{q}(0)+\partial_t\tilde{q}(\Delta t/2)\Delta t, 
\label{eq:RK2final}
\ee
where $\partial_t\tilde{q}(\Delta t/2)\Delta t$ is evaluated using 
$\tilde{q}_{\smf{trial}}(\Delta t/2)$. Each stage involves a first order 
Euler (E1) step in which the derivative is taken to be constant. 

We make one departure from standard RK2 and treat diffusive terms with an 
integrating factor. Consider an equation of the form
\be
\partial_t \tilde{q}(k) = A - \nu_n k^{2n} \tilde{q}(k),
\label{eq:modeleq}
\ee
where $A$ comprises the non-diffusive terms.  Its solution, with $A$
constant throughout the interval $\Delta t$, is
\be
\tilde{q}(\Delta t) = \left[ \tilde{q}(0) + \frac{A}{\nu_n k^{2n}}
(e^{\nu_n k^{2n}\Delta t} - 1) \right] e^{-\nu_n k^{2n}\Delta t}
\label{eq:modifiedE1}
\ee
We use this expression in place of E1 in each stage of RK2. To 
lowest order in $\nu_n k^{2n}\Delta t$, equation \refnew{eq:modifiedE1}
reduces to E1. However, it has the advantage that it yields stable solutions 
to equation \refnew{eq:modeleq} with constant $A$ for arbitrary values of 
$\nu_n k^{2n}\Delta t$ whereas E1 yields unstable solutions for
$\nu_n k^{2n}\Delta t>2$. 

\subsubsection{Dealiasing}
\label{subsubsection:dealiasing}

Bilinear terms in equations \refnew{eq:Fmomentum}, 
\refnew{eq:Finduction}, and \refnew{eq:Fpassive} are calculated by 
transforming the individual fields to real space, carrying out 
the appropriate multiplications there, and then transforming the products 
back to Fourier space. This requires $N_1N_2N_3\log N_1\log N_2\log N_3$ 
operations using the Fast 
Fourier Transform (FFT) algorithm; $(N_1N_2N_3)^2$ 
operations would be needed to carry out the equivalent convolution 
in Fourier space. 

This economy comes at the price of either a 1/3 reduction in 
resolution or an aliasing error \citep{bib:Canuto}. To appreciate this, 
consider the 1D product
\begin{eqnarray}
\widetilde{pq}(s) & = \frac{1}{N} \sum_{l} \left[
\sum_{s^\prime} \tilde{p}(s^\prime) e^{-2\pi is^\prime l/N}
\sum_{s^{\prime\prime}} \tilde{q}(s^{\prime\prime})
e^{-2\pi is^{\prime\prime} l/N}
\right] e^{2\pi isl/N} \nonumber \\
& = \frac{1}{N}
\sum_{s^\prime}\sum_{s^{\prime\prime}}
\tilde{p}(\s^\prime)\tilde{q}(s^{\prime\prime})
e^{2\pi i (s^\prime + s^{\prime\prime}) l/N}
\delta_{s, s^\prime + s^{\prime\prime} + mN},
\end{eqnarray}
where $m$ is any integer. The $m=0$ terms comprise the convolution, and the
remainder the aliasing error.  To avoid the aliasing error, we set all
Fourier components with $|s| > N/3$ to zero both before we compute the real
space fields and again after we return the bilinear terms to Fourier space. 
Truncation ensures that Fourier components of bilinear terms 
with $m\ne 0$ vanish. Its cost is the reduction of the effective spatial 
resolution from $N$ to $2N/3$.

\subsection{Tests Of The Spectral Code}

Time derivatives of field quantities computed with the spectral code
agree with those obtained from a finite difference program with an
elliptic incompressible pressure operator. Although the latter is
unstable, it offers an independent method for computing time
derivatives.

The code preserves the solinoidal character of $\vb   $ and $\bb$. To machine 
accuracy it returns $\bk\cdot\partial_t \tilde{\vb}(\bk) = 0$ and $\bk\cdot 
\partial_t \tilde{\bb}(\bk) = 0$. It also conserves energy. Provided 
$\nu_n=0$, it yields $\partial_t E = \sum_\bk \tilde{\vb}(\bk) \cdot \partial_t 
\tilde{\vb}(\bk) + \tilde{\bb}(\bk) \cdot \partial_t \tilde{\bb}(\bk) = 0$, 
again
to machine accuracy.  

Harmonic Alfv\'en waves are evolved by our spectral code in a manner consistent 
with their analytic dispersion relation.

Results obtained from a simulation of decaying hydrodynamic turbulence (Z45) run 
with our code agree with those from a more thorough simulation by 
\citet{bib:Jimenez}. Our simulation is carried out in a cubic box with $L=1.0$, 
has resolution $256^3$, kinematic viscosity $\nu=8\cdot10^{-4}$, timestep 
$\Delta t=2.5\times 10^{-4}$, and is initialized with rms velocity $v=1.0$. We 
compute components of the velocity gradient longitudinal, $\nabla_\parallel 
v_\parallel$, and transverse, $\nabla_\perp v_\parallel$, to $\vb   $ at each 
point 
in our computational box. Distribution functions, $PF_q(x)$, of each quantity, 
$q$, are compiled and moments calculated according to 
\begin{equation}
M_n = \frac{\int_{-\infty}^{\infty} x^n PF_q(x)}
{\left[ \int_{-\infty}^{\infty} x^2 PF_q(x) \right]^{n/2}}.
\label{eq:moments}
\end{equation}
These are shown in Table \ref{tab:tests}.

\medskip

\begin{deluxetable}{llllllll}
\tabletypesize{}
\tablecaption{Tests of Spectral Code \label{tab:tests}}
\tablewidth{0pt}
\tablehead{
%{} & \colhead{Simulation Z45} & {} & {} & \colhead{Jiminez et al.} & {} & {} & 
%\colhead{Gaussian}
& \multicolumn{3}{l}{Simulation Z45} &
\multicolumn{3}{l}{Jiminez et al.} & Gaussian }
\startdata
$n$ \hspace{3mm}& $v$   & $\nabla_\parallel v_\parallel$ & $\nabla_\perp 
v_\parallel$ \hspace{6mm}& $v$   & $\nabla_\parallel v_\parallel$ & 
$\nabla_\perp 
v_\parallel$  \hspace{4mm}& {}\\ 
3           & 0   & -.43 & 0     & 0    & -.50 & 0    & 0 \\
4           & 1.6 & 4.4  & 5.9   & 2.8  & 4.6  & 6.19 & 3 \\
5           & 0   & -6.5 & 0     & 0    & -8.0 & 0    & 0 \\
6           & 3.4 & 48   & 102   & 13.0 & 55   & 110  & 15 
\enddata
\end{deluxetable}

\medskip

\noindent Because our simulation is of decaying turbulence whereas
that of Jiminez et al. is forced, appropriate comparisons are restricted to 
inner scale quantities derived from components of
$\bnabla \vb$ and exclude outer scale quantities derived from
components $\vb$. With this proviso, our results are in
satisfactory agreement with theirs.

\subsection{Catalog Of Simulations}
\label{subsec:catalog}

\begin{deluxetable}{lllll}
\tabletypesize{}
\tablecaption{Simulation Parameters \label{tab:simulationparameters}}
\tablewidth{0pt}
\tablehead{
\colhead{ID} &
\colhead{$(N_\perp, N_z)$} &
\colhead{$\Delta t$} &
\colhead{$\nu_4$} & 
\colhead{Comments}
}
\startdata
F1&$ 64,256$&$4\times 10^{-4}$&$5\times10^{-37}$&${\cal P} = 2\times10^{-5}$\\
F2&$128,512$&$3\times 10^{-4} $&$5\times10^{-40}$&${\cal P} = 
2\times10^{-5}$\\
F3&$128,512$&$3\times 10^{-4}$&$5\times10^{-43}$& ${\cal P} = 
2\times10^{-5}$\\
F4&$128,512$&$3\times 10^{-4}$&$5\times10^{-43}$& ${\cal P} = 
2\times10^{-5}$\\
F5&$256,512$&$1.5\times 10^{-4}$&$5\times10^{-43}$&${\cal P} = 
2\times10^{-5}$\\
D1&$ 64,256$&$4\times 10^{-4}$&$5\times10^{-40}$& ${\cal P}=0$\\
D2&$ 64,256$&$4\times 10^{-4}$&$5\times10^{-37}$&${\cal P}=0$, 
$A\uparrow+S\downarrow$\\
D3&$ 64,256$&$4\times 10^{-4} $&$5\times10^{-37}$&${\cal P}=0$, 
$A\uparrow+A\downarrow$ \\
D4&$128,512$&$3\times 10^{-4}$&$5\times10^{-40}$&$2 < s_\perp < 4$ \\
D5&$128,512$&$3\times 10^{-4}$&$5\times10^{-40}$&$4 < s_\perp < 8$ \\
D6&$128,512$&$3\times 10^{-4}$&$5\times10^{-40}$&$8 < s_\perp < 16$\\
D7&$128,512$&$3\times 10^{-4}$&$5\times10^{-40}$&$16 < s_\perp < 32$\\
D8&$256,512$&$1.5\times 10^{-4}$&$5\times10^{-43}$&$16 < s_\perp < 32$\\
D9&$256,512$&$1.5\times 10^{-4}$&$5\times10^{-43}$&$32 < s_\perp < 64$\\
\enddata
\end{deluxetable}

\subsubsection{Simulations of Forced Turbulence}

Our basic simulations include forcing at a total power per unit mass of 
${\cal P}=2\times 10^{-5}$. Recall that $\rho=1$ and $v_A=1$. Statistically 
equal power 
is input into shear Alfv\'en and slow waves propagating in opposite directions 
along the magnetic field. Higher resolution simulations run for shorter times 
are initialized by refining lower resolution simulations run for longer times. 

Our sequence of forced simulations begins with {\bf F1} which has
resolution $64\times 64\times 256$ and runs up to $t=6.6$. Initial
condition for simulations {\bf F2}, {\bf F3}, {\bf F4} with
resolutions $128\times 128 \times 512$ are drawn from {\bf F1} at
$t=2.4,\, 4.7$, and $6.6$, respectively. These are times at which the
fluxes of oppositely directed shear Alfv\'en waves in {\bf F1}
nearly balance. The refined simulations run for an additional $\Delta
t=0.4$, long enough for small scale structure to develop up to the
dealiasing cutoff. Our highest resolution, $256\times 256\times 512$,
simulation {\bf F5} is initialized from {\bf F2} at $t=2.8$ and
run until $t=2.95$.

\subsubsection{Simulations of Decaying Turbulence}

Our simulations of decaying turbulence are designed to test specific
properties of the MHD cascades. Simulation {\bf D1} continues {\bf
F1} without forcing from t=2.8 to t=9.9. Simulations {\bf D2} and
{\bf D3} are initialized from {\bf F1} at $t=6.6$, the former by
removing the Alfv\'en down and slow up waves and the latter by
removing all slow waves. A series of simulations are initialized from
forced simulations by removing all upward propagating waves outside a
specified band while leaving the down modes unchanged. Simulations
{\bf D4}, {\bf D5}, {\bf D6}, and {\bf D7} are each
initialized from {\bf F2} at t=2.8 with the up modes band-filtered
from $2 \leq s_\perp \leq 4$, $4 \leq s_\perp \leq 8$, $8\leq s_\perp
\leq 16$, and $16 \leq s_\perp \leq 32$, respectively. Likewise,
simulations {\bf D8} and {\bf D9} are initialized from {\bf F5}
at t=2.95 with up modes band-filtered from $16 \leq s_\perp \leq 32$
and $32 \leq s_\perp \leq 64$, respectively. The former has the same
up band as {\bf D7} but twice the transverse resolution.


\begin{thebibliography}{}

\bibitem[Berezinskii(1990)]{bib:Berezinskii}
  Berezinskii, V.S. 1990, Astrophysics of Cosmic Rays(Amsterdam:North-Holland)

\bibitem[Blandford \& Eichler(1987)]{bib:Blandford}
  Blandford, R.D. \& Eichler, D. 1987, Phys. Rep. 154, 1

\bibitem[Borue \& Orsag(1994)]{bib:Borue}
  Borue, V \& Orszag, S.Z. 1995, Europhys. Lett., 29, 687 
  
\bibitem[Canuto(1988)]{bib:Canuto} Canuto, C. 1988, Spectral Methods In Fluid 	 
  Dynamics (Berlin:Springer-Verlag)

\bibitem[Chandran(2000)]{bib:Chandran}
  Chandran, B.D.G. 2000, Phys. Rev. Lett., 85, 4656

\bibitem[Cho \& Vishniac(2000)]{bib:Cho}
  Cho, J., \& Vishniac, E.T. 2000, ApJ, 539, 273    

\bibitem[Dobrowolny, Mangeney, \& Veltri(1980)]{bib:Dobro}
  Dobrowolny, M., Mangeney, A., \& Veltri, P. 1980, Phys. Rev. Lett., 45, 144

\bibitem[Goldreich \& Sridhar(1995)]{bib:GSI}
  Goldreich, P., \& Sridhar, S. 1995, ApJ, 438, 763

\bibitem[Goldreich \& Sridhar(1997)]{bib:GSII}
  Goldreich P., \& Sridhar, S. 1997, ApJ, 485, 680

\bibitem[Gruzinov(2000)]{bib:Gruzinov}
  Gruzinov, A. 2000, private communication

\bibitem[Higdon(1984)]{bib:Higdon}
Higdon, J.C. 1984, ApJ, 285, 109

\bibitem[Iroshnikov(1963)]{bib:Iroshnikov} 
  Iroshnikov, P.S. 1963, AZh, 40, 742

\bibitem[Jimenez, Wray, \& Saffman(1993)]{bib:Jimenez}
  Jimenez, J., Wray, A.A., \& Saffman, P.G. 1993, J. Fluid Mech., 255, 65

%\bibitem[Kinney \& McWilliams(1997)]{bib:kw}
%  Kinney, R., \& McWilliams, J.C. 1997, J. Plasma Phys., 57, 73

\bibitem[Kolmogorov(1941)]{bib:Kolmogorov}
  Kolmogorov, A.N. 1941, Dokl. Akad. Nauk SSSR, 30, 9

\bibitem[Kraichnan(1965)]{bib:Kraichnan}
  Kraichnan, R.H. 1965, Phys. Fluids, 8, 1385

\bibitem[Lazarian \& Vishniac(1999)]{bib:Lazarian}
  Lazarian, A. \& Vishniac, E.T. 1999, ApJ, 517, 700 

\bibitem[Lesieur(1990)]{bib:Lesieur} 
  Lesieur, M. 1990, Turbulence In Fluids (Dordrecht:Kluwer)

\bibitem[Matthaeus et al.(1998)]{bib:Matthaeus}
  Matthaeus, W.H., Oughton, S., Ghosh, S., \& Hossain, M. 1998, Phy. Rev. Lett., 
81, 2056

\bibitem[Muller \& Biskamp(2000)]{bib:Muller}
  M$\ddot u$ller, W.C. \& Biskamp, D. 2000, Phys. Rev. Lett., 84, 475

\bibitem[Ng \& Bhattacharjee(1996)]{bib:NB96}
  Ng, C.S., \& Bhattacharjee, A. 1996, ApJ, 465, 845

\bibitem[Oughton, Priest, \& Matthaeus(1994)]{bib:Oughton}
  Oughton, S., Priest, E.R., \& Matthaeus, W.H. 1994, J. Fluid Mech., 280, 95

\bibitem[Parker(1979)]{bib:Parker}
  Parker, E.N. 1979, Cosmical Magnetic Fields (Oxford:Oxford Univ. Press) 

\bibitem[Pouguet, Meneguzzi, \& Frisch(1986)]{bib:Pouquet}
  Pouquet, A., Meneguzzi, M., \& Frisch, U. 1986, Phys. Rev. A, 33, 4266

\bibitem[Rickett(1990)]{bib:Rickett}
  Rickett, B.J. 1990, ARA\&A, 28, 561

\bibitem[Shebalin, Matthaeus, \& Montgomery(1983)]{bib:Shebalin}
 Shebalin, J.V., Matthaeus, W.H., \& Montgomery, D.J.J. 1983, J. Plasma Phys., 
29, 525 

\bibitem[Sridhar \& Goldreich(1994)]{bib:SG}
  Sridhar S., \& Goldreich, P. 1994, ApJ, 432, 612

\end{thebibliography}
\end{document}